\documentclass[12pt,a4paper]{report}
\usepackage[english]{babel}
\usepackage[utf8]{inputenc}
\usepackage{enumerate}
\usepackage[T1]{fontenc}
\usepackage{amsmath}

\newcommand{\Algo}[2]{\textsc{#1}(#2)}
\usepackage{multirow}
\usepackage{multicol}
\usepackage{booktabs}
\usepackage{bm}
\usepackage{ragged2e}
\usepackage{blindtext}
\usepackage{amsfonts}
\usepackage{relsize}
\usepackage{dsfont}
\usepackage{amssymb}
\usepackage{comment}
\usepackage{picins}
\usepackage{algorithm}
\usepackage[noend]{algpseudocode}
\algnewcommand{\endIf}{\State\algorithmicend\ \algorithmicif}

\algnewcommand{\endFor}{\State\algorithmicend\ \algorithmicfor}

\algnewcommand{\endFunction}{\State\algorithmicend\ \algorithmicfunction}
\algnewcommand{\endWhile}{\State\algorithmicend\ \algorithmicwhile}

\algnewcommand\algorithmicinput{\textbf{\hskip0.5em input:}}
\algnewcommand\Input{\item[\algorithmicinput]}
\algnewcommand\algorithmicglobal{\textbf{\hskip0.5em global:}}
\algnewcommand\Global{\item[\algorithmicglobal]}
\algnewcommand\algorithmicoutput{\textbf{\hskip0.5em output:}}
\algnewcommand\Output{\item[\algorithmicoutput]}
\algrenewcommand{\algorithmiccomment}[1]{\hskip3em $/^*$\textit{#1}$^*/$}
\makeatletter
\algnewcommand{\lComment}[1]{\Statex \hskip\ALG@thistlm $/^*$\textit{#1}$^*/$}
\makeatother
\algnewcommand{\algorithmicgoto}{\textbf{go to}}%
\algnewcommand{\Goto}[1]{\algorithmicgoto~\ref{#1}}%
\usepackage{etoolbox}
\usepackage[parfill]{parskip}
\usepackage[english]{babel}
\usepackage{datetime}
\usepackage{setspace}
\usepackage{lastpage}
\usepackage{abstract}
\usepackage{graphicx}
\usepackage{wrapfig}
\usepackage[symbol]{footmisc}
\usepackage[labelsep=quad,indention=10pt]{subfig}
\usepackage[nottoc,numbib]{tocbibind}

\usepackage[table,xcdraw]{xcolor}
\definecolor{dark-red}{rgb}{0.4,0.15,0.15}
\definecolor{dark-blue}{rgb}{0.15,0.15,0.4}
\definecolor{medium-blue}{rgb}{0,0,0.5}

\newtheorem{definition}{\textsc{Definition}}
\newtheorem{remark}{\textsc{Remark}}

\newenvironment{definition*}[1][\textsc{Definition}]{\begin{trivlist}
\item[\hskip \labelsep {\bfseries #1}]}{\end{trivlist}}
 
 \newcommand{\ra}[1]{\renewcommand{\arraystretch}{#1}}
\usepackage[left=4.00cm, right=2.50cm, top=2.50cm, bottom=2.50cm]{geometry}

\begin{document}
\newdateformat{mydate}{\THEDAY-\THEMONTH-\THEYEAR}
\thispagestyle{empty}
\begin{center}
\newgeometry{top=9cm}
\singlespacing
\textbf{{\Large EXPLOITING COPLANAR CLUSTERS TO ENHANCE 3D LOCALIZATION IN WIRELESS SENSOR NETWORKS}}
\vfill
ÇAĞIRICI, ONUR
\vfill
JANUARY 2015
\thispagestyle{empty}
\end{center}

\newpage

\begin{center}

\singlespacing
\textbf{{\Large EXPLOITING COPLANAR CLUSTERS TO ENHANCE 3D LOCALIZATION IN WIRELESS SENSOR NETWORKS}}
\vfill
\textsc{a thesis submitted to\\
the graduate school of\\
natural and applied sciences of\\
izmir university of economics}
\vfill
\textsc{by}\\
ÇAĞIRICI, ONUR
\vfill
\textsc{in partial fulfillment of the requirements\\
for the degree of\\
master of science\\
in the graduate school of natural and applied sciences}
\vfill
JANUARY 2015
\thispagestyle{empty}
\end{center}

\pagenumbering{roman}
\setcounter{page}{3}
\newpage
Approval of the Graduate School of Natural and Applied Sciences

\vspace{1cm}
\flushright \underline {\hspace{2in}}
\vspace{-0.5cm}
\flushright Prof. Dr. Cüneyt GÜZELİŞ
\vspace{-0.5cm}
\flushright Director \hspace{0.5in}

\justify I certify that this thesis satisfies all the requirements as a thesis
for the degree of Master of Science.
\vspace{1cm}
\flushright \underline {\hspace{2in}}
\vspace{-0.5cm}
\flushright Prof. Dr. Turhan TUNALI
\vspace{-0.5cm}
\flushright Head of Department \hspace{0.15in}

\justify We have read the dissertation entitled \textbf{Exploiting Coplanar Clusters to Enhance 3D Localization in Wireless Sensor Networks} completed by
\textbf{Onur ÇAĞIRICI} under supervision of \textbf{Assoc. Prof. Dr. Cem EVRENDİLEK} and \textbf{Assoc. Prof. Dr. Hüseyin AKCAN}
and we certify that in our opinion it is fully adequate, in scope and in
quality, as a dissertation for the degree of Master of Science.

\underline {\hspace{2.4in}} \hspace{0.5in} \underline {\hspace{2.7in}}
\vspace{-0.5cm}
\hspace{0.3cm} Assoc. Prof. Dr. Hüseyin AKCAN \hspace{0.5in} Assoc. Prof. Dr. Cem EVRENDİLEK
\flushleft \hspace{0.8in} Co-Supervisor \hspace{2.2in} Supervisor

\vspace{0.5cm}
\textbf{Examining Committee Members} \hspace{0.8in} Date: \underline {\hspace{1.5in}}

Assoc. Prof. Dr. Hüseyin AKCAN\\
Dept. of Software Engineering, IUE \hspace{0.92in} \underline {\hspace{2in}}

Asst. Prof. Dr. Ufuk ÇELİKKAN\\
Dept. of Software Engineering, IUE \hspace{0.92in} \underline {\hspace{2in}}

Assoc. Prof. Dr. Cem EVRENDİLEK\\ 
Dept. of Computer Engineering, IUE \hspace{0.84in} \underline {\hspace{2in}}

Asst. Prof. Dr. Burkay GENÇ\\ 
Inst. Of Population Studies, Hacettepe Uni. \hspace{0.32in} \underline {\hspace{2in}}

Prof. Dr. Brahim HNICH\\ 
Dept. of Computer Engineering, IUE \hspace{0.84in} \underline {\hspace{2in}}

\newpage

\topskip0pt
\vspace*{\fill}
\begin{spacing}{2}
\textbf{``Research is what I'm doing when I don't know what I'm doing. ''}
\end{spacing}
\begin{flushright}
\textemdash\emph{Wernher Von Braun}
\end{flushright}
\vspace*{\fill}

\chapter*{Acknowledgments}
\addcontentsline{toc}{chapter}{\numberline{
}Acknowledgments}%
\justify
During the preparation process of my thesis, I have had the privilege of working with a number of people who have made my time at the Izmir University of Economics enjoyable and rewarding.

First of all, I would like to thank to my advisor, Assoc. Prof. Dr. Cem Evrendilek and my co-advisor, Assoc. Prof. Dr. Hüseyin Akcan.
They put their trusts in me and accepted me as a researcher for the research project that they investigate, which is supported by The Scientific and Technological Research Council of Turkey (TUBITAK) Career Grant no: 112E099.
Without them, I could have not found such an interesting research topic for my thesis.

I am extremely grateful to my advisor, Dr. Evrendilek who has accepted me as his thesis student and gave me the chance to study under the supervision of a great person and outstanding computer scientist. 
He has patiently mentored me through the preparation of this dissertation.
Under his guidance, I was able to learn the fundamental lessons of being a researcher. I owe every second to him that he spent on revising my draft copies.

I am very thankful for the time from Asst. Prof. Dr. Gazihan Alankuş,
who helped me to develop one of the main parts of the algorithms suggested in this thesis.

I am also very thankful to Dr. Rasmus Fonseca, who has developed a great Java library, ProGAL, which I have used to implement the algorithms in this thesis. Even though we have  never met in person, he has generously spent time on answering my e-mails about my endless questions about the tool he has developed.

I would like to thank to my colleagues Berkehan Akçay and Serhat Uzunbayır, who have never hesitated to help me in during my hard times.

Finally, I would like to express my greatest thanks to my father Ufuk, my mother Aynur and my beloved Deniz Ağaoğlu for their great support.
They have endured me during the time I was the most unbearable and always stood by me regardless of my grumpinesses.
Without their endless love and support, I would have never completed this thesis.

\chapter*{\centering \Large \textbf{ABSTRACT}}
\addcontentsline{toc}{chapter}{\numberline{}Abstract}%

\begin{center}

\vfill
EXPLOITING COPLANAR CLUSTERS TO ENHANCE 3D LOCALIZATION IN WIRELESS SENSOR NETWORKS
\vfill
Çağırıcı, Onur
\vfill
M.Sc. in Intelligent Systems Engineering\\
Graduate School of Natural and Applied Sciences
\vfill
Supervisor: Assoc. Prof. Dr. Cem Evrendilek\\
Co-Supervisor: Assoc. Prof. Dr. Hüseyin Akcan
\linebreak
January 2015, \pageref{LastPage} pages.
\vfill
\end{center}

\justify
This thesis studies range-based WSN localization problem in 3D environments that induce coplanarity.
In most real-world applications, even though the environment is 3D, the grounded sensor nodes are usually deployed on 2D planar surfaces.
Examples of these surfaces include structures seen in both indoor (\textit{e.g.} floors, doors, walls, tables etc.) and outdoor (\textit{e.g.} mountains, valleys, hills etc.) environments.
In such environments, sensor nodes typically appear as coplanar node clusters.
We refer to this type of a deployment as a \textit{planar deployment}. 
When there is a planar deployment, the coplanarity causes difficulties to the traditional range-based multilateration algorithms because a node cannot be unambiguously localized if the distance measurements to that node are from coplanar nodes.
Thus, many already localized groups of nodes are rendered ineffective in the process just because they are coplanar.
We, therefore propose an algorithm called Coplanarity Based Localization (CBL) that can be used as an extension of any localization algorithm to avoid most flips caused by coplanarity.
CBL first performs a 2D localization among the nodes that are clustered on the same surface, and then finds the positions of these clusters in 3D.
We have carried out experiments using trilateration for 2D localization, and quadrilateration for 3D localization, and experimentally verified that exploiting the clustering information leads to a more precise localization than mere quadrilateration.
We also propose a heuristic to extract the clustering information in case it is not available, which is yet to be improved in the future.

\vfill
\textit{Keywords}: Range-based localization, wireless sensor network, WSN localization in 3D, NP-Hardness

\pagebreak
\chapter*{\centering \Large \textbf{ÖZ}}
\addcontentsline{toc}{chapter}{\numberline{}Öz}%
\vfill
\begin{center}
3B'DE KABLOSUZ ALGILAYICI AĞ KONUMLAMASININ İYİLEŞTİRİLMESİ İÇİN EŞDÜZLEMSEL KÜMELERİN KULLANILMASI
\vfill
Çağırıcı, Onur
\vfill
Akıllı Mühendislik Sistemleri, Yüksek Lisans\\
Fen Bilimleri Enstitüsü
\vfill
Tez Yöneticisi: Doç. Dr. Cem Evrendilek\\
İkinci Tez Yöneticisi: Doç. Dr. Hüseyin Akcan
\linebreak
Ocak 2015, \pageref{LastPage} sayfa
\vfill
\end{center}

\justify

\begin{small}
Bu tez, 3B'de mesafe ölçümüne dayalı kablosuz algılayıcı ağları (KSA) konumlama problemini, \hyphenation{eş-düz-lem-sel-li-ği} eşdüzlemselliği tetikleyen ortamlarda inceliyor.
Gerçek hayat uygulamalarının çoğunda, ortam 3B olmasına rağmen, uçmayan algılayıcılar 2B düzlemsel yüzeyler \hyphenation{ü-ze-rin-de} üzerinde dizilirler.
Bu yüzeyler iç mekan yüzeyleri (katlar, kapılar, duvarlar, masalar \textit{vb.}) olabileceği gibi, dış mekan yüzeyleri (dağlar, vadiler, bayırlar \textit{vb.}) de olabilir.
Bu tür ortamlarda algılayıcılar tipik olarak eşdüzlemsel kümeler halinde görünürler.
Bu tip dizilime \textit{düzlemsel dizilim} adını veriyoruz.
Düzlemsel dizilimin bulunduğu ortamlarda, \hyphenation{eş-düz-lem-sel-lik} eşdüzlemsellik geleneksel mesafe ölçümüne dayalı konumlama algoritmaları için zorluklar oluşturur çünkü bir düğüm, eşdüzlemsel düğümlerden elde edilen uzaklık ölçümleriyle muğlak olmayan bir şekilde konumlanamaz.
Böylece, hali hazırda konumlanmış birçok düğüm grupları, eşdüzlemsel oldukları için etkisiz hale gelirler.
Bu nedenle, düzlemsel konuşlanma olduğunu bildiğimiz durumlarda, bu güçlükle başa çıkmak için, Coplanarity Based Localization (CBL), Türkçe adıyla Eşdüzlemsellik Tabanlı Konumlama (ETK) adında bir algoritma sunuyoruz.
Sunduğumuz algoritma herhangi bir konumlama algoritmasının uzantısı olarak kullanılabilmektedir.
ETK, ilk olarak eşdüzlemsel yüzeylerde bulunan düğüm kümelerini, aynı kümedeki diğer düğümlere göre pozisyonlarını bulmak için bir 2B konumlama algoritması kullanır ve daha sonra kümelerin 3B'de yerlerini bulur.
2B konumlama algoritması olarak \textit{trilateration}'ı ve 3B konumlama algoritması olarak quadrilateration'u kullanarak yürüttüğümüz deneylerde de gördüğümüz üzere, kümelenme bilgisini kullanmak, salt \textit{quadrilateration}'dan daha doğru sonuç veren bir konumlamaya yol açıyor.
Kümelenme bilgisinin gelmediği durumda ise, eşdüzlemsel kümeleri keşfetmeye yönelik ve geliştirilmeye açık bir sezgisel de sunuyoruz.
\end{small}
\vfill
\textit{Anahtar Kelimeler}: Uzaklık-tabanlı konumlama, kablosuz algılayıcı ağları, 3B'de KAA konumlama, NP-Zorluk.


\tableofcontents
\listoffigures
\pagebreak
\restoregeometry
\pagestyle{plain}
\chapter[Introduction]{Introduction}  \label{chap:intro}
\begin{figure}[htbp]
\centering
\includegraphics[width=0.7\linewidth]{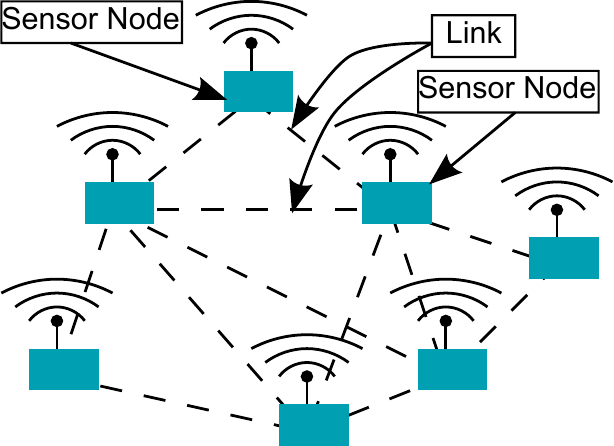}
\caption{A wireless sensor network}
\label{fig:WSNwEdgesLegend}
\end{figure}

\pagenumbering{arabic}
\justify
Recent  developments in technology pose a need for wireless sensor network (WSN) technologies to be used broadly \cite{sensornetworksurvey}.  A WSN consists of wireless sensor nodes that communicate among each other in order to complete a given task.
There are several application areas that use WSNs.
Zhong made a classification of these application areas in his thesis \cite{zhongthesis} namely, military tasks, industrial process monitoring and control, habitat and environment monitoring, health-care applications, home automation, and vehicle networks and intelligent transportation systems.
The sensor nodes usually communicate with each other to pass their gathered data to a destination. 
In Figure \ref{fig:WSNwEdgesLegend}, we see a wireless sensor network.
The rectangles represent wireless sensor nodes and the dashed lines between them represent their communication links.

Independent of the application area, the location information of wireless sensors in a WSN is crucial to improve the quality of service.
For all types of applications, the information of position can be appended to the data that is gathered from the sensor nodes, easing the traceability of the sensors, as in WSNs that works in geographical routing. 
The information of positions is able to increase the quality of the applications such as geographical based queries.
When equipped with proper devices, a sensor node can measure distance to another node in the network using the communication links shown in Figure \ref{fig:WSNwEdgesLegend}.
Finding the positions of these sensors is called \textit{WSN localization}.
If the sensor nodes are positioned only by using the distance measurements, this process is called \textit{range-based localization}. 
Even though there are  lots of studies on range-based WSN localization in 2D, \cite{zhongthesis,gpslesslowcost,dwrl,gpsfree,adhoc,changthesis,robustpositioning}, the volume of studies for 3D WSN localization is relatively small \cite{survey}.

\begin{figure}[htbp]
\centering
	\subfloat[Sensor nodes deployed onto a valley]{%
	\includegraphics[width=0.7\linewidth]{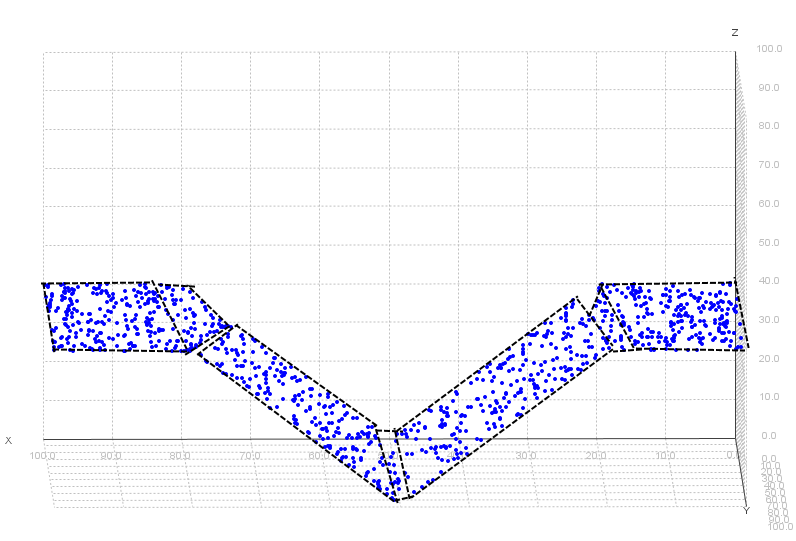}
	\label{fig:valley}%
	}
	
	\subfloat[Sensor nodes deployed onto a mountain]{%
	\includegraphics[width=0.7\linewidth]{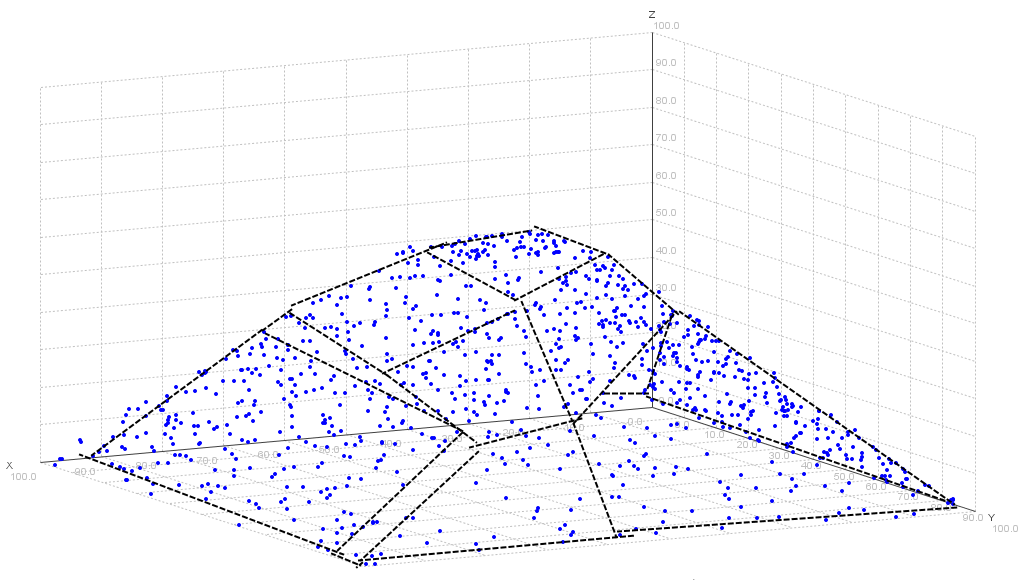}
	\label{fig:mountain}%
	}
	
	\subfloat[Sensor nodes deployed inside a building]{%
	\includegraphics[width=0.7\linewidth]{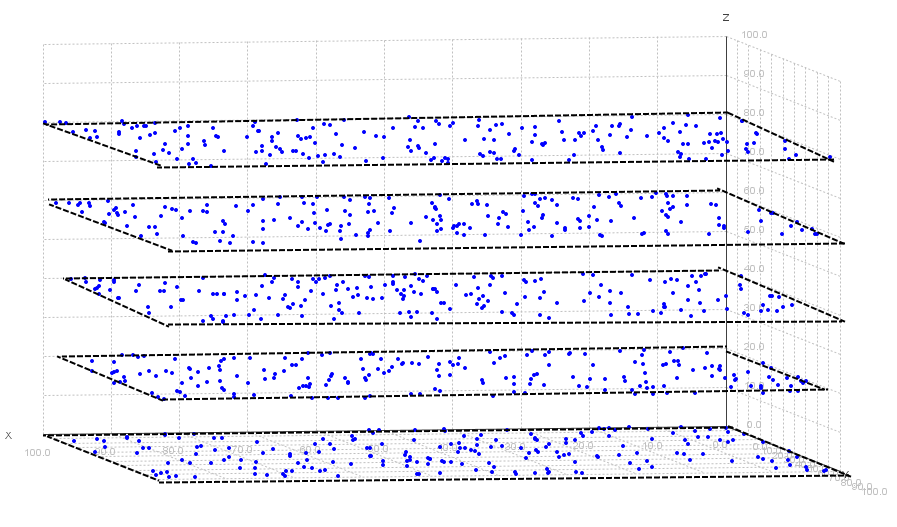}
	\label{fig:building}%
	}
\caption{Sensor nodes deployed onto a valley (a), onto a mountain (b), and inside a multi-storey building (c).}
\label{fig:planarDeployment}
\end{figure}

The physical world we live in, however, is a 3D environment. Therefore, many applications require localization in 3D.
Most of the scenarios in real-world WSN applications that need localization have usually deployments where the sensor nodes sit on planar regions to form sets of planar clusters as seen in Figure \ref{fig:planarDeployment}.
In Figure \ref{fig:valley}, there are 1000 sensor nodes deployed onto a valley.
The same number of sensors are deployed onto a mountain in Figure \ref{fig:mountain}, and inside a multi-storey building in Figure \ref{fig:building}.
It is hence observed that a method for exploiting the information of structural information in 3D is very much needed in order to improve the quality of localization achieved.
In this thesis, we study the range-based WSN localization problem in 3D environments in which sensors are deployed onto planar surfaces.

\section{Motivation}\label{sec:motivation}
Localization, in general, is defined as finding the relative coordinates of an object, with respect to a certain reference system.
The corresponding coordinates can be the relative coordinates in a room, coordinates in a building or global coordinates on the earth.
In order to find the position of an object, several techniques were developed.
A well-known method to localize an object on the earth globe is called Global Positioning System (GPS) \cite{gpsbook}.
GPS uses four or more orbiting satellites to localize an object on the earth.
In 2000, Benefon released the first cellular phone which uses an integrated GPS device \cite{benefonannual}.
With this development, GPS has become a widely-used system throughout the world.
However, there are major drawbacks of GPS, which can be listed as:

\begin{itemize}
\item The accuracy is relatively low for critical usage (up to 15 meters deviation).
\item The system has incremental costs.
\item The energy consumption of the system is excessive.
\item The system is nonfunctional in indoor environments. 
\end{itemize}

These reasons have paved the way for a localization method that works independent of GPS.
A considerable GPS-free localization system was designed by Bulusu et al. in 2000 \cite{gpslesslowcost}.
They use a frequency-based approach to localize ultra-wideband (UWB) radio nodes \cite{uwboverview}.
UWB radio nodes are able to measure distances among themselves using ultra-wide band signals \cite{uwbtutorial}. They can be used both for static localization when the nodes are immobile \cite{uwblocalization}, and dynamic localization when nodes are able to move \cite{changthesis}. This method was studied extensively, and more research is still being conducted on GPS-free localization \cite{gpsfree,dwrl,robustpositioning,techniques,evol,theory}.

In 2001, Bahl and Padmanabhan developed a very popular indoor positioning system (IPS) \cite{radar}.
The system is called RADAR (not to be confused with Radio Detection And Ranging \cite{airdefence}) and its working principle is similar to GPS but was designed for one floor in a building.
After RADAR was developed, several studies were made in this area \cite{ipssurvey,wirelessipssurvey}.
The system uses three \textit{base stations} in order to localize objects and track their movements.
Hence, the installment of the system is expensive for a multi-storey building.
Besides, power consumption of the base stations increases with the number of the objects being tracked.
Thus, developing an energy efficient solution becomes critical for the succesful operation of IPSs.

A well-known decentralized system, called a \textit{wireless sensor network} (WSN) \cite{fundamentalsofwsn} has been used in order to achieve a scalable, efficient and low-cost localization for indoor environments. WSN is a type of \textit{wireless ad hoc network} \cite{adhoc,adhocmobilewsn} and uses low-cost devices called \textit{wireless sensor nodes}.
Localization can be done using the wireless sensor nodes either by utilizing the pairwise distances among the sensor nodes, called range-based localization \cite{dwrl,radar,theory,changthesis} or just by using the connectivity information of the sensor nodes, called range-free localization \cite{zhongthesis,gpslesslowcost,military5}.

Range-based WSN localization is a method to obtain the relative positions of the sensor nodes with respect to a number of nodes with known positions by using the available pairwise distances among the sensor nodes.
While measuring distances, sensor nodes use methods such as time of arrival (TOA) \cite{adhoc}, time difference of arrival (TDOA) \cite{dynamicfinegrained}, or received signal strength (RSS) \cite{vander2009design}.
Although range-based WSN localization is proven to be an NP-Hard problem \cite{analysis}, Eren \textit{et. al.} \cite{rigidity} showed that, if certain conditions are satisfied, range-based localization can be done in polynomial time, using trilateration which uses three distance measurements to localize a node. However, when there are errors in distance measurements which is the case in real life, localization of a network by trilateration becomes intractable \cite{trilat}.
Despite being intractable, trilateration is still used frequently \cite{survey,dwrl,uwblocalization}.

\begin{figure}[h]
\centering
\includegraphics[width=\linewidth]{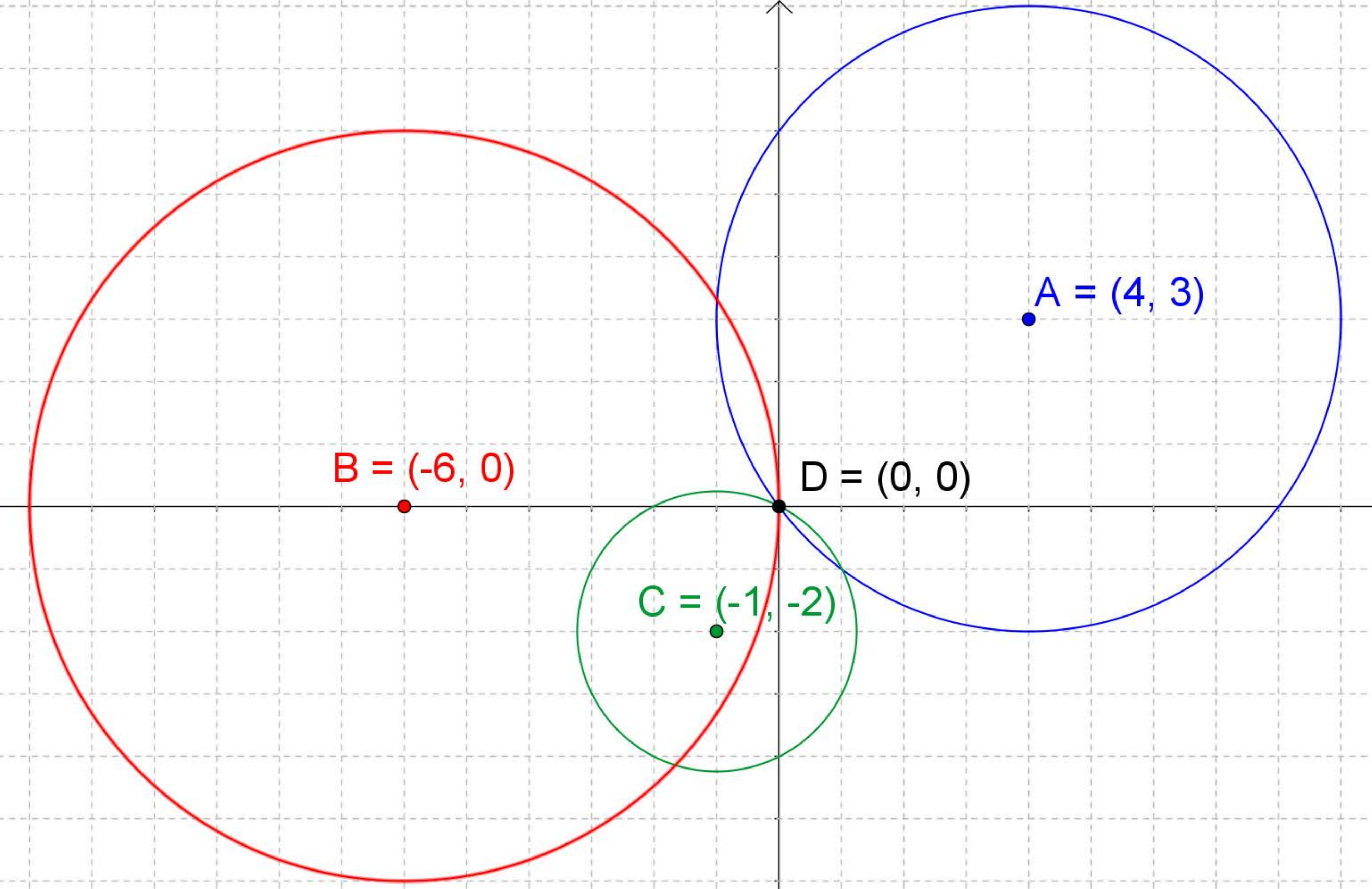}
\caption{A node being localized by three localized nodes in 2D}
\label{fig:accurateMeasurements}
\end{figure}

In Figure \ref{fig:accurateMeasurements}, we see a node $D$, being localized by three nodes $A$, $B$ and $C$.
The circles in the figure represent the distance between the unlocalized node and the localized nodes.
The coordinates of $D$ can be determined by computing the intersection point of these circles.

\begin{figure}[!h]
\centering
\includegraphics[width=0.9\linewidth]{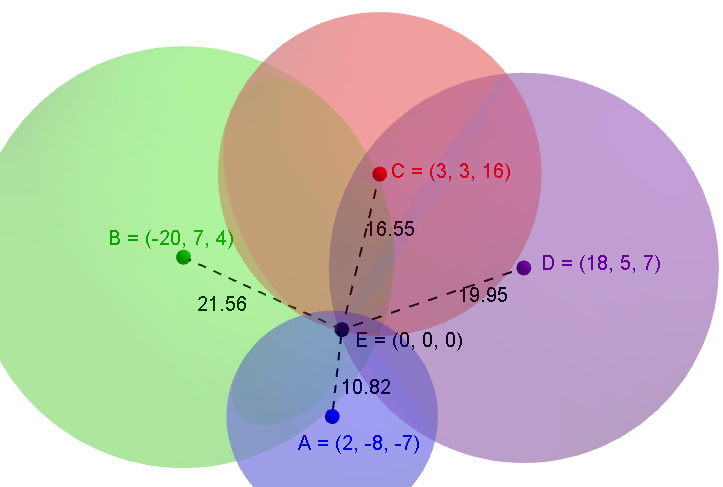}
\caption{A node being localized by four localized nodes in 3D}
\label{fig:4spheres}
\end{figure}

In 3D, quadrilateration can be used to localize WSN nodes. 
Analogous to trilateration in 2D, quadrilateration uses four distance measurements from four already localized non-coplanar nodes to localize a fifth node.
In Figure \ref{fig:4spheres}, we see the localization of an unlocalized node $E$, using the distance measurements from four localized nodes $A$, $B$, $C$ and $D$.

\begin{figure}[h]
\centering
\includegraphics[width=0.8\linewidth]{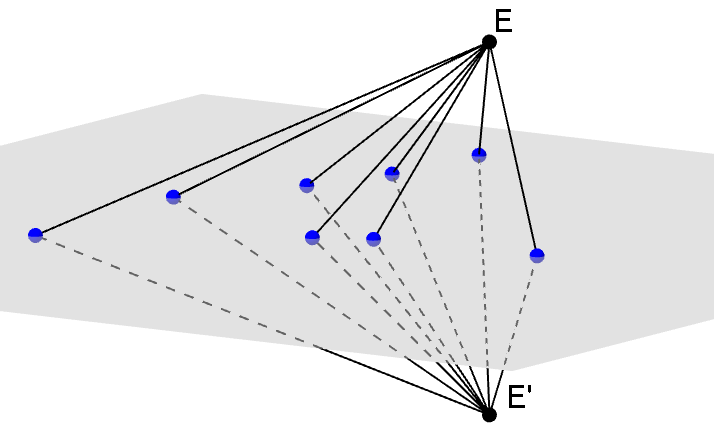}
\caption{The unlocalized node $E$ cannot be localized unambiguously because of coplanarity.}
\label{fig:coplanarityFail}
\end{figure}

When the localized nodes are on the same plane, \textit{i.e.} coplanar, however, no matter how many distance measurements are available, a node cannot be unambiguously located unless it is also on the same plane as the others.
In Figure \ref{fig:coplanarityFail}, there are many distance measurements from many localized nodes to $E$.
Since the localized nodes sit on the same plane, there are always two possible positions for the unlocalized node.
One is $E$, and the other is the reflection of $E$ about the plane that the others sit on, $E'$.
Since $E$ is not coplanar with the localized nodes, there is no way of telling $E'$ from $E$.

In most real-world applications, sensor nodes are deployed on planar surfaces.
Examples of these surfaces include structures seen in both indoor (\textit{e.g.} floors, doors, walls, tables etc.), and outdoor (\textit{e.g.} mountains, valleys, hills etc. recall Figure \ref{fig:planarDeployment}) environments.
In such environments, sensor nodes typically appear as coplanar node clusters.
We call this type of a deployment as a \textit{planar deployment}. 
We try to extract and utilize such information in order to achieve a better range-based WSN localization than the traditional range-based localization methods, such as trilateration \cite{rigidity}, quadrilateration \cite{techniques}, and multilateration \cite{dynamicfinegrained} when the environment is particularly known to follow a planar deployment.

\section{Contributions}\label{sec:contributions}

Our main contributions in this thesis are:

\begin{enumerate}
\item We show that when the nodes are deployed on planar surfaces in 3D, as is mostly the case in real-world applications, quadrilateration does not work as expected due to a massive amount of coplanar sensors.
\item We propose an algorithm to utilize the coplanarity, if the sensor nodes are known to follow a planar deployment pattern.
We call this algorithm Coplanarity Based Localization (CBL).
\item We show that CBL performs a more precise localization than mere quadrilateration when the information of coplanar node clusters are known apriori.
\item When the clustering information is unavailable, we propose a heuristic algorithm that tries to extract the coplanar clusters.
Although our first attempt to extract clustering information is yet to be improved, we also clearly highlight the need for an algorithm to extract such information accurately in 3D localization.
\end{enumerate}

\section{Organization of the Thesis}\label{sec:organization}
The organization of this thesis is as follows: In Chapter \ref{chap:intro}, we present the introduction.
In Chapter \ref{chap:background}, we give the basic background and knowledge for range-based wireless sensor network localization.
In Chapter \ref{chap:coplanaritybasedlocalization}, we present CBL, and a heuristic to extract coplanar clusters.
Finally in Chapter \ref{chap:conclusion}, we conclude this thesis.

\chapter{Background and Terminology} \label{chap:background}
In this chapter, we present basic background and terminology for WSN localization.

A WSN graph is a graph $G = (V,E)$, such that each vertex $v \in V$ corresponds to a sensor node.
We assume that each sensor node has the same sensing range, denoted by $R$, and an edge $(v,w) \in E$ exists if and only if $d(v,w) \leq R$ where $d(v,w)$ denotes the Euclidean distance between $v$ and $w$.
This type of a graph is called a \textit{unit-disk graph} in 2D, and a \textit{unit-ball graph} in 3D.

In our problem, a WSN graph can contain three types of nodes; namely seed, localized, and unlocalized nodes.
The coordinates of a \textit{seed node} is known apriori. 
The network is usually localized based upon the positions of such seed nodes.
If a node is not a seed node, but localized subsequently based onto the positions of the seed nodes, it is called a \textit{localized} node.
We say that a node is \textit{unlocalized} if its position is unknown.

Localizing a WSN graph $G$ means assigning a position to each vertex $v \in V$ in $d \in {2,3}$ dimensions, so that the distances given by the edge weights are all satisfied.
This operation is referred to as finding a \textit{point formation} of $G$ \cite{pointformationEren,pointformation}.
The point formation of a graph $G$ is denoted by $G_{\mathbb{F}}$.
For instance, let us consider a WSN graph with the vertex set $V = \{v_1,\dots,v_5\}$, and the edge set as given in Figure \ref{fig:edgeSet}.

\begin{figure}
\begin{center}
$d(v_1,v_2) = 2.00000$\\
$d(v_1,v_3) = 3.60555$\\
$d(v_1,v_4) = 3.00000$\\
$d(v_2,v_3) = 2.23606$\\
$d(v_2,v_4) = 3.60555$\\
$d(v_2,v_5) = 5.00000$\\
$d(v_3,v_4) = 3.16227$\\
$d(v_3,v_5) = 3.16227$\\
$d(v_4,v_5) = 2.82842$
\end{center}
\caption{An example edge set of a graph}
\label{fig:edgeSet}
\end{figure}

In Figure \ref{fig:pointFormation}, a possible point formation of the given graph in 2D is shown with respect to the edge set given in Figure \ref{fig:edgeSet}.

\begin{figure}[htbp]
\begin{displaymath}
{\mathbb{F}_G} =
\left([v_1, v_2, v_3, v_4, v_5],
 \begin{bmatrix}
  0, 0 \\
  0, 2 \\
  2, 3 \\
  3, 0 \\
  5, 2
  \end{bmatrix}\right)
\end{displaymath}
\caption{Point formation of the graph with the edge set given in \ref{fig:edgeSet}}
\label{fig:pointFormation}
\end{figure}

\begin{figure}[htbp]
\centering
\includegraphics[width=\linewidth]{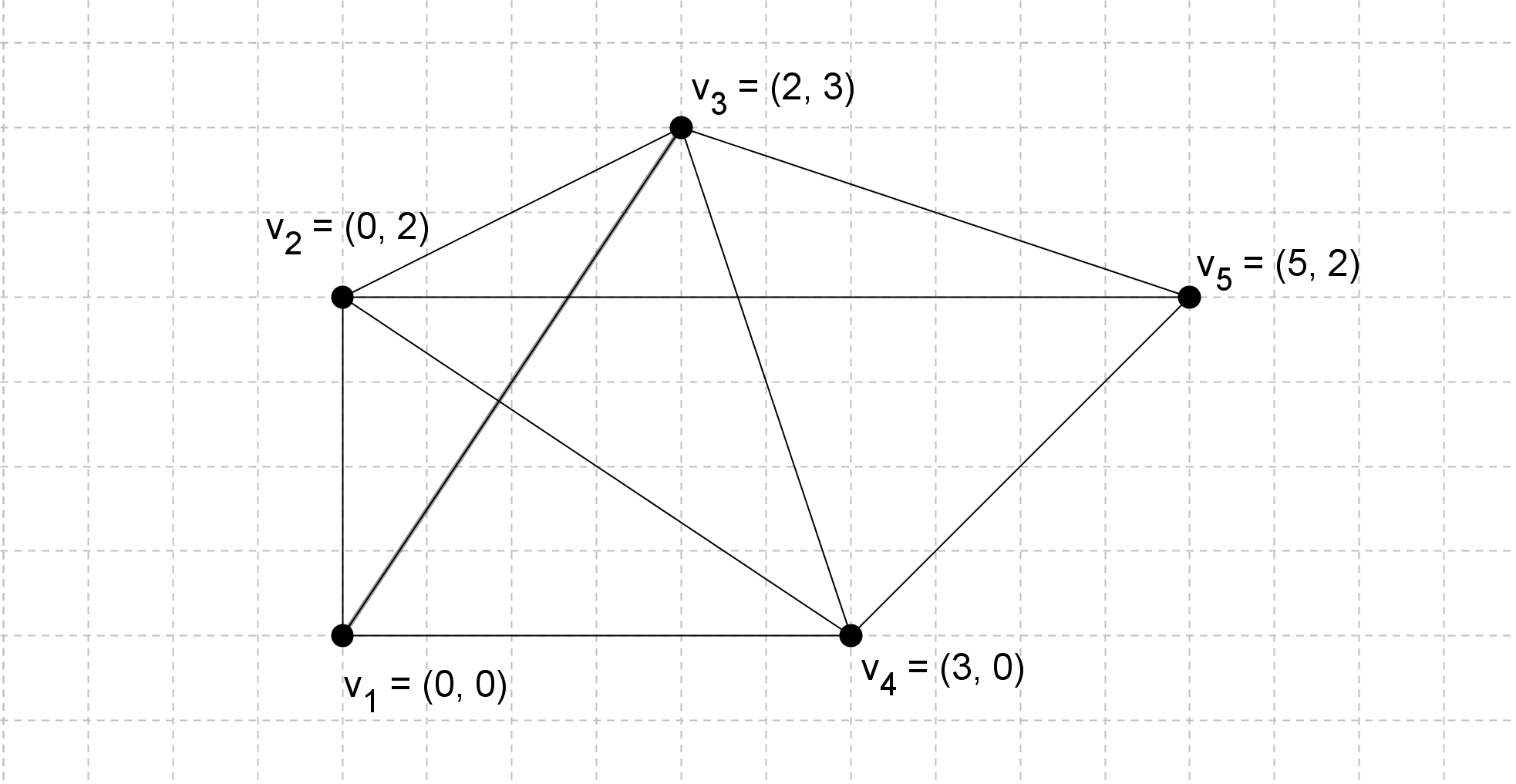}
\caption{A point formation of a graph with the edge set given in Figure \ref{fig:edgeSet}}
\label{fig:pointFormationVisual}
\end{figure}

In Figure \ref{fig:pointFormationVisual}, 2D coordinates are assigned to the vertices of the graph whose point formation is given in Figure \ref{fig:pointFormation}.

If we fix the positions of any three nodes in Figure \ref{fig:pointFormationVisual}, then unique positions can be assigned to the rest.
If there is a unique point formation of a graph $G$ in $d$ dimensions, then $G$ is called \textit{globally rigid} in $d$ dimensions.
Global rigidity is both sufficient and necessary for a graph to be localizable in 2D \cite{theory}, and can be tested in polynomial time \cite{conditionsforuniquegraph,connectedRigidity,connelly}.
Although global rigidity is a necessary condition is 3D as well \cite{laman}, the sufficiency for a graph to be localizable in 3D is still an open problem \cite{connelly}.
Saxe \cite{saxe} proved that localization is an NP-Hard problem for all dimensions even when the network is known to be localizable.
	
\begin{figure}[htbp]
\centering
\includegraphics[width=0.6\linewidth]{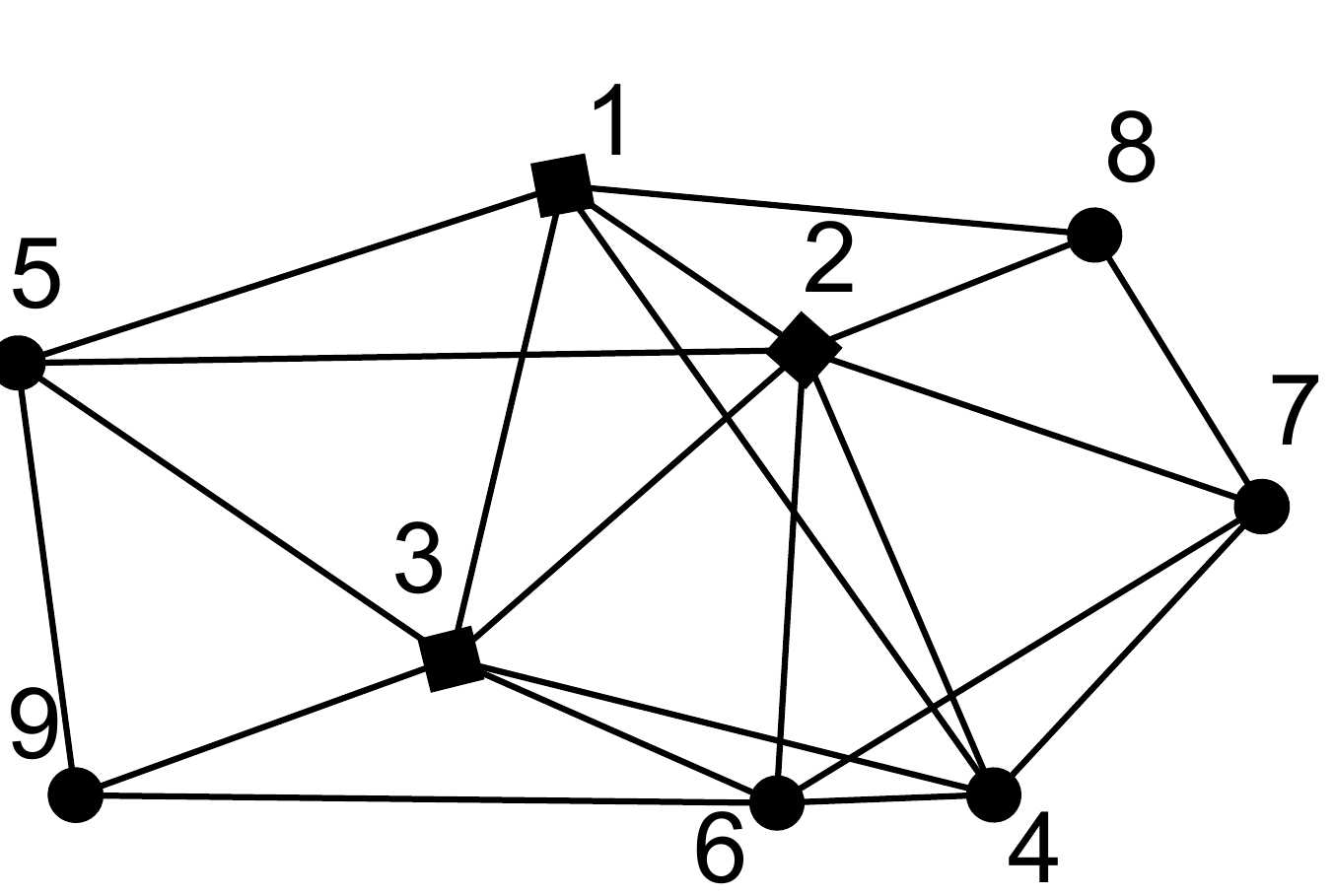}
\caption{A graph with trilateration ordering}
\label{fig:trilaterationOrdering}
\end{figure}

\textit{Trilateration} \cite{rigidity} is a localization algorithm that uses distance measurements from three non-collinear nodes to localize an unlocalized node in 2D.
If a WSN graph can be fully localized using trilateration, we say that this graph has a \textit{trilateration ordering} \cite{rigidity,theory}.
In Figure \ref{fig:trilaterationOrdering}, we see a graph with a trilateration ordering where the vertices are labeled 1 through 9.
The nodes 1, 2 and 3 are picked as seed nodes and marked with squares.

The whole graph in Figure \ref{fig:trilaterationOrdering} can be localized in six steps with the following ordering
\begin{align*}
1,2,3 \rightarrow 4\\
1,2,3 \rightarrow 5\\
2,3,4 \rightarrow 6\\
2,4,6 \rightarrow 7\\
1,2,7 \rightarrow 8\\
3,5,6 \rightarrow 9\\
\end{align*}

Notice that we cannot localize the graph if the seed nodes have no common neighbors. For instance, if we pick $3$, $5$ and $9$ as seed nodes, then we cannot localize any nodes since they have no common neighbors.
In order to localize the maximum possible number of nodes \textit{i.e.} to reach the maximum recall percentage possible, the algorithm tries every possible node triplet as the seed when a trilateration ordering is not specified.
Even though trilateration is a polynomial time algorithm, if the distance measurements are not fully accurate due to environmental noise, localizing a graph with trilateration is NP-Hard \cite{nphard} even when a trilateration ordering is given as part of the input.
In 3D, a similar algorithm, \textit{quadrilateration} \cite{survey,techniques} can be used to localize a WSN that uses distance measurements from four non-coplanar sensors to localize an unlocalized sensor in 3D, which is also a polynomial time algorithm.

Since we need to use non-coplanar nodes to avoid ambiguities, we should detect if four nodes lie on the same plane.
In order to detect coplanarity, we can form a tetrahedron using the pairwise distances between four points $a$, $b$, $c$ and $d$. 
The volume of the tetrahedron can be found by \textit{Cayley-Menger determinant} \cite{cayley} of the matrix given below.

\[M = \left[ \begin{array}{ccccc}
0 		& d(a,b)^2 &	d(a,c)^2 & d(a,d)^2 & 1	\\
d(a,b)^2 &		0 & d(b,c)^2 & d(b,d)^2 & 1	\\
d(a,c)^2 & d(b,c)^2 &		0	& d(c,d)^2 & 1	\\
d(a,d)^2 & d(b,d)^2 & d(c,d)^2	&		0 & 1	\\
1 		&		1 &			1&		1 & 0\end{array} \right]\]
where $d()$ corresponds to the Euclidean distance between nodes. Hence, the volume of a tetrahedron can be computed as follows.

\begin{align*}
\mathcal{V}_{abcd} = \dfrac{\sqrt{\text{det}(M)}}{288}
\end{align*}
where $\mathcal{V}_{abcd}$ denotes the volume of the tetrahedron formed by the points $a$, $b$, $c$, $d$, and given pairwise distances.

\begin{figure}[h!]
\centering
\includegraphics[width=0.5\linewidth]{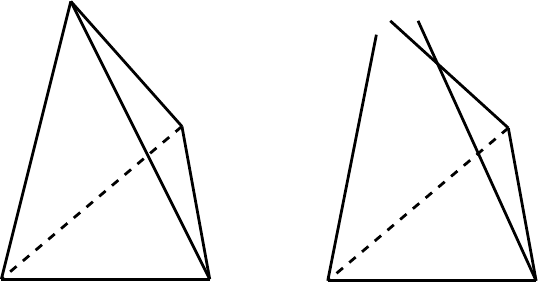}
\caption{A complete (on the left), and an incomplete (on the right) tetrahedron }
\label{fig:tetrahedron}
\end{figure}

If $\text{det}(M) > 0$, then $\mathcal{V}_{abcd} > 0$, indicating that the points are not coplanar.
If $\text{det}(M) = 0$, then $\mathcal{V}_{abcd} = 0$, indicating that the points are coplanar.
If $\text{det}(M) < 0$, then $\mathcal{V}_{abcd} \notin \mathbb{R}$, indicating that the tetrahedron is \textit{incomplete}.
An incomplete tetrahedron means that one of its corners is "open".
In Figure \ref{fig:tetrahedron}, we see two tetrahedra, one is complete and the other is incomplete.
  
\begin{figure}
\centering
\includegraphics[width=0.5\linewidth]{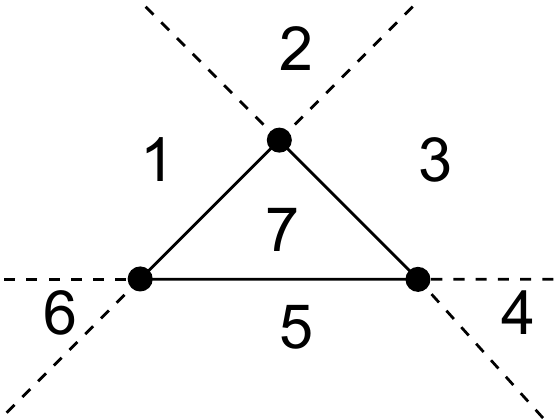}
\caption{Seven topological regions that a node can be with respect to three localized nodes}
\label{fig:7regions}
\end{figure}

Let us assume that a node $D$ is to be localized by three localized nodes $A$, $B$ and $C$.
The node $D$ is said to be \textit{flip} in 2D, if its computed position is in a different topological region than its actual position with respect to the nodes $A$, $B$ and $C$.
If a node is localized to a different topological area with respect to its three beacons, this situation is referred to as a \textit{flip} \cite{flipanalysis}.
These topological areas can be obtained by drawing three lines passing through each pair of localized nodes, as seen in Figure \ref{fig:7regions}.

\begin{figure}[h]
\centering
\includegraphics[width=\linewidth]{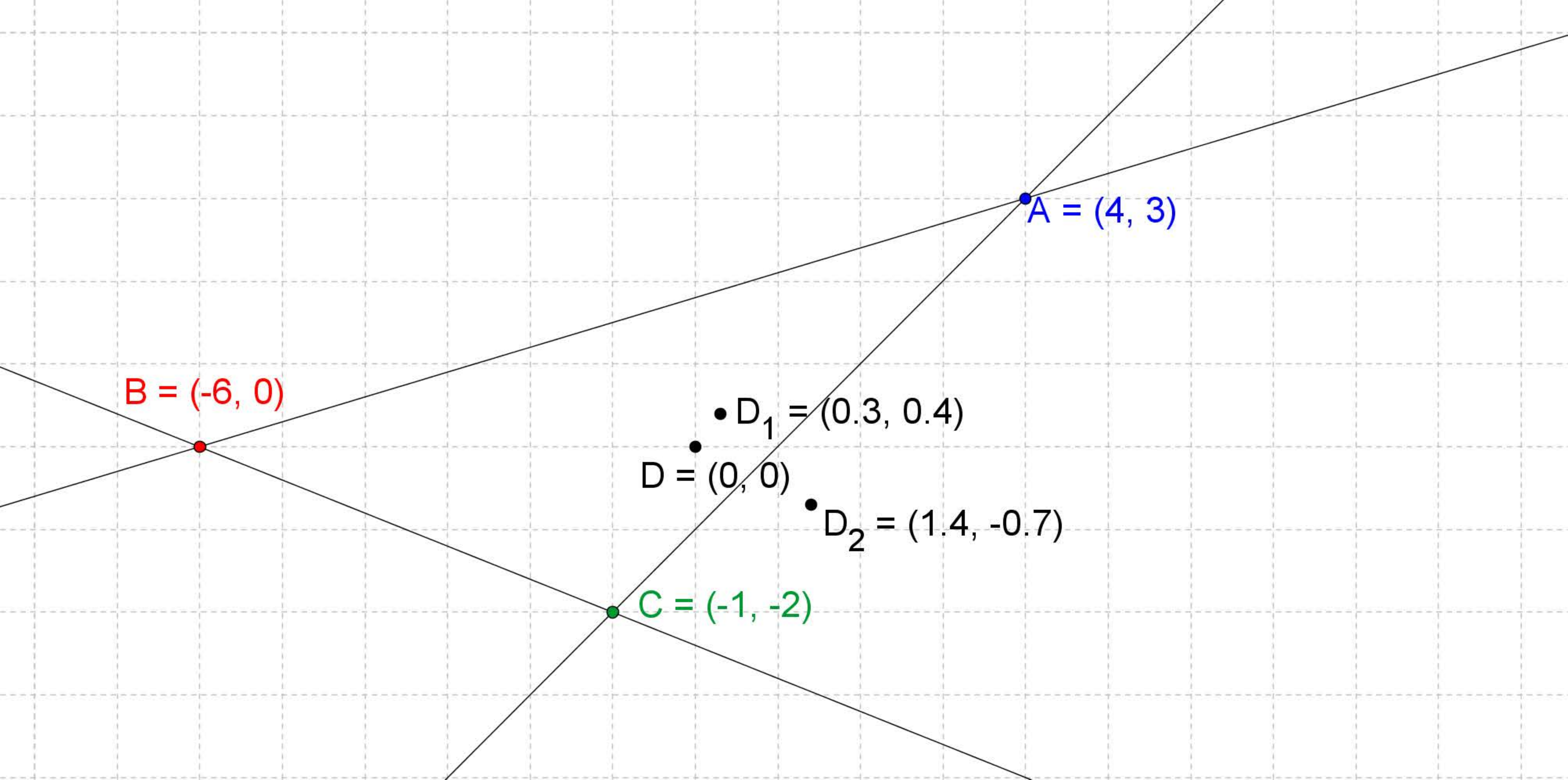}
\caption{Three possible positions for the node in three different topological regions}
\label{fig:flip}
\end{figure}

The drawn lines divide the plane into seven topological areas.
In Figure \ref{fig:flip} a flip ambiguity is demonstrated.
$A$, $B$ and $C$ are the localized nodes.
The dashed lines divide the 2D plane into seven regions.
Three possible positions $D_1$, $D_2$ and $D_3$ are inside three different topological regions with respect to $A$, $B$ and $C$.

The same concept can be used to define a flip in 3D.
Note that each triplet in a localized node quadruplet define a plane, dividing the 3-space into fifteen regions with four planes.
We say that a node is flipped in 3D if its estimated position is inside a different region than its original position with respect to the four localized nodes used to localize it.

\begin{figure}[htbp]
	\centering
	\subfloat[An example of a planar deployment as seen from a particluar angle]{%
		\includegraphics[width=\linewidth]{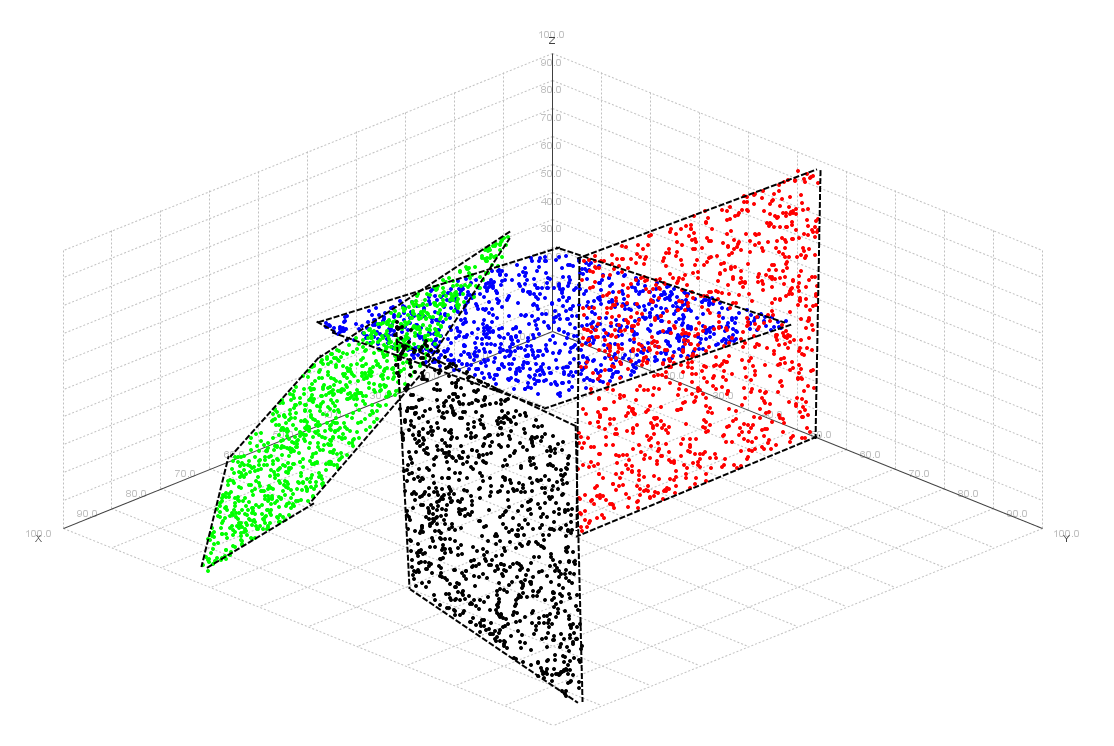}
		\label{fig:planarDeploymentExample1}%
	}
	
	\subfloat[The same planar deployment as seen from another angle.]{%
		\includegraphics[width=0.8\linewidth]{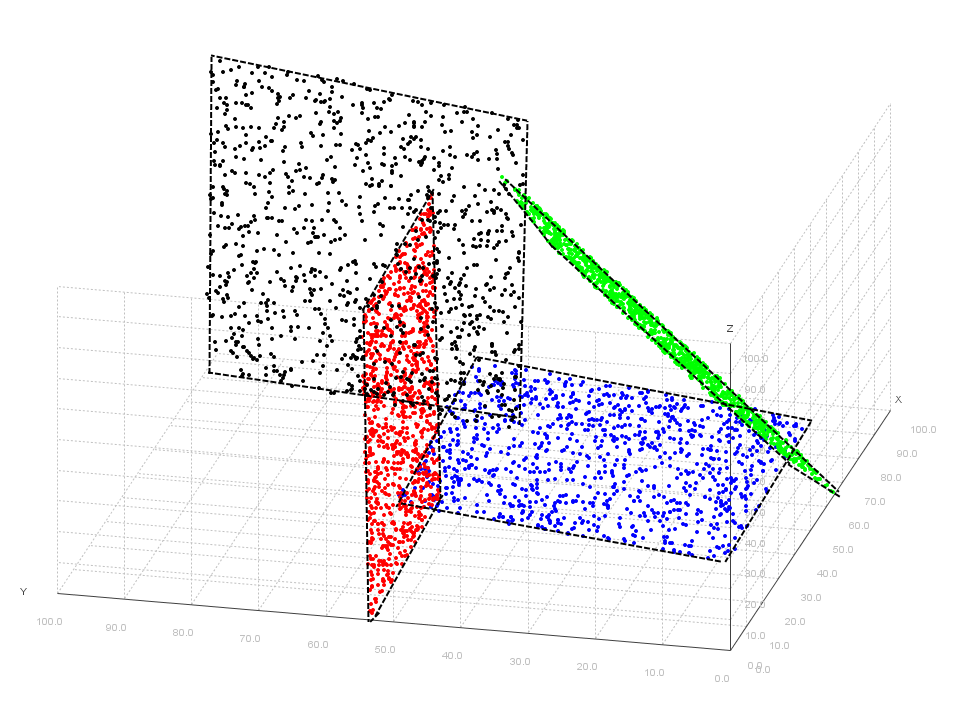}
		\label{fig:planarDeploymentExample2}%
	}
	\caption{Planar deployment}
	\label{fig:planarDeploymentExamples}
\end{figure}

Figure \ref{fig:planarDeploymentExamples} demonstrates a planar deployment of 4000 nodes that are distributed into four coplanar clusters.
In Figures \ref{fig:planarDeploymentExample1} and \ref{fig:planarDeploymentExample2}, we see the clusters from two different angles of views.
Each cluster contains 1000 nodes and indicated with a different color.
If two neighbor nodes are in the same cluster, they are called \textit{coplanar neighbors}.
Otherwise, we refer to them as \textit{interplanar neighbors}.
Correspondingly, an edge between two coplanar neighbors is called a \textit{coplanar edge}, and otherwise, an \textit{interplanar edge}.
A group of nodes that are in the same cluster cannot be used to localize one of their common interplanar neighbors unambiguously.

\begin{figure}[htbp]
	\centering
	\subfloat[An example of random deployment with the same number of nodes as seen from the same angle of view as in \ref{fig:planarDeploymentExample1}]{%
		\includegraphics[width=0.8\linewidth]{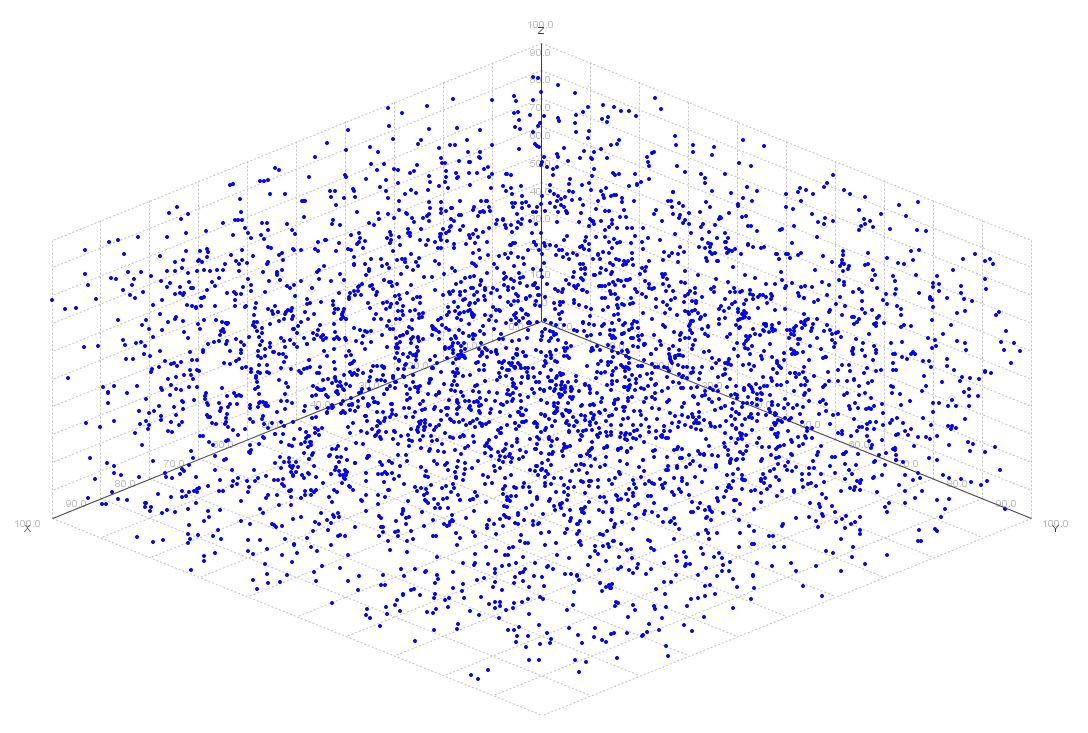}
		\label{fig:randomDeploymentExample1}%
	}
	
	\subfloat[The same random deployment as seen from the same angle of view as in \ref{fig:randomDeploymentExample2}]{%
		\includegraphics[width=0.8\linewidth]{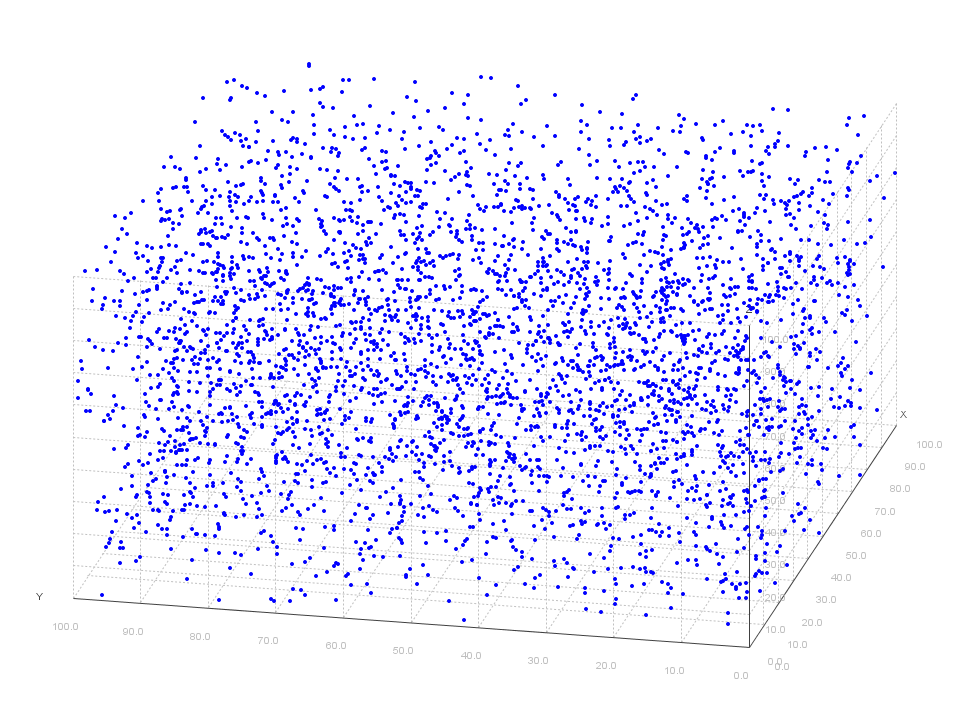}
		\label{fig:randomDeploymentExample2}%
	}
	\caption{Random deployment}
	\label{fig:randomDeploymentExamples}
\end{figure}

Figure \ref{fig:randomDeploymentExample1} and \ref{fig:randomDeploymentExample2} demonstrate the corresponding views when an equal of nodes are randomly distributed.
The figures show the difference between a planar deployment, and a random deployment in 3D.
When there is a random deployment, the nodes are seen as a cloud of points.
However, in planar deployment, the planar structures are more distinguishable to the human eye.
Moreover, during the localization process, a single node has a relatively more number of coplanar neighbors than its interplanar neighbors.

One can notice that every three nodes define a plane in 3D. 
However, the planar deployments that we seek to exploit are not particularly these plane formations with so few nodes in each cluster.
In order to indicate how planar the deployment is, we define a metric called \textit{planarity factor}, denoted by $\mu$, and calculate it as $1 - \dfrac{k^2}{n}$, where $k$ is the number of the clusters and $n$ is the number of sensor nodes.
It is basically the number of clusters divided by the number of nodes on each cluster and subtracted from 1.

\chapter{Coplanarity Based Localization} \label{chap:coplanaritybasedlocalization}

In this chapter we introduce the localization problem that we deal with, and propose an algorithm to solve it.
There are two main problems that we tackle when the sensor nodes are known to follow a planar deployment pattern.
The first problem is \textit{localizing the coplanar clusters}, in case we have the clustering information, namely, we know which coplanar cluster does each node belong to.
This information can be obtained when airdropping the sensor nodes onto planar surfaces with a known map. If the sensor nodes are dropped in clusters, even though we may not know the exact positions of the sensor nodes, we will know the planar surfaces that the nodes are on.

The second problem that we address arises when the clustering information is either lost or not available.
In this case, an additional effort is needed to discover the coplanar node clusters.
Only after the clustering information is made available, CBL can exploit it.
We refer to this second problem as \textit{extracting the coplanar clusters} or \textit{planar clustering}.

In Section \ref{sec:assumptions}, we give the assumptions and preliminaries for the algorithms that we present.
In Section \ref{sec:trilatquad}, we explain the algorithms that we use for 2D and 3D localization, namely, trilateration and quadrilateration.
In Section \ref{sec:localizingclusters}, we present a framework for range-based localization of the coplanar node clusters either found previously or given as part of input.
If the clustering information is not given, we describe a heuristic algorithm in Section \ref{sec:extractingclusters}, to find the coplanar node clusters.

\section{Assumptions and Preliminaries} \label{sec:assumptions}
We make the following assumptions while presenting the algorithms:

\begin{itemize}
\item WSN graph $G = (V,E)$ is a global variable and can be reached by any function.

\item Each sensor node $n$ has a \textit{neighbor list} that contains the nodes inside the sensing range of $n$.
Any subsets of the neighbors of $n$ can be accessed by using the notation $n.\textit{Neighbors}(\{n_i, n_j, n_k, \dots\})$.

\item A coplanar cluster $C_i = (V_i, E_i)$ is a subgraph of a WSN graph $G = (V,E)$ where $V_i \subset V$, $E_i \subset E$, and $\forall a \neq b \in V_i, (a,b) \in E \Leftrightarrow (a,b) \in E_i$.
The coplanar clusters are stored inside a set called \textit{CSet}, defined as a global variable.

\item The cluster of a sensor node $n$ can be accessed by using the notation $\textit{ClusterOf}(n)$

\item Each sensor node $n$ has two positions, namely its \textit{local position} and its \textit{global position} which are denoted by $n.\textit{LPos}$ and $n.\textit{GPos}$ respectively.
Global position is the coordinates in 3D.
Local position is in 2D, and relative to the local positions of its coplanar neighbors.
If a node $n$ in a cluster $C_i$ is localized with respect to nodes $\{C_i \setminus n\}$ only, we say that the node is \textit{locally positioned}.
Determining the global position of a node means the that node is \textit{globally positioned}.

\item All sensor nodes except the seed nodes are initially unlocalized in both 2D and 3D.

\item After performing 2D localization on a coplanar cluster $C_i = (V_i, E_i)$, we are able to obtain the 2D point formation of the sensors in that cluster.

\item Although CBL can be used as an extension of any localization algorithm, we assume that trilateration and quadrilateration are used for 2D and 3D localization respectively.
These two algorithms are presented in Section \ref{sec:trilatquad}.

\item When the distance measurements are noisy, we assume that a certain value is added to the measured distance values based on the \textit{error magnitude}.
The error is modeled as the summation of a high probability small noise and a low probability large noise as experimentally gathered \cite{dwrl,realworld1,realworld2}.
The small noise is a Gaussian random process with
mean  $N(f(R), \mathcal{E}/100)$ where $R$ is the sensing range of the sensors, $f(R) = 0.022 ln(1+R) - 0.038$ and $\mathcal{E}$ is the magnitude of error.
The large noise value is selected with a uniform random process between $\pm P\%$ of the wireless range with probability 0.05 where $P \in [0,10]$.
It is assumed that we have the knowledge of the error magnitude.
\end{itemize}

\begin{figure}[htbp]
\centering
\includegraphics[width=\linewidth]{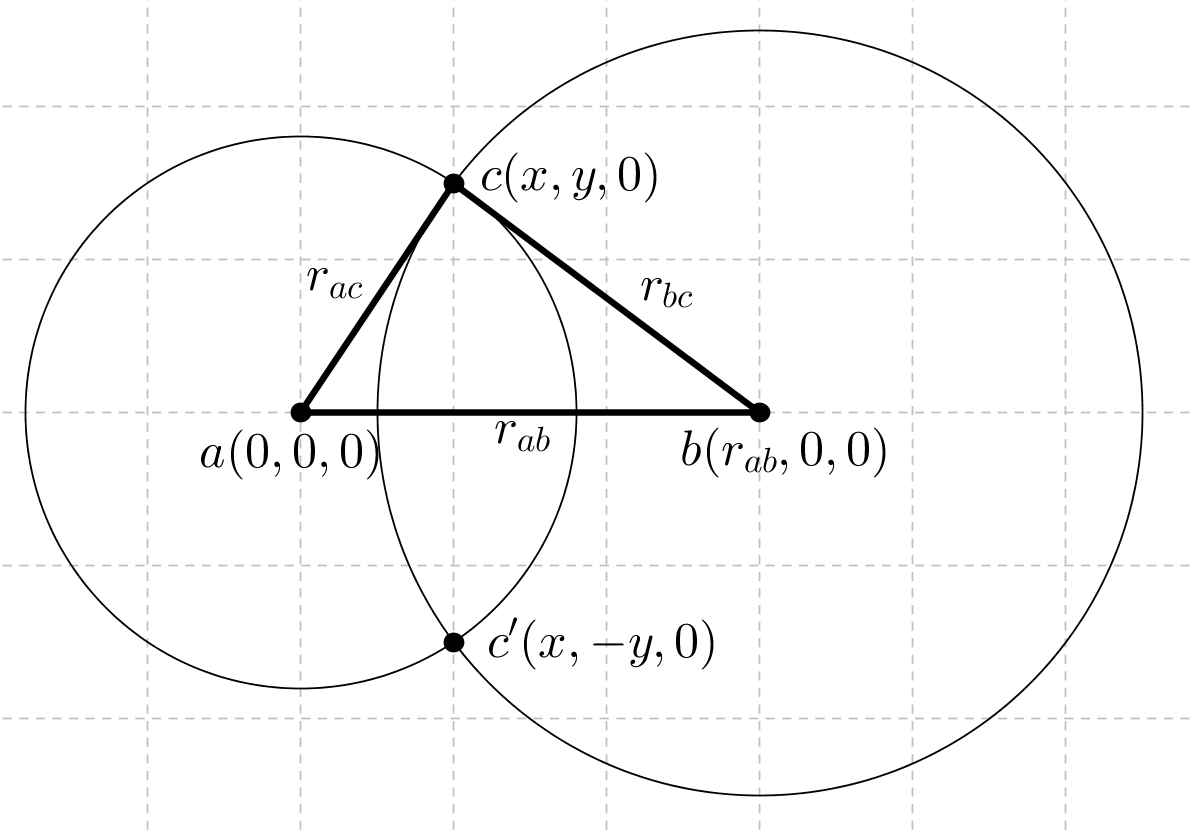}
\caption{Assigning positions to the seed nodes}
\label{fig:seedPick}
\end{figure}

While running trilateration among a coplanar node cluster, we pick in that cluster three seed nodes with a known distance between any two.
Since we do not know any of the coordinates initially, we construct a triangle on the $z=0$ plane using the pairwise distances among these three nodes.
Figure \ref{fig:seedPick} demonstrates three nodes $a$, $b$ and $c$ picked as the seed nodes.
The pairwise distances are denoted by $r_{ab}$, $r_{ac}$ and $r_{bc}$.

The first node, $a$ is placed on the origin.

\begin{align}
a.\textit{LPos} \leftarrow (0,0,0)
\end{align}

Then, we place $b$ on the positive side of the $x$-axis, at distance $r_{ab}$ to $a$.

\begin{align}
b.\textit{LPos} \leftarrow (r_{ab},0,0)
\end{align}

In Figure \ref{fig:seedPick}, we see three nodes $a$, $b$ and $c$.
As given in (1) and (2), we assign the coordinates to $a$ and $b$.
The coordinates of $c$ is yet to be computed and shown as $(x,y)$.

After assigning coordinates to $a$ and $b$, we compute the two intersections as follows.

\begin{align}
x^2 + y^2 = r_{ac}^2\\
(x-r_{ab})^2 + y^2 = r_{bc}^2
\end{align}

If we combine (3) and (4), then we have

\begin{align}
x^2 - 2*r_{ab}*x + r_{ab}^2 - x^2 &= r_{bc}^2 - r_{ac}^2\\
x &= \dfrac{r_{ab}^2 - r_{bc}^2 + r_{ac}^2}{2*r_{ab}}
\end{align}

Using (5), we compute $x$ coordinate of the node $c$ as in (6).
From (3), we obtain;

\begin{align}
y^2 &= r_{ac}^2 - x^2\\
y &= \pm\sqrt{\dfrac{4*r_{ab}^2*r_{ac}^2 - (r_{ab}^2 - r_{bc}^2 + r_{ac}^2)^2}{4*r_{ab}^2}}
\end{align}

Notice that there are two possible values for the $y$ coordinate.
We pick the positive value.
As a result, given three seeds $a$, $b$ and $c$ with the pairwise distances, we assign the following coordinates to the seeds:

\begin{align*}
a.\textit{LPos} &\leftarrow (0,0,0)\\
b.\textit{LPos} &\leftarrow (r_{ab},0,0)\\
c.\textit{LPos} &\leftarrow (x,y,0)
\end{align*}
where
\begin{align*}
x &= \dfrac{r_{ab}^2 - r_{bc}^2 + r_{ac}^2}{2*r_{ab}}\\
y &= \dfrac{\sqrt{4*r_{ab}^2*r_{ac}^2 - (r_{ab}^2 - r_{bc}^2 + r_{ac}^2)^2}}{2*r_{ab}}
\end{align*}

\begin{figure}
\centering
\includegraphics[width=\linewidth]{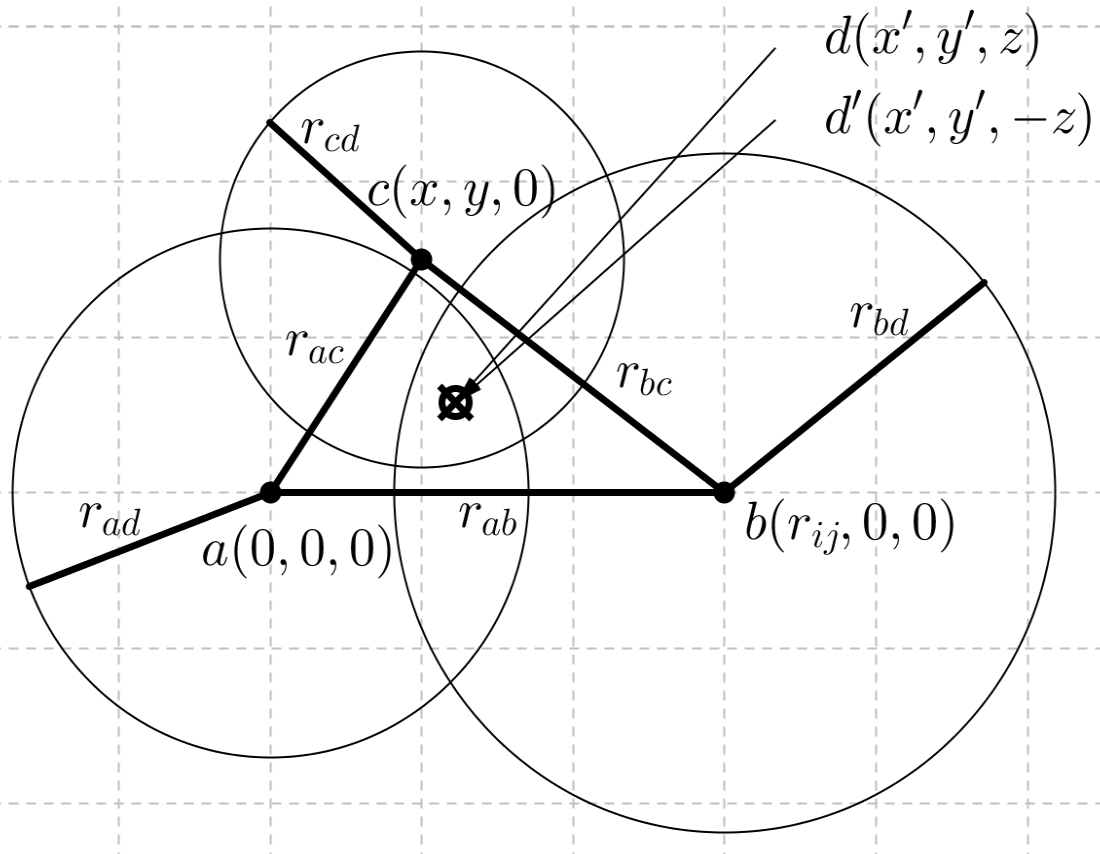}
\caption{The projection of three spheres with centers $a$, $b$, $c$ and $d$ to the $z=0$ plane.}
\label{fig:seedPick3}
\end{figure}

Before performing quadrilateration, we pick four seed nodes, $a$, $b$, $c$ and $d$.
Instead of forming a triangle, we form a tetrahedron in 3D.
Hence, we take the intersection of three spheres.
We place $a$, $b$ and $c$ onto $z=0$ plane by using the computations (1) to (8).
Figure \ref{fig:seedPick3} demonstrates the projection of the spheres on $z=0$ plane.
The centers of the spheres are $a$, $b$ and $c$.
The pairwise distances are the radii of the spheres and denoted by $r_{ab}$, $r_{ac}$, and $r_{ad}$.

In order to assign coordinates $(x',y',z)$ to $d$, we make the following computations.

\begin{align}
r_{ad}^2 &= x'^2 + y'^2 + z^2\\
r_{bd}^2 &= (x' - r_{ab})^2 + y'^2 + z^2\\
r_{cd}^2 &= (x'-x)^2 + (y'-y)^2 + z^2
\end{align}

If we subtract two equations (9) - (10), then we have

\begin{align}
r_{ad}^2 - r_{bd}^2 &= (x')^2 - (x' - r_{ab})^2\\
					&= 2*x'*r_{ab} - r_{ab}^2
\end{align}

\noindent then we can compute $x'$ and $y'$ as follows.

\begin{align}
x' &= \dfrac{r_{ad}^2 - r_{bd}^2 + r_{ab}^2}{2*r_{ab}}\\
y' &= \dfrac{r_{ad}^2 - r_{cd}^2 - (x')^2 + (x'-x)^2 + y^2}{2*y}
\end{align}

Using the Pythagorean theorem, we can compute $z$

\begin{align}
z &= \pm \sqrt{r_{ad}^2 - (x')^2 - (y')^2}
\end{align} 

Similar to 2D case, we pick the positive value among two possible values for $z$ coordinate.
Hence, following the steps (1) to (10) we 

\begin{align*}
a.\textit{GPos} &\leftarrow (0,0,0)\\
b.\textit{GPos} &\leftarrow (r_{ab},0,0)\\
c.\textit{GPos} &\leftarrow (x,y,0)\\
d.\textit{GPos} &\leftarrow (x',y',z)
\end{align*}

\noindent where

\begin{align*}
x' &= \dfrac{r_{ad}^2 - r_{bd}^2 + r_{ab}^2}{2*r_{ab}}\\
y' &= \dfrac{r_{ad}^2 - r_{cd}^2 - (x')^2 + (x'-x)^2 + y^2}{2*y}\\
z &= \sqrt{r_{ad}^2 - (x')^2 - (y')^2}
\end{align*}

When three spheres intersect, there are two possible values for the $z$ coordinate, leaving us with two symmetrical points with respect to the plane through $a$, $b$ and $c$.
If we are localizing the point $n$ in 3D, we need an extra distance measurement from a point which is not coplanar with $a$, $b$ and $c$, to choose between two symmetrical points.
The process of obtaining two candidate points, and then arbitrarily picking one is referred to as \textit{semi-localization} of a node.

If we use a fourth distance to another localized node $d$, then we are able to pick the correct point $p$ between $p_1$ and $p_2$, assuming that the distances are accurate.
However, if $d$ is coplanar with $a$, $b$ and $c$, both of the points will satisfy the distance constraint and we will not be able to tell if $p_1$ or $p_2 \neq p_1$ is the correct position for $n$.

\begin{align}
   p = \left\{
     \begin{array}{ll}
     \textsc{null} &: \|\overrightarrow{p_1d}\| = \|\overrightarrow{p_2d}\|\\
       p_1 &: \|\overrightarrow{p_1d}\| = d(n,d)\\
       p_2 &: \|\overrightarrow{p_2d}\| = d(n,d)
     \end{array}
   \right.
\end{align}

\noindent where $d(n,d)$ is the Euclidean distance between the unlocalized node $n$ and the fourth localized node $d$ and $\|\overrightarrow{p_1d}\|$ denotes the magnitude of the vector drawn from $p_1$ to the global position of $d$.

The last step assigns a unique value to the global position of unlocalized node $n$.
Thus, we call this process \textit{unique-localization} of a node in 3D.

\begin{definition}[Unique-localization of a node]
A node $n$ is said to be uniquely localized with respect to four other localized nodes $a$, $b$, $c$ and $d$ such that $a \neq b \neq c \neq d$, if there is only one possible position of $n$ with respect to the global positions of $a$, $b$, $c$ and $d$.
\end{definition}

Assigning a position to a single node is similar in trilateration and quadrilateration.
When a node has two candidate positions, which can be obtained by intersecting two circles in 2D and intersecting three spheres in 3D, a point can be picked by using (17).
However, if we consider that the distance measurements might be noisy, a point qualifies as a solution if it is inside a certain margin with respect to the distances as shown in Figure \ref{func:findUniquePos} where $p_1$ and $p_2$ are the candidate points, $r$ is the distance to the last node used to eliminate one of the solutions, and $\mathcal{E}$ is the error magnitude.

\begin{figure}[h!]
	\begin{align*}
	\sigma(p_1, p_2, r) = \left\{
	\begin{array}{ll}
	\textsc{null}, &\text{if } \big|\|\overrightarrow{p_1d}\| - r\big| \leq r*\mathcal{E} \bigwedge \big|\|\overrightarrow{p_2d}\| - r\big| \leq r*\mathcal{E}\\
	p_1, &\text{if } \big|\|\overrightarrow{p_1d}\| - r\big| \leq r*\mathcal{E}\\
	p_2, &\text{if } \big|\|\overrightarrow{p_2d}\| - r\big| \leq r*\mathcal{E}\\
	\textsc{null}, &\text{otherwise} 
	\end{array}
	\right.
	\end{align*}
	\caption{Picking a position among two candidate points}
	\label{func:findUniquePos}
\end{figure}

\section{Trilateration and Quadrilateration}
\label{sec:trilatquad}

In this section, we present the algorithms that we choose to  perform 2D and 3D localization.
We use trilateration to find the local coordinates of the nodes with respect to their coplanar neighbors and quadrilateration to find the global positions of the nodes in a WSN graph.

In Figure \ref{alg:trilateration}, we present trilateration algorithm.
we first initialize the point formation $\mathbb{F}_\text{best}$, that stores the best 2D point formation of the nodes in line \ref{trilatFbest}.
Then, we iterate on each fully-connected non-collinear node triplet $(a,b,c)$ from line \ref{trilatIterateSeed} through line \ref{trilatIterateSeedEnd}.
In line \ref{trilatFinit}, we initialize $\mathbb{F}$ that stores the 3D point formation of $G$ with respect to each seed triplet $(a,b,c)$.
In line \ref{trilatFormTriangle}, we determine the local positions of $a$, $b$ and $c$ by forming a triangle as explained in equations (1) - (8).
Then, we add $a$, $b$ and $c$ into a queue data structure $Q_{\textit{localized}}$ that keeps track of the localized nodes.
From line \ref{trilatQnotEmpty} through line \ref{trilatEndWhile}, we obtain the 2D point formation $\mathbb{F}$ of the given graph $G$ with respect to the seed nodes $a$, $b$ and $c$.
In line \ref{trilatDequeue}, we remove the node $i$ from $Q_{\textit{localized}}$ and notify its neighbors.
Notifying a node in this context means increasing its localized neighbor count by one.
In line \ref{trilatCountge3}, we check if $j$ has are more than three localized neighbors.
Hence, as last two steps, we check if $j$ is collinear with these neighbors in line \ref{trilatNonCollinear} and try to localize the node in line \ref{trilatCanBeLocalized}.
If so, then we add $j$ into the queue of localized nodes in line \ref{trilatAddIntoQ}.
When there are no more nodes left to be localized, we check if all the nodes in the graph are localized.
If so, then we return the obtained point formation in line \ref{trilatAllLocalized}.
In line \ref{trilatFisBetter}, we check if the obtained point formation $\mathbb{F}$ contains more nodes than the best point formation $\mathbb{F}_{\text{best}}$.
If so, then we declare the new best point formation as the current one in line \ref{trilatFisBetter}.
In line \ref{trilatReturnBest}, we return the best point formation with respect to the number of nodes.

In Figure \ref{alg:quadrilateration}, we present quadrilateration algorithm.
we first initialize the point formation $\mathbb{F}_\text{best}$, that stores the best 3D point formation of the nodes in line \ref{quadFbest}.
Then, we iterate on each fully-connected and non-coplanar node triplet $(a,b,c,d)$ from line \ref{quadIterateSeed} through line \ref{quadIterateSeedEnd}.
In line \ref{trilatFinit}, we initialize $\mathbb{F}$ that stores the 3D point formation of $G$ with respect to each seed quadruplet $(a,b,c,d)$.
In line \ref{quadFormTetrahedron}, we determine the global positions of $a$, $b$, $c$ and $d$ by forming a tetrahedron as explained in equations (1) - (16).
Then, we add $a$, $b$, $c$ and $d$ into a queue data structure $Q_{\textit{localized}}$ that keeps track of the localized nodes.
From line \ref{quadQnotEmpty} through line \ref{quadEndWhile}, we obtain the 3D point formation $\mathbb{F}$ of the given graph $G$ with respect to the seed nodes $a$, $b$, $c$ and $d$.
In line \ref{quadDequeue}, we remove the node $i$ from $Q_{\textit{localized}}$ and notify its neighbors.
Notifying a node in this context means increasing its localized neighbor count by one.
In line \ref{quadCountge4}, we check if $j$ has are more than four localized neighbors.
Hence, as last two steps, we check if $j$ is coplanar with these neighbors in line \ref{quadNonCoplanar} and try to localize the node in line \ref{trilatCanBeLocalized}.
If so, then we add $j$ into the queue of localized nodes in line \ref{quadAddIntoQ}.
When there are no more nodes left to be localized, we check if all the nodes in the graph are localized.
If so, then we return the obtained point formation in line \ref{quadFisBetter}.
In line \ref{trilatFisBetter}, we check if the obtained point formation $\mathbb{F}$ contains more nodes than the best point formation $\mathbb{F}_{\text{best}}$.
If so, then we declare the new best point formation as the current one in line \ref{quadAllLocalized}.
In line \ref{trilatReturnBest}, we return the best point formation with respect to the number of nodes.

\begin{figure}[htbp]

\subfloat[Trilateration]{
\fbox{\begin{minipage}{\columnwidth}
\Input{Graph $G = (V,E)$}
\Output{2D point formation of $G$}

\Algo{Trilateration}{$G$}
\begin{algorithmic}[1]
\State $\mathbb{F}_\text{best} \gets ([\text{  }],[\text{  }])$ \Comment{Best point formation} \label{trilatFbest}
\For{each non-collinear and fully-connected $\{a,b,c\} \subseteq V_i$} \label{trilatIterateSeed}
\State $\mathbb{F} \gets  ([\text{  }],[\text{  }])$ \label{trilatFinit}
	\State Form triangle using $a$, $b$, $c$  \Comment{Equations (1) - (8)} \label{trilatFormTriangle}
	\State $Q_{\textit{localized}} \gets [a,b,c]$ \label{trilatInitQ}
	\While{$Q_{\textit{localized}}$ is not empty} \label{trilatQnotEmpty}
		\State $i \gets \textit{dequeue}(Q_{\textit{localized}})$\label{trilatDequeue}
		\State add $\left(i, i.\textit{LPos}\right)$ into $\mathbb{F}$ \label{trilatAddIntoF}
		\For{all neighbors $j$ of $i$} \label{trilatIterateOnNeighbors}
			\If{$(\texttt{++}j.\textit{Count}) \geq 3$} \label{trilatCountge3}
				\If{$j.\textit{LocalizedNeighbors}(1,2,\textit{last})$ are not collinear} \label{trilatNonCollinear}
					\If{$j$ can be localized in 2D} \label{trilatCanBeLocalized}
						\State $\textit{enqueue}(j,Q_{\textit{localized}})$ \label{trilatAddIntoQ}
					\EndIf
				\EndIf
			\EndIf
		\EndFor
	\EndWhile
	\endWhile \label{trilatEndWhile}
	\If{$|\mathbb{F}^1| = |V|$} 
		\Return $\mathbb{F}$ \label{trilatAllLocalized}
	\EndIf
	\If{$|\mathbb{F}^1| > |\mathbb{F}_\text{best}^1|$}
		$\mathbb{F}_\text{best} \gets \mathbb{F}$ \label{trilatFisBetter}
	\EndIf
\EndFor
\endFor \label{trilatIterateSeedEnd}
\State \Return $\mathbb{F}_\text{best}$ \label{trilatReturnBest}
\end{algorithmic} \end{minipage}}
\label{alg:trilateration}
}

\subfloat[Quadrilateration]{
\fbox{\begin{minipage}{\columnwidth}
\Input{Graph $G = (V,E)$}
\Output{3D point formation of $G$}

\Algo{Quadrilateration}{$G$}
\begin{algorithmic}[1]
\State $\mathbb{F}_\text{best} \gets ([\text{  }],[\text{  }])$ \Comment{Best point formation} \label{quadFbest}
\For{each non-coplanar and fully-connected $\{a,b,c,d\} \subseteq V_i$} \label{quadIterateSeed}
	\State $\mathbb{F} \gets  ([\text{  }],[\text{  }])$ \label{quadFinit}
	\State Form tetrahedron using $a$, $b$, $c$, $d$  \Comment{Equations (1) - (16)} \label{quadFormTetrahedron}
	\State $Q_{\textit{localized}} \gets [a,b,c,d]$ \label{quadInitQ}
	\While{$Q_{\textit{localized}}$ is not empty} \label{quadQnotEmpty}
		\State $i \gets \textit{dequeue}(Q_{\textit{localized}})$\label{quadDequeue}
		\State add $\left(i, i.\textit{GPos}\right)$ into $\mathbb{F}$ \label{quadAddIntoF}
		\For{all neighbors $j$ of $i$} \label{quadIterateOnNeighbors}
			\If{$(\texttt{++}j.\textit{Count}) \geq 4$} \label{quadCountge4}
				\If{$j.\textit{LocalizedNeighbors}(1,2,3,\textit{last})$ are not coplanar} \label{quadNonCoplanar}
					\If{$j$ can be localized in 3D} 
						\State $\textit{enqueue}(j,Q_{\textit{localized}})$ \label{quadAddIntoQ}
					\EndIf
				\EndIf
			\EndIf
		\EndFor
	\EndWhile
	\endWhile \label{quadEndWhile}
	\If{$|\mathbb{F}^1| = |V|$} 
		\Return $\mathbb{F}$ \label{quadAllLocalized}
	\EndIf
	\If{$|\mathbb{F}^1| > |\mathbb{F}_\text{best}^1|$}
		$\mathbb{F}_\text{best} \gets \mathbb{F}$ \label{quadFisBetter}
	\EndIf
\EndFor
\endFor \label{quadIterateSeedEnd}
\State \Return $\mathbb{F}_\text{best}$ \label{quadReturnBest}
\end{algorithmic} \end{minipage}}
\label{alg:quadrilateration}
}
\caption{Trilateration (a) and quadrilateration (b) algorithms}
\label{alg:trilatquad}
\end{figure}

\begin{remark} \label{rem:trilatquad}
For a given graph $G = (V,E)$, finding a trilateration ordering consists of $O(|V|^3)$ phases in the worst case.
As three distance measurements from three non-collinear nodes are sufficient to localize a node, $O(|E|)$ time is spent in each phase by using a queue data structure keeping the track of localized nodes.
This can be accomplished by keeping track of the number of localized neighbors of every node and inserting a node when the count hits a score of three.
An effective three is the smallest count $\geq 3$ with three non-collinear neighbors.
On reaching the count three for the first time check collinearity.
From then on, every new increment check only the new with any two previous.
Therefore, it terminates after at most after $O(|V|^3*|E|)$ steps.
Quadrilateration, on the other hand, tries every possible node quadruplet as seed in order to achieve its highest possible localization percentage.
Therefore, it runs in $O(|V|^4*|E|)$ time in the worst case.
\end{remark}

We use CBL as the extension of the algorithms presented in Figures \ref{alg:trilateration} and \ref{alg:quadrilateration}.
In the following section, we conduct simulations to test trilateration and quadrilateration to observe the effect of error on the precision of localizations achieved by both algorithms.

\pagebreak
\subsection{Experimental Evaluation of Trilateration and Quadrilateration}

In this section, we conduct experiments to test the performance trilateration and quadrilateration when there is a random deployment.
For the experiments with trilateration, we deploy 100 sensors randomly in a $100 \times 100$ units square with a uniform distribution.
For the experiments with quadrilateration, we deploy the same number of sensors randomly inside a $100 \times 100 \times 100$ units cube with a uniform distribution.

\begin{figure}[htbp]
\centering
\includegraphics[width=\linewidth]{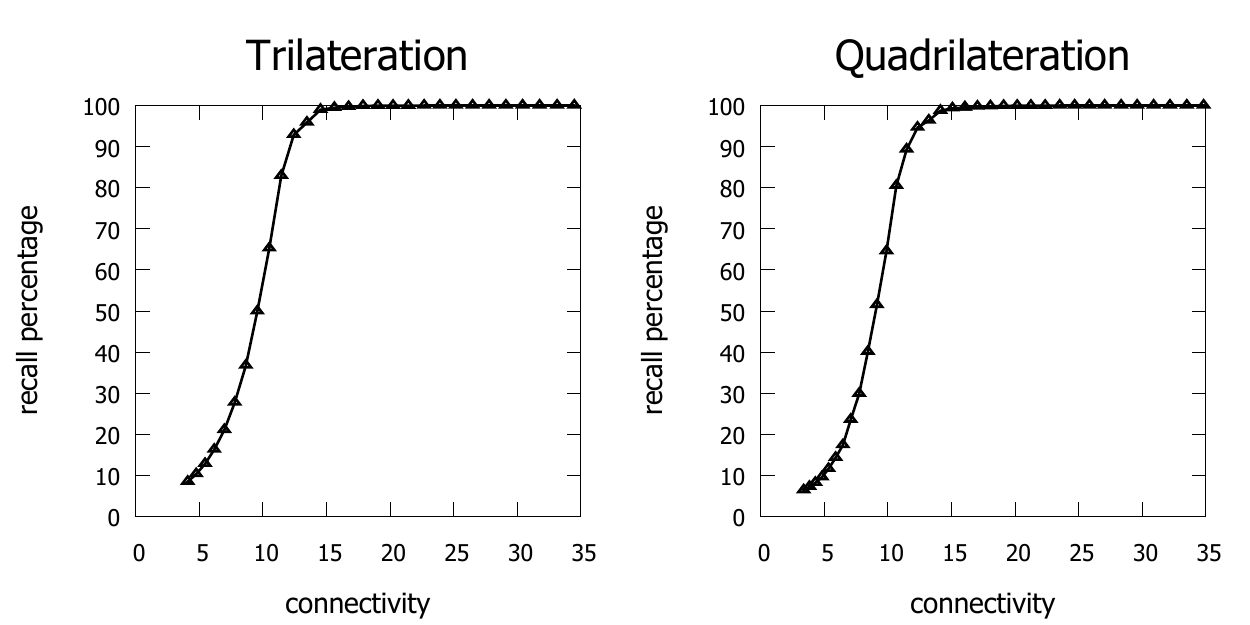}
\caption{The recall percentages of trilateration (a) and quadrilateration (b) with respect to increasing node connectivity}
\label{fig:quadVStrilatRecall}
\end{figure}

We take the average node connectivity \textit{i.e.} average number of neighbors per node as the control value for the tests.
We run 1000 tests for each connectivity value, and report the average of the recall percentage.
In Figure \ref{fig:quadVStrilatRecall}, we test both algorithms with noiseless range measurements, and observe the changes in recall percentages with respect to the increasing node connectivity.

\begin{figure}[htbp]
	\centering
	\includegraphics[width=\linewidth]{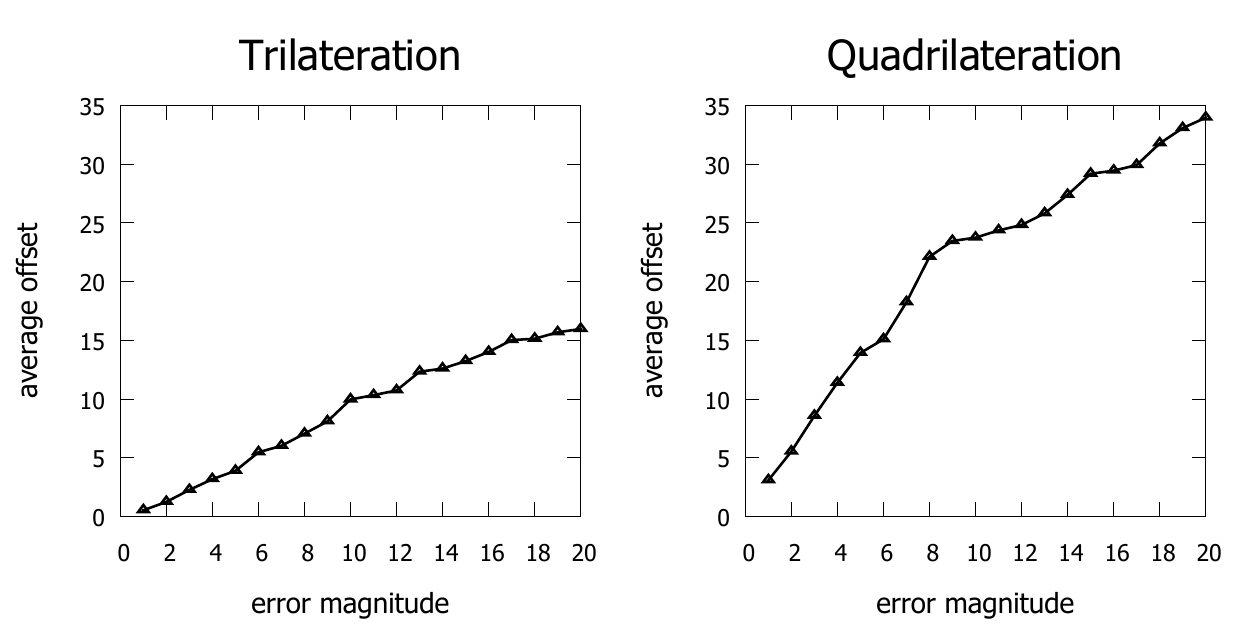}
	\caption{The average offsets of trilateration (a) and quadrilateration (b) with respect to increasing error magnitude when the average node connectivity is 15}
	\label{fig:quadVStrilatOffset}
\end{figure}

The figure tells us that both trilateration and quadrilateration require an average connectivity of 15 neighbors per node.
Now, let us check the average offsets of both algorithms with increasing error magnitude.
We ifc the average connectivity at 15 neighbors per node, which is the value for both algorithms to reach their maximum recall percentage.
For trilateration, we use robust triangles \cite{reducing} to reduce the number of the flips.
In Figure \ref{fig:quadVStrilatOffset}, we see the average offsets for both algorithms with increasing error magnitude.

The figure tells us that the error has more impact in 3D compared to 2D.
Therefore, we expect that when we limit the use of quadrilateration by running trilateration to find the local positions of the nodes, the average offset will drop.

\section{Localizing the Coplanar Node Clusters}
\label{sec:localizingclusters}

In this section, we tackle the problem of localizing a WSN in 3D where there is a planar deployment, and the information on the coplanar clusters is given apriori.
We first define the problem of localizing the coplanar node clusters, and then propose an algorithm called Coplanarity Based Localization (CBL) to solve the corresponding problem.

\subsection{Problem Definition} \label{sec:problemdefinition}

The problem of localizing clusters is finding the 3D point formation of a WSN graph $G = (V,E)$ which is partitioned into $k$ subgraphs within each of which the nodes are coplanar.
These partitions are called \textit{coplanar clusters} and stored in a set $\textit{CSet} = \{C_1 = (V_1, E_1), C_2 = (V_2, E_2), \dots, C_k = (V_k, E_k)\}$ with the following properties.

\begin{align*}
&\bigcup\limits_{i \in [1,k]} V_i = V\\
&\bigcup\limits_{i \in [1,k]} E_i \subset E\\
&\bigcap\limits_{i \in [1,k]} V_i = \emptyset\\
&\bigcap\limits_{i \in [1,k]} E_i = \emptyset\\
& \forall a \neq b \in V_i, (a,b) \in E \Leftrightarrow (a,b) \in E_i
\end{align*}

Two nodes $v,w$ in the vertex set $V_i$ of a coplanar cluster $C_i = (V_i, E_i)$ are called \textit{coplanar neighbors}, and the edge $(w,v)$ is called a \textit{coplanar edge}.
In an effort to localize $G$ in 3D, we first localize the nodes in each cluster $C_i$ in 2D with respect to their coplanar neighbors and then find the positions of the coplanar clusters in 3-space.
The coplanarity causes difficulties for traditional multilateration methods because of flip ambiguities.
Therefore, we propose a localization algorithm with potential to increase the precision of the localization when there is a planar deployment.
This algorithm is called Coplanarity Based Localization (CBL). 
CBL obviously assumes the existence of coplanar node clusters given as the part of the input.
Our approach has two main advantages over the known 3D localization techniques:
\begin{enumerate}[i)]
\item We break the WSN localization problem of size $n$ in 3D into $k$ 2D localization problems.
\item By using the clustering information, we avoid using coplanar nodes to localize another node that may lead to a flip ambiguity.
\end{enumerate}
The algorithm that we propose to find the 3D point formation of a WSN graph $G$ is presented in Section \ref{sec:thecbl}

\subsection{The CBL Algorithm} \label{sec:thecbl}

In this section, we present the proposed algorithm CBL to localize a WSN in 3D when there is a planar deployment.
CBL is presented as an extension of quadrilateration with the ability to exploit the information on coplanar clusters.
It should be noted that CBL is orthogonal to the localization algorithm used at its core.
In other words, it will work with algorithms other than quadrilateration.

As the clustering information is assumed to be available, we can apply known 2D localization techniques in each cluster.
We use trilateration presented in Figure \ref{alg:trilateration} as the 2D localization algorithm.
Localizing a coplanar cluster means computing the equation of the plane that the cluster is on.
Thus, it is enough to find the global positions of three locally positioned nodes from a cluster to compute the equation of the plane.
These three nodes are called \textit{support nodes} of the cluster.
Using the global positions of the support nodes, all the remaining locally positioned nodes in a coplanar cluster can be globally positioned by a simple transformation with respect to the support nodes as soon as the support nodes are globally positioned.

\begin{figure}[htbp]
\centering
\includegraphics[width=\linewidth]{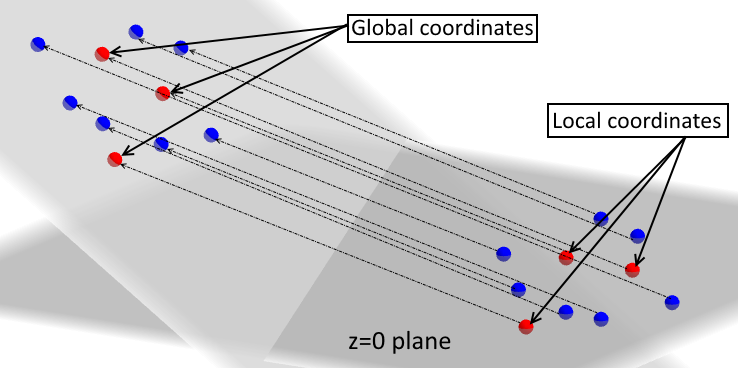}
\caption{Transformation of a coplanar cluster}
\label{fig:transformation}
\end{figure}

In Figure \ref{fig:transformation}, we see an example transformation of a coplanar cluster.
Both the local and the global positions of the support nodes are shown in red they are marked with arrows.
The dashed arrows indicate the transformation from $z=0$ plane to the actual plane that the cluster is on.

The transformation process is as follows.
Given three support nodes $a$, $b$ and $c$, let us denote the local coordinates of these nodes as $a_{\text{local}}$, $b_{\text{local}}$ and $c_{\text{local}}$ respectively.
The global coordinates of these nodes are denoted by $a_{\text{global}}$, $b_{\text{global}}$ and $c_{\text{global}}$ respectively.
In order to build a transformation matrix, we use a common coordinate system.
Hence, we pick the centroid of the local coordinates, $CL$, as the origin of local coordinate system and the centroid of the global coordinates, $CG$, as the origin of the global coordinate system.
The $x$-axis of the local coordinate system is

\begin{align}
\overrightarrow{x_{\text{local}}} = \dfrac{\overrightarrow{(CL)(a_{\text{local}})}}{\|\overrightarrow{(CL)(a_{\text{local}})}\|}
\end{align}
where $\|\overrightarrow{(CL)(i_{\text{local}})}\|$ denotes the magnitude of the vector drawn from the origin of the local coordinate system to the local coordinates of $i$.
Then, we find a vector that is orthogonal to $x$-axis by first finding

\begin{align}
\vec{v} = \dfrac{\overrightarrow{(CL) (b_{\text{local}}) } }{\|\overrightarrow{(CL) (b_{\text{local}}) }\|}
\end{align}

This vector is used to determine the $z$-axis of the local coordinate system, which is orthogonal to $\vec{x}$ and $\vec{v}$.

\begin{align}
\overrightarrow{z_{\text{local}}} = \dfrac{\overrightarrow{x_{\text{local}}} \times \vec{v}}{\|\overrightarrow{x_{\text{local}}} \times \vec{v}\|}
\end{align}

Finally, we compute the $y$-axis of the local coordinate system, which is certainly orthogonal to $x$-axis and $z$-axis.

\begin{align}
\overrightarrow{y_{\text{local}}} = \overrightarrow{x_{\text{local}}} \times \overrightarrow{z_{\text{local}}}
\end{align}

The matrix of the local coordinate system is called \textit{input matrix} and computed as follows.

\begin{align}
M_{\text{input}} = 
\begin{bmatrix}
x_{\text{local}}^1 & y_{\text{local}}^1 & z_{\text{local}}^1\\
x_{\text{local}}^2 & y_{\text{local}}^2 & z_{\text{local}}^2\\
x_{\text{local}}^3 & y_{\text{local}}^3 & z_{\text{local}}^3\\
\end{bmatrix}
\end{align}
where the columns of the matrix are the transpositions of $\overrightarrow{x_{\text{local}}}$, $\overrightarrow{y_{\text{local}}}$ and $\overrightarrow{z_{\text{local}}}$

The global coordinate system is computed similar to steps (18) - (22).

\begin{align}
\overrightarrow{x_{\text{global}}} &= \dfrac{\overrightarrow{(CG) (a_{\text{global}}) }}{\|\overrightarrow{ (CG) (a_{\text{global}}) }\|}\\
\vec{v} &= \dfrac{\overrightarrow{(CG) (b_{\text{global}}) }}{\|\overrightarrow{(CG) (b_{\text{global}}) }\|}\\
\overrightarrow{z_{\text{global}}} &= \dfrac{\overrightarrow{x_{\text{global}}} \times \vec{v}}{\|\overrightarrow{x_{\text{global}}} \times \vec{v}\|}\\
\overrightarrow{y_{\text{global}}} &= \overrightarrow{x_{\text{global}}} \times \overrightarrow{z_{\text{global}}}\\
M_{\text{output}} &= 
\begin{bmatrix}
x_{\text{global}}^1 & y_{\text{global}}^1 & z_{\text{global}}^1\\
x_{\text{global}}^2 & y_{\text{global}}^2 & z_{\text{global}}^2\\
x_{\text{global}}^3 & y_{\text{global}}^3 & z_{\text{global}}^3\\
\end{bmatrix}
\end{align}

The transformation matrix is computed using $M_{\text{input}}$ and $M_{\text{output}}$ by making the following matrix multiplication.

\begin{align}
\left[M_{\text{transform}}\right] = \left[M_{\text{output}}\right]\left[M_{\text{input}}\right]^T
\end{align}

\noindent where $[M_{\text{input}}]^T$ denotes the transposition of the input matrix.

After we build the transformation matrix, we apply the transformation to each locally positioned node $n$ with local coordinates $n_{\text{local}}$.
We compute the global coordinates of node $n_{\text{global}}$ by applying the transformation, we use the origins of the local and the global coordinate systems, $CL$ and $CG$.

\begin{align}
n_{\text{global}} = \left[M_{\text{transform}}\right]\cdot\left(\overrightarrow{n_{\text{local}}} - \overrightarrow{CL}\right) + \overrightarrow{CG}
\end{align}
where the origins $CL$ and $CG$ are treated as vectors from the point $(0,0,0)$.

At the beginning of the localization process, after a cluster is picked as seed, the nodes in that cluster cannot be used to perform a unique localization of a node.
Therefore, we are only left with the option of semi-localization of a node.
If a cluster is localized based on a semi-localized support node, we refer to that cluster as a \textit{semi-localized cluster}.

\begin{definition}[Semi-localization of a cluster]
A coplanar cluster $C_i = (V_i, E_i)$ is said to be semi-localized if one of the support nodes in $V_i$ is semi-localized while the other two are uniquely localized.
\end{definition}

Once we are alone with a seed cluster and a semi-localized cluster, we can localize a third cluster as given by the following definition.

\begin{definition}[Rigid-localization of a cluster] \label{def:rigid}
A cluster $C = (V_i, E_i)$ is said to be rigidly localized if all the three support nodes in $V_i$ are uniquely localized.
\end{definition}

\begin{figure}[htbp]
\centering
\subfloat[Semi-localization of a cluster]{
\includegraphics[width=0.65\linewidth]{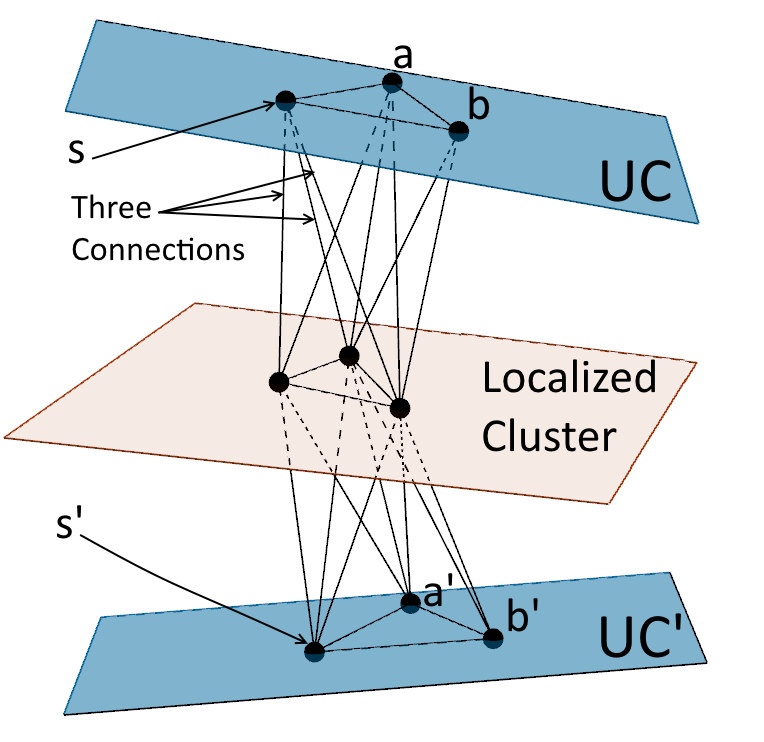}
\label{fig:semi}%
}
\vfill
\subfloat[Rigid-localization of a cluster]{
\includegraphics[width=0.65\linewidth]{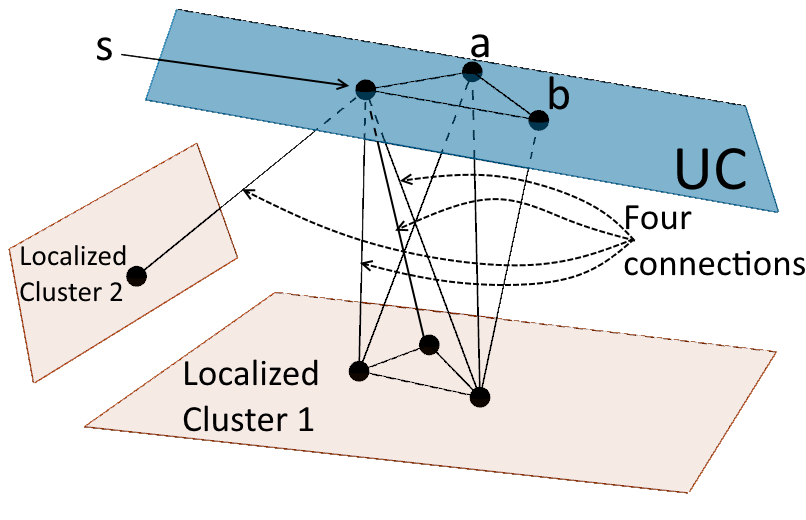}
\label{fig:rigid}
}
\caption{Semi-localization (a), and rigid-localization (b) of an unlocalized cluster}
\label{fig:semirigid}%
\end{figure}

In Figure \ref{fig:semi}, we see the semi-localization of an unlocalized coplanar cluster $UC$.
The support nodes of $UC$ are indicated by letters $s$, $a$ and $b$. 
After picking one of the possible positions for the first support node, denoted by $s$ and $s'$, the remaining support nodes $a$ and $b$ can be uniquely localized with respect to $s$ and three other localized nodes from localized cluster.

\begin{remark} \label{rem:semi}
The only possible combinations of positions of the support nodes in Figure \ref{fig:semi} are either $(s,a,b)$ or $(s',a',b')$, which are reflections around the localized cluster.
\end{remark}

\begin{remark} \label{rem:rigid}
The rigid-localization of a cluster cannot be performed unless the number of already localized clusters is $m \geq 2$, and one coplanar cluster is on a different plane than at least one of the remaining $m-1$.
\end{remark}

In Figure \ref{fig:rigid}, we see that the support node $s$ has one extra localized neighbor which is not coplanar with the rest.
Therefore, $s$ can be uniquely localized, determining unique positions for its coplanar neighbors. In order to localize all the nodes, we have to perform rigid-localization of all the remaining clusters.

\begin{figure}[htbp]
\centering
\fbox{\begin{minipage}{\columnwidth}
\Input{WSN graph $G = (V,E)$, coplanar cluster set $\textit{CSet} = \{C_1, C_2, \dots, C_k\}$}
\Output{3D point formation of $G$}

\Algo{CBL}{$G$, \textit{CSet}}
\begin{algorithmic}[1]
\State find local positions of the nodes using trilateration \label{cblFindLocal}
\State $\mathbb{F}_\text{best} \gets ([\text{  }],[\text{  }])$ \Comment{Best point formation} \label{cblFbest}
\For{all $C_{\textit{seed}} \in \textit{CSet}$} \label{cblSeed}
	\For{each $C_{\textit{semi}} \neq C_{\textit{seed}} \in \textit{CSet}$} \label{cblSemi}
		\If{$C_\textit{semi}$ can be semi-localized with respect to $C_\textit{seed}$} \label{semiCanBe}
		\lComment{See Figure \ref{fig:semi}}
			\State $Q_{\textit{localized}} \gets [C_{\textit{seed}}, C_{\textit{semi}}]$ \label{cblInitQ}
			\State $\mathbb{F} \gets  ([\text{  }],[\text{  }])$ \label{cblFinit}
			\While{$Q_{\textit{localized}}$ is not empty} \label{cblWhile}
				\State $C_i \gets $ \textit{dequeue($Q_{\textit{localized}}$)} \label{cblDequeue}
				\For{all nodes $v \in V_i$} \label{cblIterateNodes}
					\If{$v$ is globally positioned} \label{cblIsLocallyPositioned}
						\State add $(v,v.\textit{GPos})$ into $\mathbb{F}$ \label{cblAddIntoF}
						\For{all neighbors $w$ of $v$} \label{cblIterateNeighbors}
							\lComment{$C_w \gets \textit{ClusterOf}(w)$} 
							\If{$\texttt{++}w.\textit{Count}$ changes the state of $C_w$ to rigid} \label{cblRigidLocalized}
							\lComment{$C_w$ can be rigid-localized (see Figure \ref{fig:rigid})}
							\State $\textit{enqueue}(j,Q_{\textit{localized}})$ \label{cblAddIntoQ}
							\EndIf
						\EndFor
						\endFor \label{cblIterateNeighborsEnd}
					\EndIf
					\endIf
				\EndFor
				\endFor
			\EndWhile
			\endWhile \label{cblEndWhile}
			\If{all the coplanar clusters are localized} 
				\Return $\mathbb{F}$ \label{cblAllLocalized}
			\EndIf
			\If{$|\mathbb{F}^1| > |\mathbb{F}_\text{best}^1|$}
				$\mathbb{F}_\text{best} \gets \mathbb{F}$ \label{cblFisBetter}
			\EndIf
		\EndIf
		\endIf
	\EndFor
	\endFor \label{cblEndFor}
\EndFor
\endFor
\Return $\mathbb{F}_\text{best}$ \label{cblReturnBest}
\end{algorithmic} \end{minipage}}
\caption{CBL algorithm}
\label{alg:cbl}
\end{figure}

We present CBL in Figure \ref{alg:cbl}.
The algorithm takes a WSN graph $G = (V,E)$ as the input and returns the estimated 3D point formation of $G$.
In order to localize the maximum possible number of nodes \textit{i.e.} to reach the maximum recall percentage possible, the algorithm tries every possible coplanar cluster pair as the seed cluster $C_{\textit{seed}}$ and semi-localized cluster $C_{\textit{semi}}$.
In line \ref{cblFindLocal}, we find the local positions of the nodes with respect to their coplanar neighbors.
We initialize the point formation $\mathbb{F}_{\text{best}}$ that stores the best 3D point formation of the nodes in line \ref{cblFbest}.
Then, we iterate on each coplanar cluster $C_{\textit{seed}}$ from line \ref{cblSeed} through line \ref{cblEndFor}.
At each iteration, we pick $C_{\textit{seed}}$ as the seed cluster.
In line \ref{cblSemi}, we iterate on the coplanar clusters except the seed cluster.
If we find a cluster $C_{\textit{semi}}$ that can be semi-localized with respect to the seed cluster in line \ref{semiCanBe}, then we add $C_{\textit{seed}}$ and $C_{\textit{semi}}$ into a queue data structure $Q_{\textit{localized}}$ that keeps track of the localized nodes in line \ref{cblInitQ}.
Then, we initialize $\mathbb{F}$ that stores the 3D point formation of $G$ with respect to each $(C_{\textit{seed}}, C_{\textit{semi}})$ pair.
When we find a seed and a semi-localized cluster, we perform rigid-localization for the remaining coplanar clusters, from line \ref{cblWhile} through line \ref{cblEndWhile}.
In line \ref{cblDequeue}, we remove the cluster $C_i = (V_i, E_i)$ from $Q_{\textit{localized}}$.
We iterate on each node $v$ in the vertex set $V_i$ of cluster $C_i$ in line \ref{cblIterateNodes}.
If $v$ is globally positioned, then we add $v$ and its global position into $\mathbb{F}$.
Then, we notify the neighbors $w$ of $v$ from line \ref{cblIterateNeighbors} through line \ref{cblIterateNeighborsEnd}.
Notifying a node in this context means increasing its localized neighbor count by one.
In line \ref{cblRigidLocalized}, we first increase the localized neighbor count of $w$ and then check if the updated value leads to the rigid-localization of the coplanar cluster of $w$.
If so, then we add the cluster of $w$ into $Q_{\textit{localized}}$ in line \ref{cblAddIntoQ}.
When there are no more coplanar clusters left to be localized, we check if all the clusters in the graph are localized.
If so, then we return the obtained point formation in line \ref{cblAllLocalized}.
In line \ref{cblFisBetter}, we check if the obtained point formation $\mathbb{F}$ contains more nodes than the best point formation $\mathbb{F}_{\text{best}}$.
If so, then we declare the new best point formation as the current one.
In line \ref{cblReturnBest}, we return the best point formation with respect to the number of nodes.
					
\begin{remark}
By the same reasoning given in Remark \ref{rem:trilatquad}, given a graph $G = (V,E)$ and $k$ coplanar clusters, CBL works in $O(k^2*|E|)$ time in the worst case.
\end{remark}

\subsection{Experimental Evaluation of CBL} \label{sec:cblexperiments}

In this section, we present the experimental results on the performance of CBL in various environments.
We conduct our simulations for the environments that show different types of characteristics in both planar clustering and localization phases.
These characteristics are basically modeled by the planarity factor $\mu$, calculated by dividing the number of coplanar clusters by the average number of nodes in each cluster and subtracting the result from 1.

\begin{align*}
\mu = 1 - \frac{k^2}{n}
\end{align*}
where $k$ is the number of nodes and $n$ is the number of clusters.

We measure the quality of a localization with respect to the changes in recall percentage (\textit{i.e.} the ratio of the localized nodes), and precision error (\textit{i.e.} positional offset per node).
The control parameter for the tests is picked as the magnitude of the environmental noise. 
It is verified experimentally that using the available information on coplanar clusters leads to a more precise localization than mere quadrilateration.
While evaluating the experimental results, we refer to the tests where information of clusters is utilized as \textbf{CBL}.
When the network is localized by mere quadrilateration, we refer to these type of tests as \textbf{quadrilateration}. In mere quadrilateration, the clustering information is not used even though it is available.

We conduct simulations using Java \cite{java} as the programming language and Eclipse IDE \cite{eclipse} as the development tool.
We developed an in-house simulator to simulate the environment by placing static nodes with a uniform distribution probability in a $100 \times 100 \times 100$ unit cubic volume.

We use two main types of deployments:
the deployments where the clusters are intersecting, and non-intersecting.
In these two types of deployments, we use various number nodes, and various number of clusters.
A deployment is named after the combination of the number of coplanar clusters, and the number of sensor nodes on each coplanar cluster separated by a dash in between.
We also use the letter I or D at the beginning in order to indicate if the clusters are intersecting or disjoint.
In Table \ref{table:deployments}, we give the deployments that we use along with the planarity factor of the deployment.
We divide the planarity factors of the deployments into three.
Namely, \textit{low}, \textit{average} and \textit{high}.
We say that the planarity factor is \textit{high} when $\mu \in \{0.988,0.985,0.980,0.960\}$, \textit{average} when $\mu \in \{0.920,0.900,0.840\}$ and \textit{low} when $\mu \in \{0.680,0.360,0.100,-0.800\}$.

\begin{table*}[h!]
\centering
\ra{1.3}
\begin{tabular}{@{}cccccccc@{}}\toprule
\multicolumn{2}{c}{\textbf{High}} & \phantom{ab}& \multicolumn{2}{c}{\textbf{Average}}& \phantom{ab}&
\multicolumn{2}{c}{\textbf{Low}}\\
\cmidrule{1-2} \cmidrule{4-5} \cmidrule{7-8}
\textbf{Deployment}& \textbf{$\mu$} && \textbf{Deployment}& \textbf{$\mu$} && \textbf{Deployment}& \textbf{$\mu$}\\ \midrule

I-3-200& 0.988 && I-4-100& 0.960 && D-16-50& 0.680\\
\rowcolor[HTML]{EFEFEF} 
I-3-270& 0.985 && D-8-100& 0.920 && D-16-25& 0.360\\
I-4-200& 0.980 && I-5-50& 0.900 && D-27-30& 0.100\\
\rowcolor[HTML]{EFEFEF} 
I-5-160& 0.968 && D-8-50& 0.840 && D-27-15& -0.800\\
\bottomrule
\end{tabular}
\caption{The names of the sub-type deployments with respect to the number of clusters and the number of nodes per cluster}
\label{table:deployments}
\end{table*}

The disjoint planar deployment is simulated with respect to the number of coplanar clusters, $k$.
In order to create disjoint clusters, we first divide the original cube into $k$ sub-cubes.
In each sub-cube, we generate a random plane using the plane equation $ax + by + cz + d = 0$ and scatter equal number of nodes on these planes.

\begin{figure}[htbp]
\centering
\subfloat[8 sub-cubes]{
\includegraphics[width=0.3\linewidth]{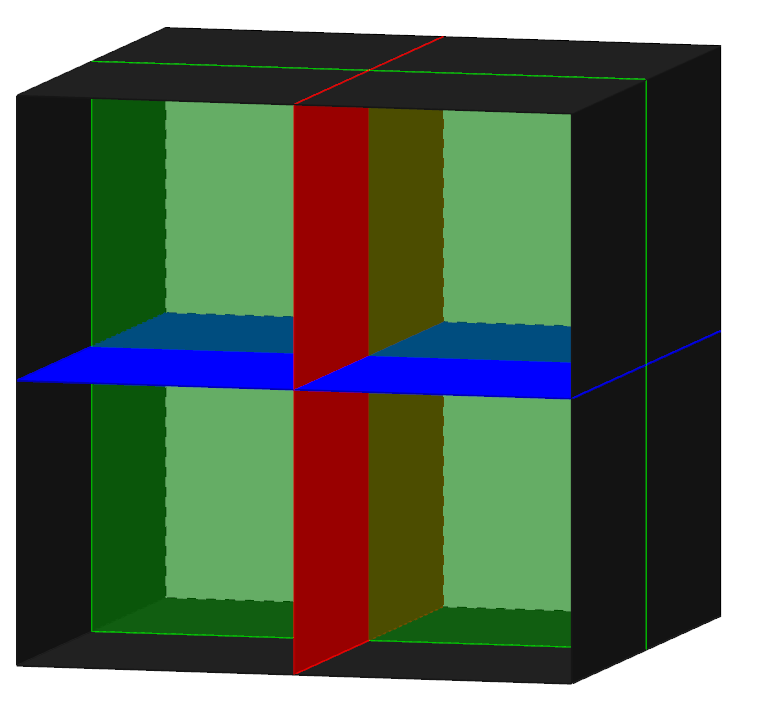}
\label{fig:8cubicles}
}
\subfloat[9 sub-cubes]{
\includegraphics[width=0.3\linewidth]{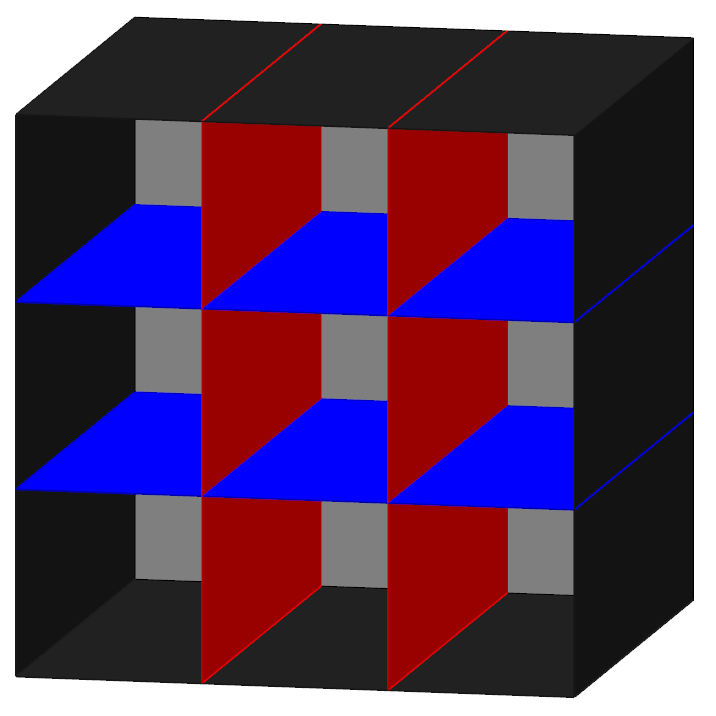}
\label{fig:9cubicles}
}\hfill
\caption{Slicing a cube with perpendicular planes}
\end{figure}

In Figure \ref{fig:8cubicles}, we see a cube divided into eight sub-cubes.
In Figure \ref{fig:9cubicles}, the same cube is divided into nine sub-cubes

In our simulations, we use unit ball graph as the sensing model and assume that the network graph is always connected.
We denote the sensing range by $R$.
In order to model noisy range measurements, we use empirically gathered noise data \cite{realworld1,realworld2}. 
Each edge $(v,w) \in E$ in $G=(V,E)$ is modified with respect to a random number generated using a Gaussian random distribution with $N(f(R), \mathcal{E}/100)$ where $\mathcal{E}$ is the magnitude of the error and $R$ is the sensing range.
$f(R)$ is defined as follows;

\begin{align*}
f(R) = 0.022 ln(1+R) - 0.038
\end{align*}

The large noise value is selected as $\mp P\%$ of the sensing range where $P$ is generated with a uniform random process between 0 and 10 units with probability 0.05.

\begin{figure}[htbp]
	\centering
	\subfloat{
		\includegraphics[width=0.8\linewidth]{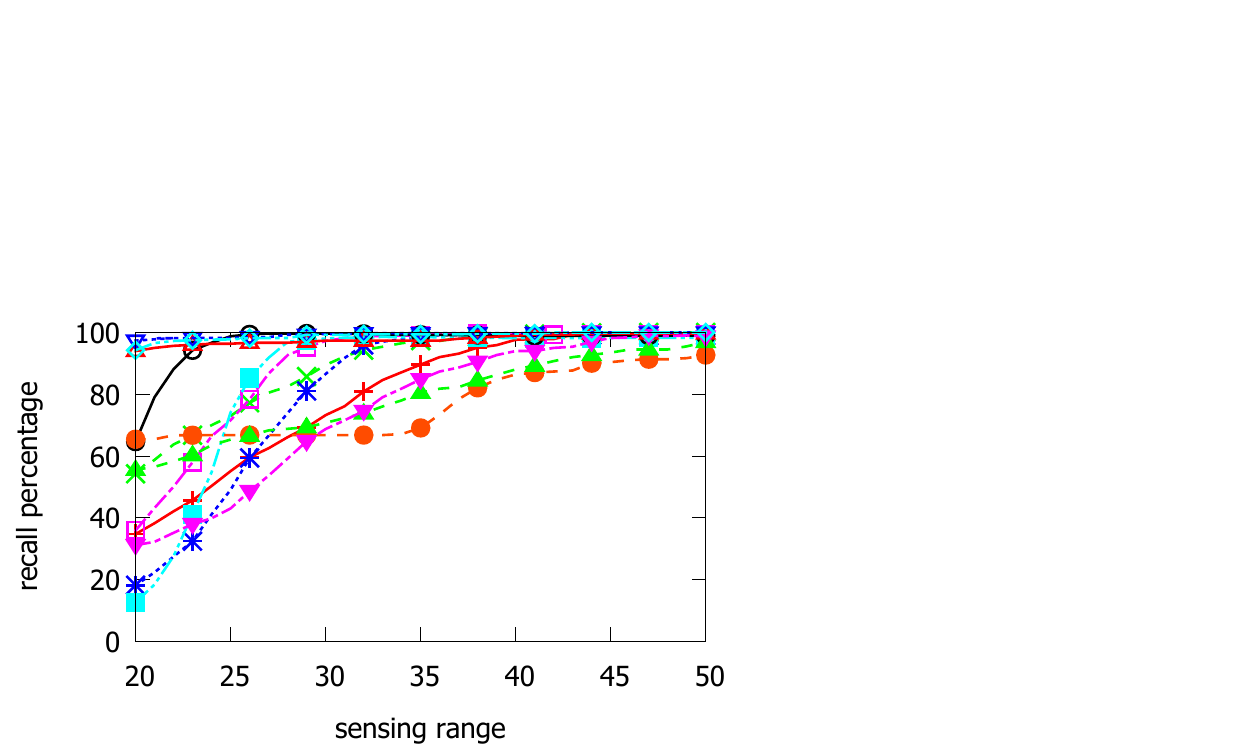}
	}
	
	\subfloat{
		\includegraphics[width=0.8\linewidth]{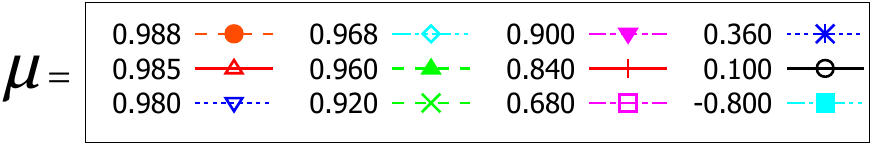}
	}
	\caption{Recall percentage of CBL with noiseless distance measurements.}
	\label{fig:recallCBL}
\end{figure}

First, let us present the recall percentage of CBL with noiseless distance measurements in Figure \ref{fig:recallCBL}.
Based on the recall percentages, we pick 40 units as the sensing range for the tests with noisy range measurements, where over $80\%$ of the nodes are localized for all values of $\mu$.

\begin{figure}[htbp]
\centering
\subfloat[Average offsets]{
	\includegraphics[width=\linewidth]{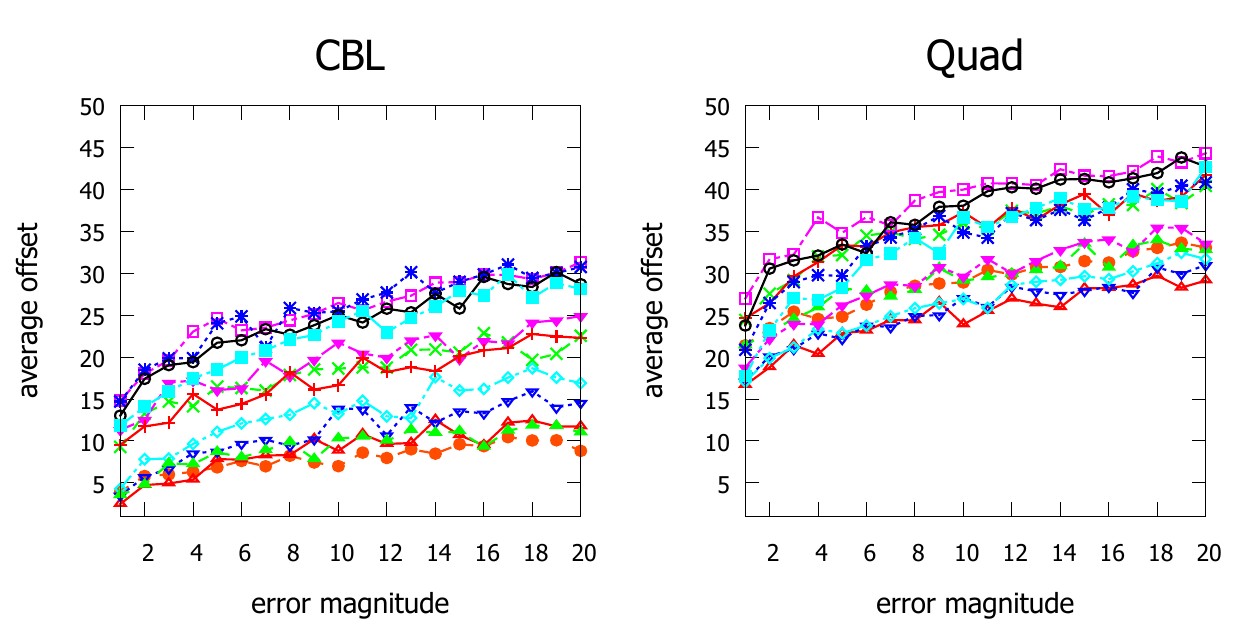}
	\label{fig:cblVSquadOffset}
}

\subfloat[Recall percentages]{
	\includegraphics[width=\linewidth]{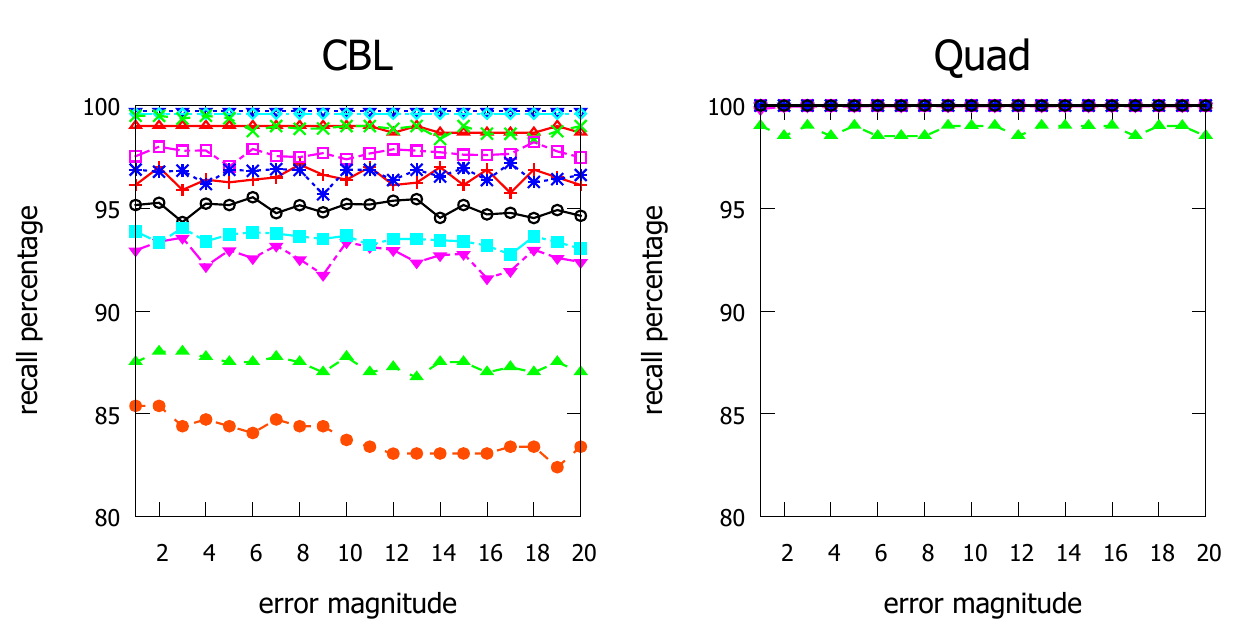}
	\label{fig:cblVSquadRecall}
	
}

\caption{The change in average offsets (a), and recall percentages (b) of CBL and quadrilateration with respect to increasing error magnitude and various planarity factors when the sensing range is 40 units}
\label{fig:CBLvsQuad}

\end{figure}

In Figure \ref{fig:cblVSquadOffset} we see the average offsets and in Figure \ref{fig:cblVSquadRecall} we see the recall percentages when the sensing range is 40 units and the error magnitude takes values in range between 1 and 20.
The figure tells us that exploiting the clustering information reduces the average offset of the localization.
Even though the recall percentage of CBL is lower than quadrilateration, we can say that instead of localizing a node imprecisely, CBL performs a more precise localization than quadrilateration.
Quadrilateration, on the other hand, is able to localize more nodes with more precision errors.

\begin{figure}[htpb]
	\centering
	\subfloat[CBL]{
		\includegraphics[width=0.8\linewidth]{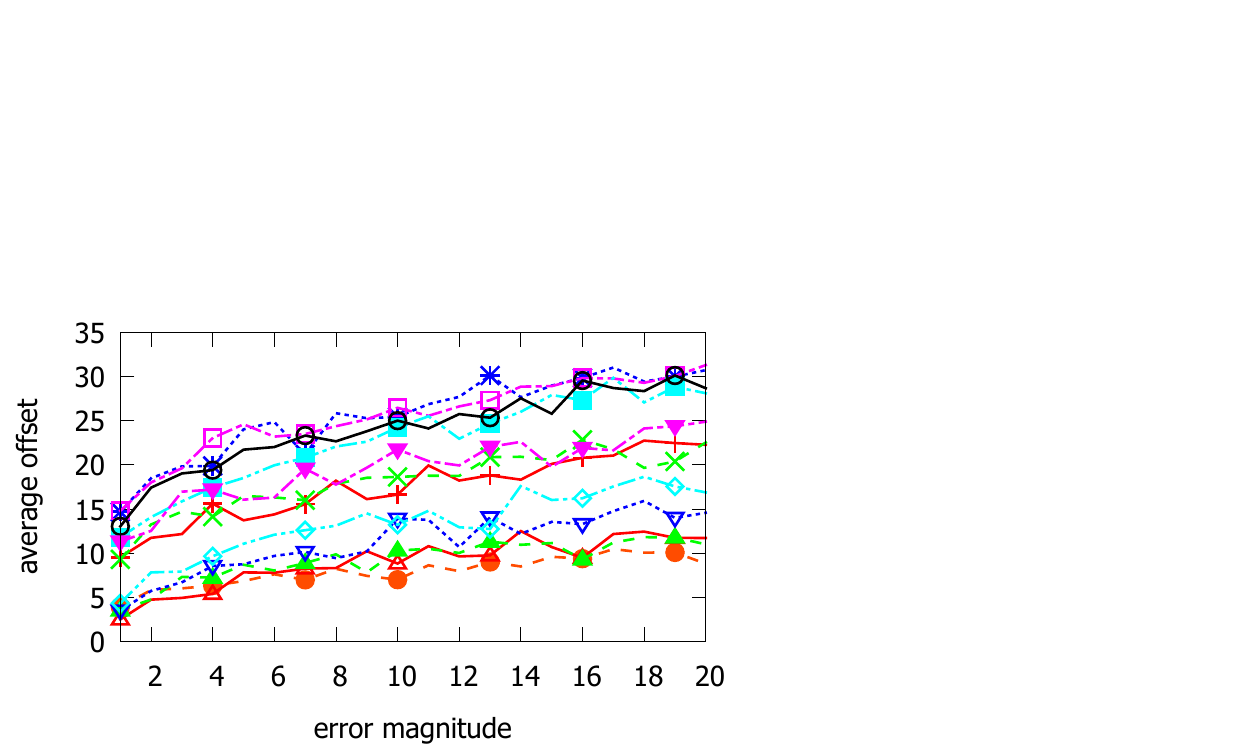}
		\label{fig:cblOffset}
	}
	
	\subfloat[Quadrilatereation]{
		\includegraphics[width=0.8\linewidth]{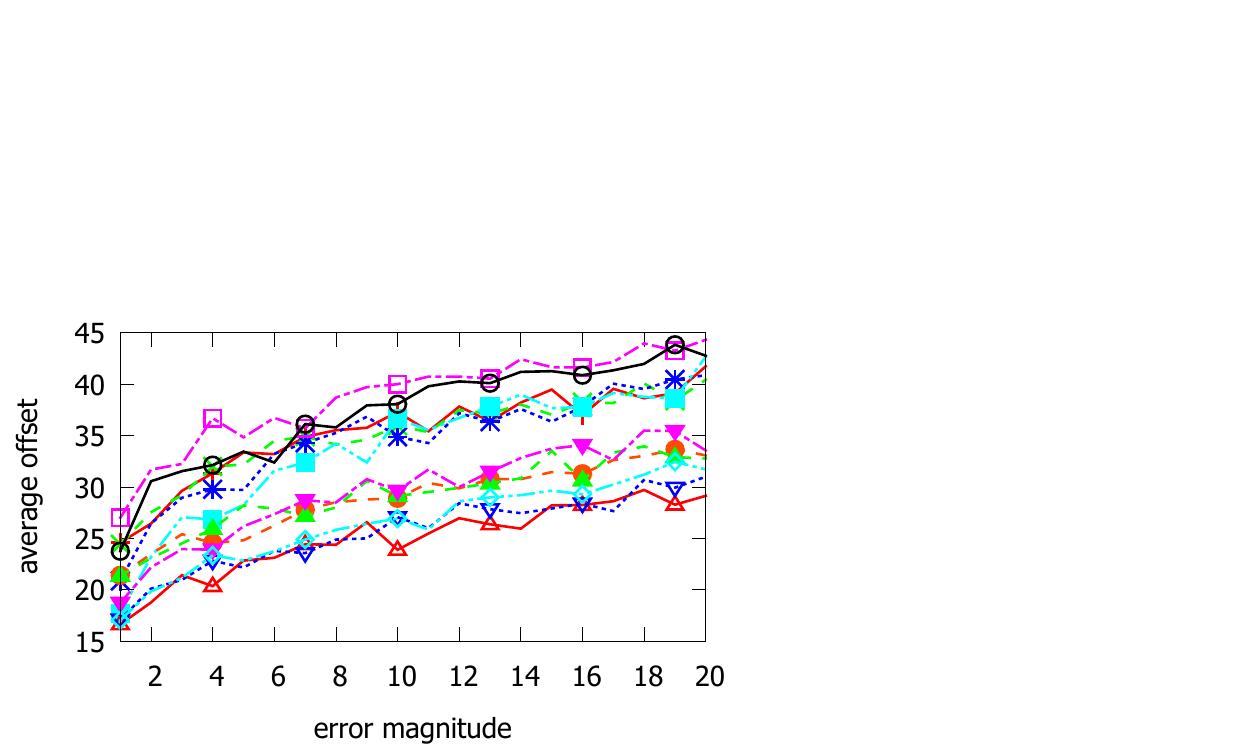}
		\label{fig:quadOffset}
	}
	
	\subfloat{
		\includegraphics[width=0.8\linewidth]{./legend}
	}
	
	\caption{A closer view of the plots in Figure \ref{fig:cblVSquadOffset}}
	\label{fig:cvsqOffsets}
\end{figure}

For a better comparison, we give the offset plots with a bigger scale in Figure \ref{fig:cvsqOffsets}.
In Figure \ref{fig:cblOffset}, we see the average offset of CBL.
In Figure \ref{fig:quadOffset}, we see the average offset of quadrilateration.
We see that ignoring the available clustering information increases the offset by 10 units per node in general
In Figure \ref{fig:cblOffset}, we see the average offset of CBL.
When the planarity factor is high, CBL causes less offset than the other cases.
Even though CBL is effected by the increasing error magnitude, the average offset stays under 20 units per node when $\mu \geq 0.960$.
When the planarity factor is average, CBL localizes the network almost as precise as the low planarity factor case.
However, when the error magnitude increases, the difference between two cases becomes more obvious.
When the planarity factor is low, CBL causes the highest average offset with an offset of around 30 units per node.
When the planarity factor is average, the average offset ranges between 15 units per node and 25 units per node.

\section{Extracting the Coplanar Clusters} \label{sec:extractingclusters}

In this section, we deal with the case where there is a planar deployment and the information on coplanar clusters are not available in 3D WSN localization problem.
First, we define the problem of extracting the coplanar clusters and then we propose a heuristic algorithm to solve the defined problem.

\subsection{Problem Definition}
\label{sec:problemdefinitionextract}

Given a set of points in 3D with no position information, the problem of extracting the coplanar clusters is to cover these points with $k$ planes based only on the available pairwise distances.
Given a WSN graph $G = (V,E)$, the points correspond to the nodes in $V$ and the pairwise distances are the edge set $E$.
We refer to the partitions $\{C_1 = (V_1, E_1), C_2 = (V_2, E_2), \dots, C_k = (V_k, E_k)\}$ as \textit{coplanar clusters}.
Each node included into a coplanar cluster is called an \textit{on-plane node} while the rest is called as \textit{off-plane node}s.
In 2D, a variation of this problem is presented as "\textit{Covering a Set of Points with a Minimum Number of Lines}" in \cite{maximumcoverage}.
The problem presented assumes that the coordinates of the points are known.
Even though with known positions, it is proved to be inapproximable in \cite{coverageapprox}.
Instead of the positions, we are only able to use pairwise distances.
Considering that these distances might be inaccurate, it is conjectured that extracting the coplanar clusters is also a difficult problem.

For checking the coplanarity of four nodes, we use Cayley-Menger determinant (for the details on Cayley-Menger determinant usage, see Chapter \ref{chap:background}).
Trying to extract the coplanar clusters by using noisy pairwise distance measurements have two main challenges.

\begin{enumerate}[i)]

	\item Even though the points are coplanar, because of the inaccurate distances, Cayley-Menger determinant gives either positive or an imaginary number as its result. Therefore, we set a volume threshold $\kappa$ to accept the results less than some value.
	As a result of experimental verification, this value is set as $6*ln(\mathcal{E}+1)$ where $\mathcal{E}$ is the error magnitude.
	
	\item When the distance between a group of interplanar neighbors are too close, they might be interpreted as coplanar because of the volume threshold. In order to avoid this, also use a value called \textit{hop distance} denoted by $\theta$.
	The edges whose values are less than $\theta$ are unavailable during the extension process.
	This value is set to half the average edge weight, and increased by one at each iteration during the extension phase of the coplanar clusters.
\end{enumerate}

\subsubsection{Setting the Volume Threshold $\kappa$}

As the distance measurements might possibly be inaccurate, searching for a perfect plane, would be a waste of effort.
Therefore, we set a threshold value, $\kappa$, for the volume of the tetrahedron formed by the six distance measurements among four nodes $\{a,b,c,d\}$.
We say that a node subset $\{a,b,c,d\}$ are coplanar if $\mathcal{V}_{abcd} \leq \kappa$, where $\mathcal{V}_{abcd}$ is the volume of the tetrahedron formed by $a$, $b$, $c$ and $d$ as given by Cayley-Menger determinant.

After finding a set of four coplanar nodes $\{a,b,c,d\}$, we create a coplanar cluster $C_i = (\{a,b,c,d\})$.
Then we try to add more nodes into $C_i$ by finding other nodes that are also coplanar with three of the nodes inside that cluster.
This process is called \textit{extending} that cluster.
We add a node $n$ into a cluster $C_i$ if $V_{nabc} \leq \kappa$ such that $\{a,b,c\} \subset C_i$.
When there is no node left to be added into $C_i$, we search for another cluster $C_j \neq C_i$, and then extend $C_j$ as described above.
We continue finding and extending clusters until there are no more off-plane nodes left.
The whole operation is called \textit{extracting the coplanar clusters} or \textit{planar clustering}.

While extracting the coplanar node clusters, the volume threshold value needs to be set carefully.
Otherwise, a coplanar cluster might be expanded to include nodes belonging to a completely different coplanar cluster.
This situation arises particularly when coplanar clusters are deployed on many planes are very close to each other.
In this case, it becomes a major issue to tell whether the variations in the volume of tetrahedra stem from the errors in the distance measurements or picking nodes in the vicinity of the intersection of actually different planes.
In order to set a volume threshold, 10000 tests are run by placing four points $a,b,c,d$ in a unit square.
We take the average difference between actual volume $\mathcal{V}_{abcd}^{\text{act}}$ and the volume computed with the noisy distance measurements $\mathcal{V}_{abcd}^{\text{comp}}$.
As a result, the volume threshold $\kappa$ is defined as $6*ln(1+\mathcal{E})$ where $\mathcal{E}$ is the error magnitude.

\begin{figure}[htbp]
	\centering
	\subfloat[A tetrahedron whose volume is zero]{%
		\includegraphics[width=0.7\linewidth]{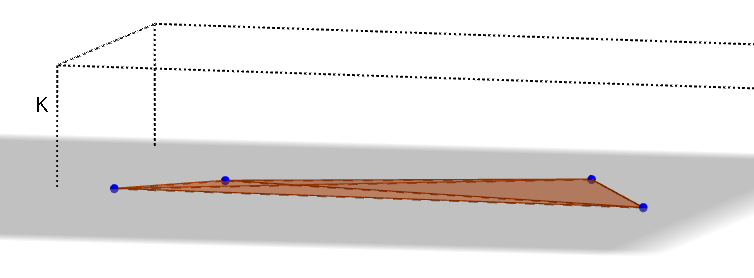}
		\label{fig:volumethresholdcoplanar}%
	}
	
	\subfloat[A tetrahedron whose volume is between zero and the volume threshold]{%
		\includegraphics[width=0.7\linewidth]{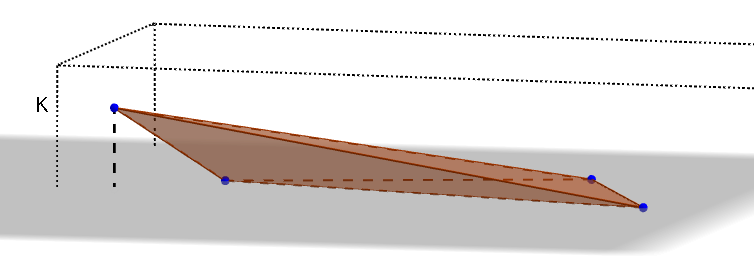}
		\label{fig:volumethresholdinside}%
	}
	
	\subfloat[A tetrahedron whose volume is greater than the volume threshold]{%
		\includegraphics[width=0.7\linewidth]{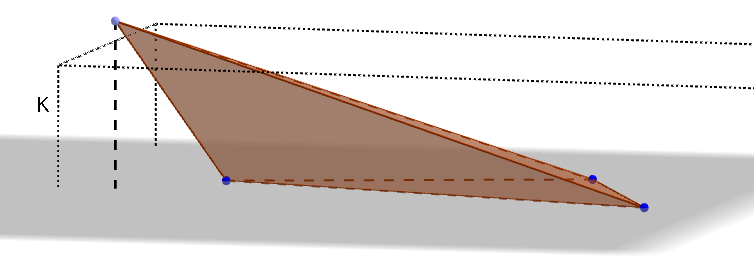}
		\label{fig:volumethresholdoutside}%
	}
	\caption{A coplanar group of nodes (a), a non-coplanar group of nodes, interpreted as coplanar (b), a non-coplanar group of nodes, not interpreted as coplanar (c)}
	\label{fig:volumethreshold}
\end{figure}

Figure \ref{fig:volumethreshold} demonstrates three examples for the usage of volume threshold.
In Figure \ref{fig:volumethresholdcoplanar}, there are four coplanar nodes.
The volume of the tetrahedron formed by these nodes is zero.
When the distance measurements are noisy, the same tetrahedron might be formed differently, as seen in Figure \ref{fig:volumethresholdinside}.
Even though the nodes in Figure \ref{fig:volumethresholdinside} are not coplanar, they are treated as coplanar because the volume of the formed tetrahedron is less than the volume threshold.
In Figure \ref{fig:volumethresholdoutside}, the volume of the tetrahedron formed by four nodes is bigger than the threshold.
Thus, the four nodes are interpreted as non-coplanar nodes.

\begin{figure}[htbp]
	\centering
	\subfloat[]{
		\includegraphics[width=\linewidth]{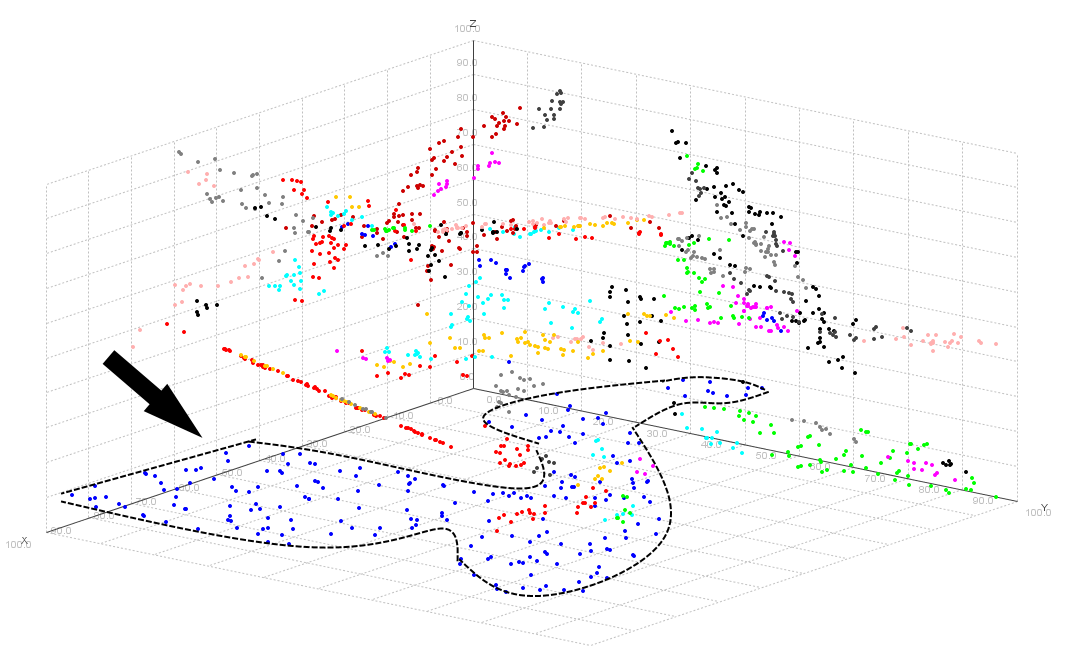}
		\label{fig:wca1}
	}
	
	\subfloat[]{%
		\includegraphics[width=0.5\linewidth]{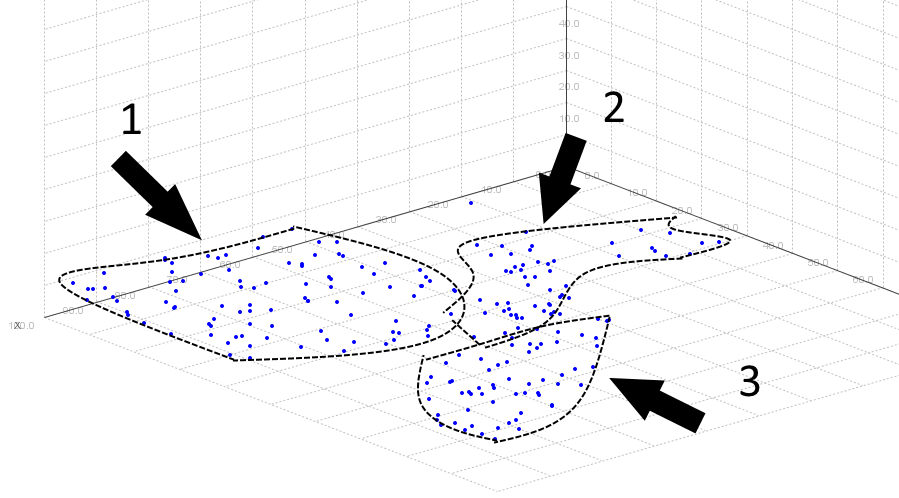}
		\label{fig:wc1}%
	}
	\subfloat[]{%
		\includegraphics[width=0.5\linewidth]{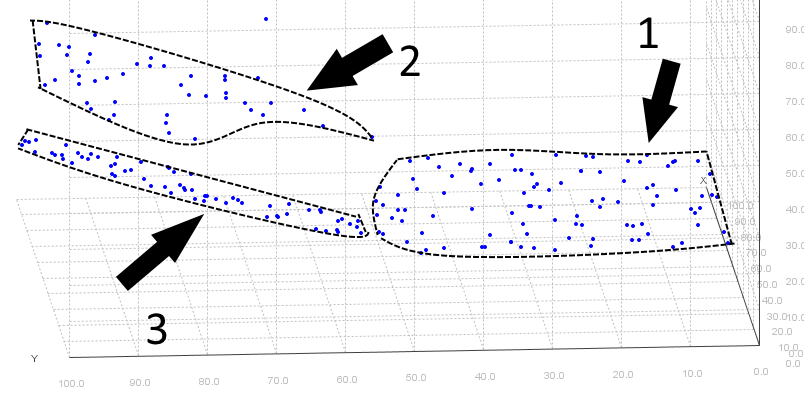}
		\label{fig:wc2}%
	}
	\caption{An example of inappropriate expansion from two angles of view}
	\label{fig:wrongClustering}
\end{figure}

If a group of coplanar nodes are within the communication range of nodes from another coplanar cluster, the pairwise distances between interplanar neighbors might be too small.
Thus, the volumes of the formed tetrahedra could be smaller than that specified by the volume threshold, even though the nodes are in different coplanar clusters.
In Figure \ref{fig:wca1}, we see a planar deployment where there are 16 coplanar node clusters.
Each different color indicates a coplanar cluster found.
The cluster that is pointed at by a black arrow consists of nodes from actually three different coplanar node clusters.
This kind of clustering leads to very high offsets since all the nodes in the found cluster will be localized imprecisely both in 2D and in 3D.
Figure \ref{fig:wrongClustering} demonstrates the improper expansion from two angles of view.
In Figure \ref{fig:wc1}, three black arrows with numbers indicate three different clusters.
In Figure \ref{fig:wc2}, we see the same deployment from a different angle of view.

This type of extensions occur when the volume threshold is greater than zero and the interplanar edges are used frequently.
In order to limit the usage of interplanar edges, we take the following precautions.

\begin{figure}[htbp]
\centering
\includegraphics[width=\linewidth]{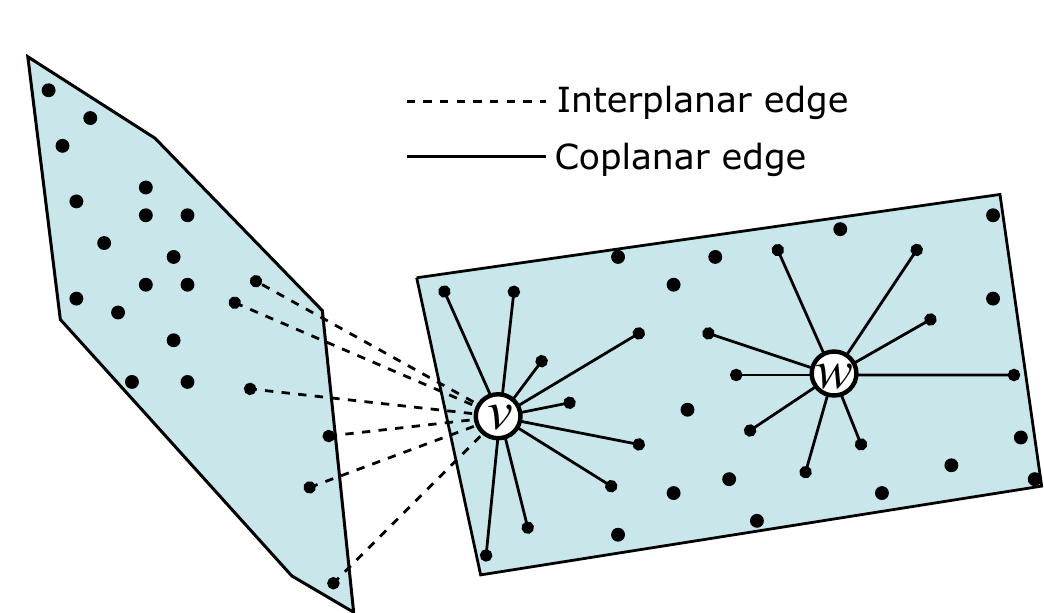}
\caption{Two nodes from a coplanar cluster, and their neighbors}
\label{fig:connections}
\end{figure}

\begin{itemize}
\item We sort the nodes by their connectivities and seek for a set of four coplanar nodes among the nodes with the lowest connectivity. Picking the nodes with smaller number of neighbors reduces the possibility to use interplanar connections while extending a cluster. As seen in Figure \ref{fig:connections}, the node $v$ has more neighbors since it is closer to another cluster while node $w$ has coplanar neighbors only. 
The interplanar edges are indicated with dashed lines and coplanar edges are indicated with straight lines.
\item We ignore the edges with higher weights than a pre-defined value called \textit{hop distance} and denoted by $\theta$.
Notice that in Figure \ref{fig:connections}, the interplanar edges are usually longer than the coplanar edges. Thus, by setting a hop distance, we reduce the possibility of checking the coplanarity with interplanar neighbors.
\end{itemize}

\subsubsection{Setting the Hop Distance $\theta$}
\begin{figure}[htbp]
	\centering
	\subfloat[$\theta = 10$]{%
		\includegraphics[width=0.7\linewidth]{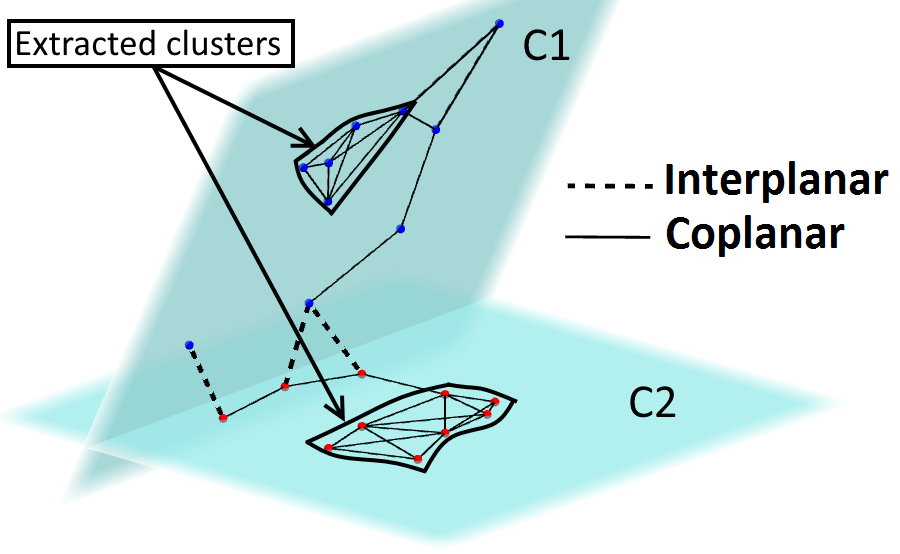}
		\label{fig:10hop}%
	}
	
	\subfloat[$\theta = 20$]{%
		\includegraphics[width=0.7\linewidth]{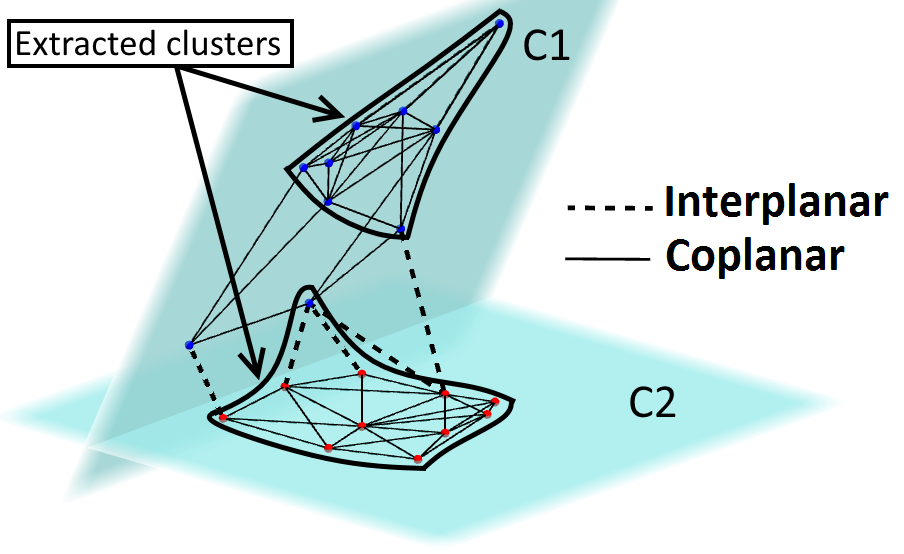}
		\label{fig:20hop}%
	}
	
	\subfloat[$\theta = 30$]{%
		\includegraphics[width=0.7\linewidth]{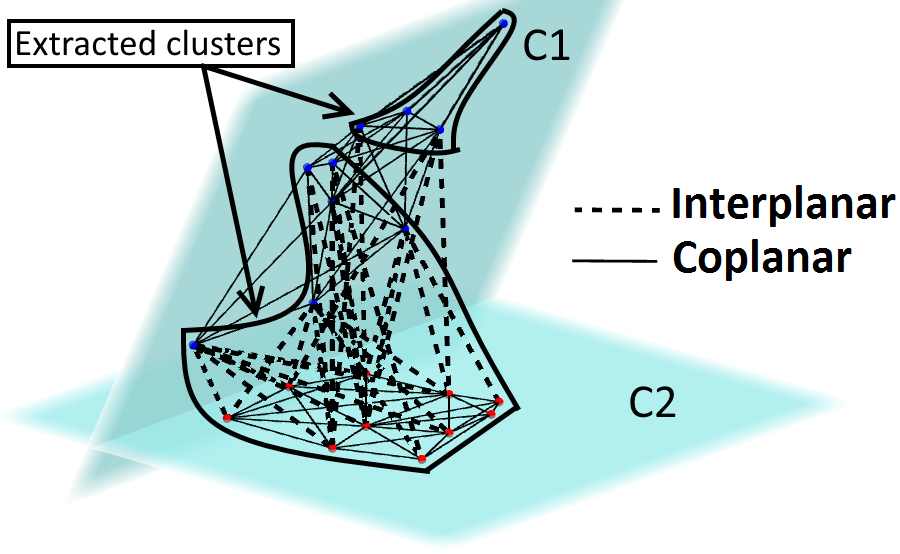}
		\label{fig:30hop}%
	}	
	\caption{Filtering edges with respect to changing hop distance}
	\label{fig:hopdistance}
\end{figure}

As well as the volume threshold, the hop distance is also very important to achieve an accurate extraction of coplanar clusters.
Let us explain the effect of $\theta$ with an example.
In Figure \ref{fig:hopdistance}, we see two coplanar clusters, denoted by $C1$ and $C2$, with ten nodes on each.
The coplanar edges are denoted by straight lines and interplanar edges are denoted by dashed lines.
In Figure \ref{fig:10hop}, the hop distance is set as 10 units.
Therefore, while extending a cluster, the distance values greater that 10 units are ignored in order to avoid the extension demonstrated in Figure \ref{fig:wrongClustering}.
However, seven out of twenty nodes are interpreted as off-plane because of insufficient connectivity, even though they are in a coplanar cluster.
In Figure \ref{fig:20hop}, the hop distance is set as 20 units.
In this case, only two nodes are interpreted as off-plane.
Notice that one of the nodes in $C1$ has three interplanar neighbors, and it forms a tetrahedron whose volume is less than the volume threshold.
Hence, it was included into wrong cluster.
In Figure \ref{fig:30hop}, the hop distance is set to 30 units.
Because of dense interplanar edges, after a small number of nodes from $C1$ is included into $C2$, the extension of $C2$ continues including the nodes in $C1$.

\begin{figure}[htbp]
\centering
	\subfloat[Sensor nodes deployed on one plane]{
	\includegraphics[width=\linewidth]{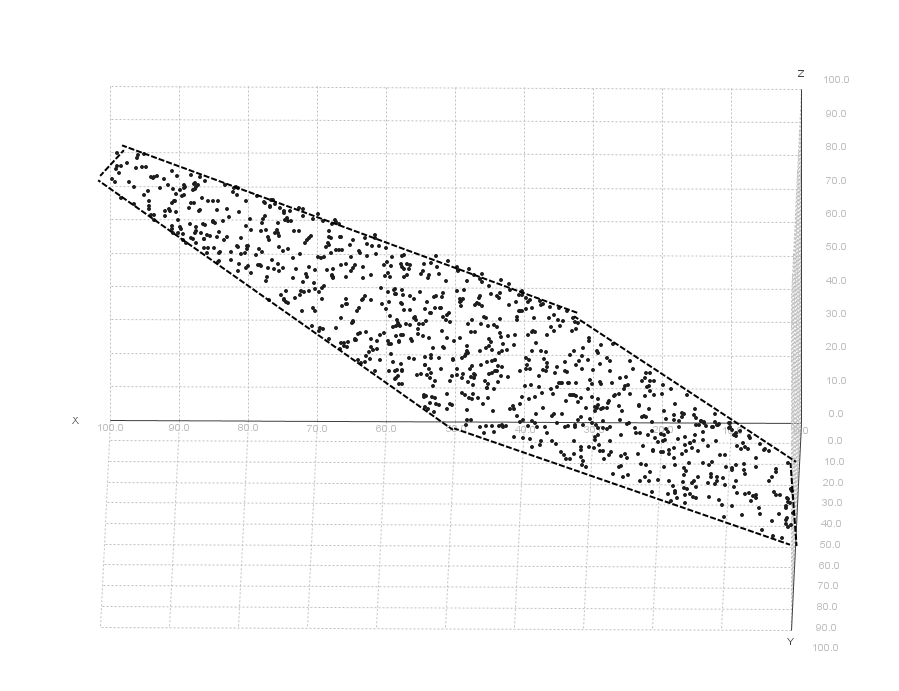}
	\label{fig:originalC3D}%
	}
	
	\subfloat[2D view of the coplanar cluster]{%
	\includegraphics[width=0.5\linewidth]{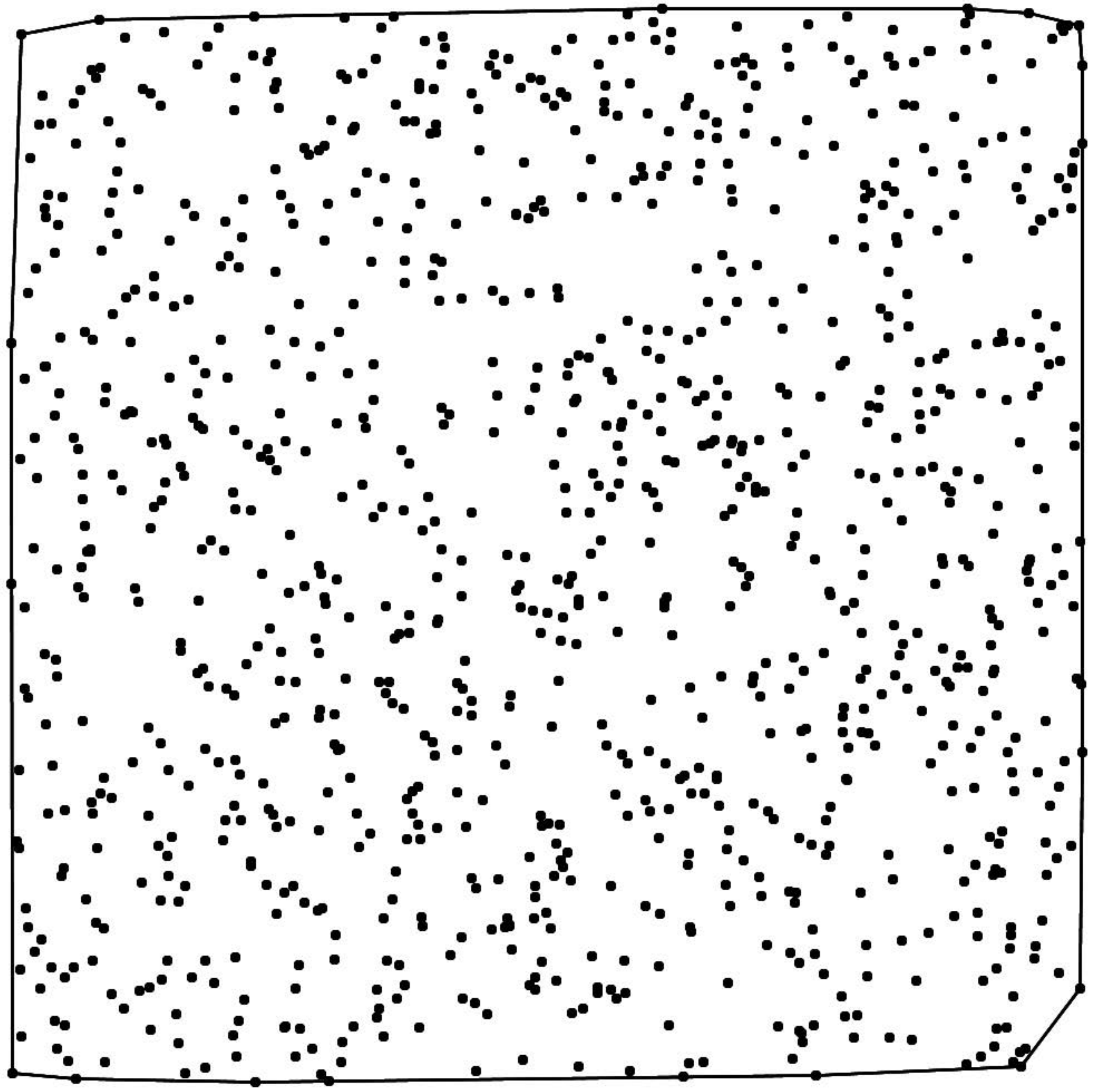}
	\label{fig:originalC}%
	}
	\subfloat[Extracted clusters out of the original one]{%
	\includegraphics[width=0.5\linewidth]{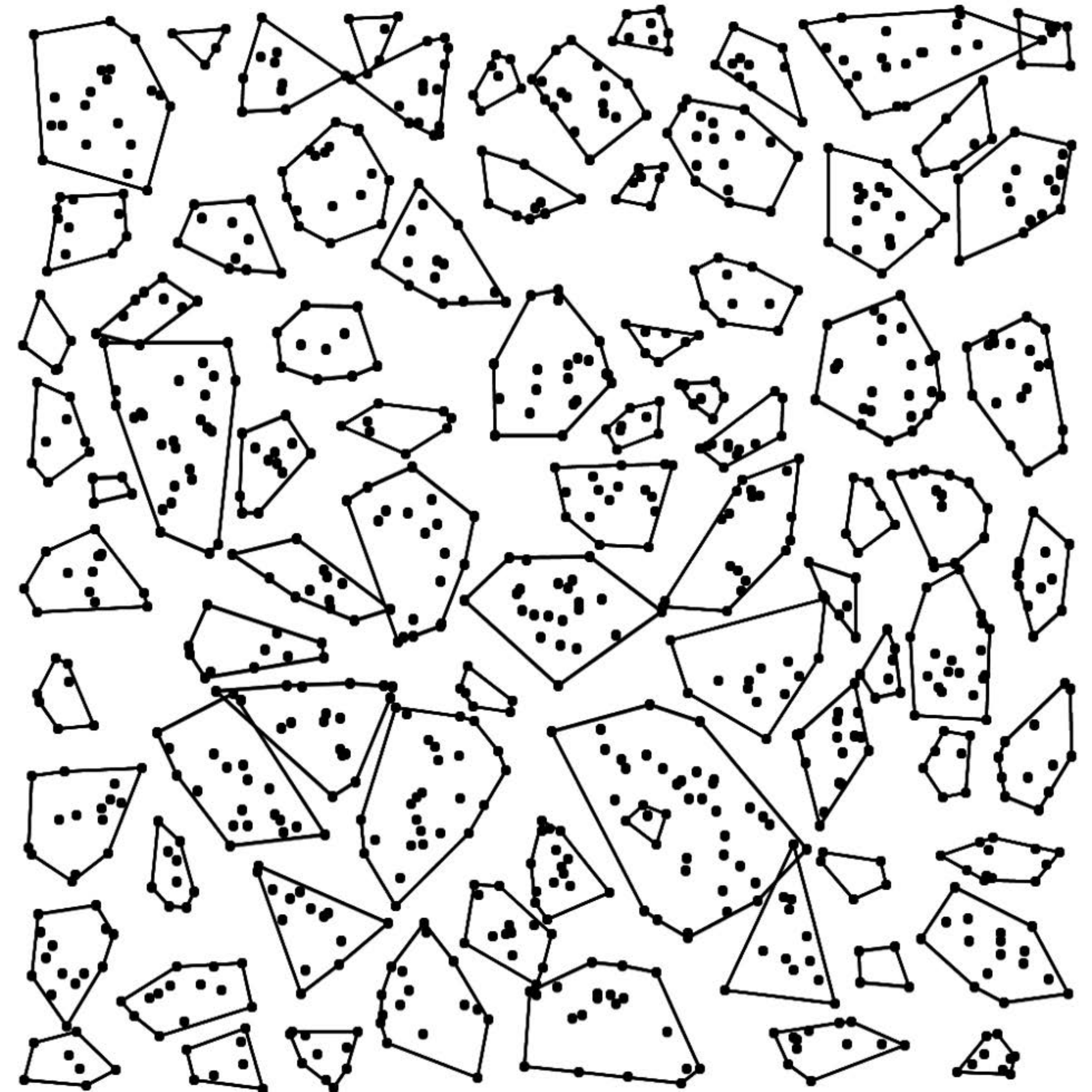}
	\label{fig:foundC}%
	}
\caption{Planar clustering of a coplanar node set}
\label{fig:originalVSfound}
\end{figure}

Considering the deployment given in Figure \ref{fig:hopdistance}, we can say that for this specific deployment, 20 units is the best hop distance.
However, without any information on the deployment, setting the hop distance as an arbitrary value may end up with extracting $k' \gg k$ coplanar clusters where $k'$ is the number of extracted clusters and $k$ is the number of actual clusters.
An example of this case is demonstrated in Figure \ref{fig:originalVSfound} with a coplanar cluster with 1000 nodes. 
Assume that all the nodes are deployed on a single plane \textit{i.e.} there is only one coplanar cluster, as seen in Figure \ref{fig:originalC3D}.
In Figure \ref{fig:originalC}, we see the 2D view of the coplanar cluster shown in Figure \ref{fig:originalC3D}.
If we know that there is only one coplanar cluster, then we are able to run 2D localization algorithm on that cluster and determine the global positions of all nodes the network.
Otherwise, we have to extract the coplanar clusters.
Considering the noisy distance measurements, if we set a hop distance very small, then we end up with extracting excessive number of clusters out of one.
In Figure \ref{fig:foundC}, we see the extracted clusters out of the original one when the hop distance is 10 units.

\pagebreak
To sum up, there are two types of improper extensions that we have to avoid.
\begin{enumerate}[i)]
	\item \textit{Over-extension:} The extension of a coplanar cluster to include more than one cluster, as seen in Figure \ref{fig:wrongClustering}.
	
	\item \textit{Short-extension:} The extension of a coplanar cluster that covers only a tiny part of the actual coplanar cluster, as seen in \ref{fig:originalVSfound}.
\end{enumerate}

We use the volume threshold $\kappa$ and the hop distance $\theta$ in order to avoid improper extensions.
$\kappa$ is set once at the beginning and is not changed during the extraction process.
Finding the best value for $\theta$ leads to the extraction of the actual coplanar clusters, which is very difficult if we do not have information on the deployment.
Therefore, we set the hop distance as half the average edge weight in the WSN graph $G=(V,E)$.

\begin{align*}
\theta = \dfrac{\mathlarger\sum\limits_{(v,w) \in E} d(v,w)}{2*|E|}
\end{align*}
\noindent where $E$ is the edge set and $|E|$ is the number of edges in the graph.

After one round of extensions, if there are still off-plane nodes, then we increase $\theta$ by one to include more nodes into the coplanar clusters.
When $\theta$ is increased, that means some unavailable edges become available for use while extending clusters.
The algorithm that we propose to find the 3D point formation of a WSN graph $G$ is presented in Section \ref{sec:heuristictoextract}.

\subsection{A Heuristic to Extract Coplanar Clusters} \label{sec:heuristictoextract}
In this section, we give our heuristic algorithm to extract the coplanar node clusters in a given WSN graph.
We assume that the number of coplanar clusters is unknown as well as the number of sensors in each cluster.
We present our heuristic as a two part algorithm in \ref{alg:planarClustering}.
In Figure \ref{alg:extractClustersOverall}, we give the overall algorithm to extract the coplanar clusters.
The algorithm tries to extract coplanar clusters by using the available pairwise distances in WSN graph $G=(V,E)$ and adds the extracted clusters into coplanar cluster set \textit{CSet}.
Since the algorithm only modifies the global variables, it does not take any input.
The threshold value $\kappa$ and the hop distance $\theta$ are set inside the algorithm.
Figure \ref{alg:extendCluster} shows the extension process of a coplanar cluster.

Remember that a coplanar cluster is denoted by $C_i = (V_i, E_i)$. 
We present the heuristic by assuming that as a node is placed into a coplanar cluster $C_i$, it is added into the vertex set $V_i$, and the coplanar edges of this node are added into the edge set $E_i$.

\begin{figure}[htbp]
\subfloat[Overall planar clustering algorithm]{
\fbox{\begin{minipage}{\columnwidth}
\begin{algorithmic}[1]
\Input{WSN graph $G = (V,E)$}
\Output{Coplanar cluster set \textit{CSet}}
\Function{ExtractClusters}{$G$}
\State $\kappa \gets 6*ln(\mathcal{E} + 1)$ \Comment{$\mathcal{E}$ is the error magnitude} \label{setKappa}
\State $\theta \gets \dfrac{\sum\limits_{(v,w) \in E} d(v,w)}{2*|E|}$ \Comment{Average edge weight divided by two} \label{setTheta}
\lComment{$E$ is the edge set, and $|E|$ is the number of edges in $G = (V,E)$ }
\State $\textit{OffPlane} \gets V$ \label{allOffPlane}
\State sort the nodes in \textit{OffPlane} by connectivity in ascending order \label{sortOffPlane}
\State $i \gets 0$ \label{clusterIndex}
\For{Each fully-connected $\{a,b,c,d\} \subset \textit{OffPlane}$} \label{pcInitialFor}
	\If{$\mathcal{V}_{abcd} \leq \kappa$} \label{pcClusterFound}
		\State $C_i \gets \{a,b,c,d\}$ \label{pcCreateCluster}
		\State \Call{ExtendCluster}{$C_i$, \textit{OffPlane}, $\theta$, $\kappa$} \label{pcExtendClusterInitial}
		\State add $C_i$ into \textit{CSet} \label{pcAddIntoCSet}
		\State $i \gets i+1$ \label{increaseIndex}
	\EndIf
	\endIf
\EndFor
\endFor \label{pcInitialForEnd}
\While{($\textit{OffPlane} \neq \emptyset$) AND ($\theta \leq \textit{sensing range}$)} \label{pcSecondExtensionWhile}
	\State$\theta \gets \theta + 1$ \label{pcUpdateTheta}
	\For{Each $C_i \in \textit{CSet}$} 
		\Call{ExtendCluster}{$C_i$, \textit{OffPlane}, $\theta$, $\kappa$} \label{pcExtendCluster}
	\EndFor
	
\EndWhile
\endWhile \label{pcSecondExtensionEndWhile}
\EndFunction
\endFunction
\end{algorithmic} \end{minipage}}
\label{alg:extractClustersOverall}
}

\subfloat[Extending a coplanar cluster]{
\fbox{\begin{minipage}{\columnwidth}
\begin{algorithmic}[1]
\Function{ExtendCluster}{$C_i$, \textit{OffPlane}, $\theta$, $\kappa$}
\Repeat \label{ecRepeat}
	\State $\textit{Newbies} \gets \emptyset$ \label{ecNewbies}
	\For{Each $\{a,b,c\} \subset C_i$} \label{ecIterateTriplets}
		\For{Each off-plane node $d \in \textit{OffPlane}$} \label{ecFilterEdgesFor}
			\If{($d(a,d) \leq \theta$) AND ($d(b,d) \leq \theta$) AND ($d(c,d) \leq \theta$)}
			\lComment{$d(a,d)$ denotes the Euclidean distance between $a$ and $d$} \label{ecFilterEdgesCondition}
				\If{$\mathcal{V}_{abcd} \leq \kappa$} \label{ecVolumeLess}
					add $d$ into \textit{Newbies} \label{ecAddIntoNewbies}
				\EndIf
			\EndIf
		\EndFor
	\EndFor
	\State $C_i \gets C_i \cup \textit{Newbies}$ \label{ecAaddNewbiesIntoC}
	\State $\textit{OffPlane} \gets \textit{OffPlane} \setminus \textit{Newbies}$ \label{ecRemoveFromOffPlane}
\Until{$\textit{Newbies} = \emptyset$} \label{ecUntil}
\EndFunction
\endFunction
\end{algorithmic}  \end{minipage}}
\label{alg:extendCluster}
}
\caption{Heuristic to extract coplanar clusters}
\label{alg:planarClustering}
\end{figure}

The algorithm given in Figure \ref{alg:extractClustersOverall} uses the WSN graph $G = (V,E)$ and the coplanar cluster set \textit{CSet} which are defined as global variables.
Notice that \textit{CSet} is empty at the beginning.
In line \ref{setKappa}, we set the value of the volume threshold.
In line \ref{setTheta}, the hop distance is set.
In line \ref{allOffPlane}, we mark all the nodes in the vertex set of WSN graph as off plane by adding them into the set called \textit{OffPlane}.
Then, we sort the nodes in \textit{OffPlane} with respect to the number of neighbors they have from lowest to highest in line \ref{sortOffPlane}.
In line \ref{clusterIndex}, we set an integer value $i$ as zero.
This value will be used as the indices of the found clusters later on.
In order to find a cluster, we iterate on each node quadruplet $(a,b,c,d)$ in the off-plane node set in line \ref{pcInitialFor}.
We check if these four nodes form a tetrahedron with a volume smaller than the volume threshold in line \ref{pcClusterFound}.
If so, then we create a cluster in line \ref{pcCreateCluster}.
Then, we call the function to extend a coplanar cluster \textsc{ExtendCluster($C_i$, \textit{OffPlane}, $\theta$, $\kappa$)} where $C_i$ is the cluster created in line \ref{pcCreateCluster}, \textit{OffPlane} is the set of off-plane nodes, $\theta$ is the hop distance set and $\kappa$ is the volume threshold.
After extension of the cluster is done, we add the cluster into the coplanar cluster set in line \ref{pcAddIntoCSet} and then increase $i$ by one in line \ref{increaseIndex}.
After we find a cluster and finish extending it, we search for a new cluster.
When all the fully-connected off-plane node quadruplets are iterated and no clusters are found, we move onto the next phase.
Because of the limited edges, it is possible that some of the nodes are left out while finding and extending clusters.
Therefore, from line \ref{pcSecondExtensionWhile} through line \ref{pcSecondExtensionEndWhile} we keep on extending the found clusters.
This second extension phase continues until either the increasing hop distance is equal to the sensing range or there are no more off-plane nodes left.
In line \ref{pcUpdateTheta}, we increase the hop distance by one unit.
Of course, the increment might be different if the nodes are distributed into a cube with a different volume.
After updating the hop distance, we iterate on and extend each coplanar cluster in the cluster set in line \ref{pcExtendCluster}.

The extension process of a coplanar cluster is presented in Figure \ref{alg:extendCluster}.
The algorithm takes four parameters.
Namely, a coplanar cluster to be extended $C_i$, the set of off-plane nodes \textit{OffPlane}, the hop distance $\theta$ and the volume threshold $\kappa$.
We search for off-plane nodes that is coplanar with any three of the nodes in the coplanar cluster from line \ref{ecRepeat} through line \ref{ecUntil}.
In line \ref{ecNewbies}, we create a set called \textit{Newbies} to store the new nodes that will be added into the coplanar cluster.
In line \ref{ecIterateTriplets}, we iterate on each node triplet $(a,b,c)$ in the coplanar cluster $C_i$.
In line \ref{ecFilterEdgesFor}, we iterate on each off-plane node $d$ and we check if $d$ is closer to $a$, $b$ and $c$ than the hop distance in line \ref{ecFilterEdgesCondition}.
We assume that if a pairwise distance is not available \textit{i.e.} two sensor nodes cannot sense each other, the distance value is infinite.
If all three pairwise distances between $d$ and $a$, $b$, $c$ are less than $\theta$, then we check if the volume $\mathcal{V}_{abcd}$ is less than the allowed volume threshold $\kappa$.
If all the conditions are satisfied, then we add $l$ into \textit{Newbies} in line \ref{ecVolumeLess}.
When all the triplets in the cluster are iterated, we add the found coplanar nodes into coplanar cluster $C_i$ in line \ref{ecAaddNewbiesIntoC}.
We also remove them from off-plane node set in line \ref{ecRemoveFromOffPlane}.
The algorithm runs until there are no new nodes to add into the coplanar cluster that is being extended.

\begin{remark}
In the algorithm presented in Figure \ref{alg:extendCluster},
it might seem redundant to iterate on each triplet in line \ref{ecIterateTriplets} at each iteration of extraction.
However, the new nodes added into the coplanar cluster in line \ref{ecAaddNewbiesIntoC} might have common off-plane neighbors with a node that is already in the cluster.
Since we do not want to skip that off-plane neighbor, we iterate all the triplets all over again.
\end{remark}

\subsection{Experimental Evaluation of Planar Clustering}

In this section, we present the experimental results on the performance of CBL, run after the heuristic that we propose in Figure \ref{alg:extendCluster} with the same setup used in Section \ref{sec:cblexperiments}.
We have shown that CBL localizes the given network more precisely in Section \ref{sec:cblexperiments}.
In this section, we assume that the clustering information is not present to test the proposed heuristic to extract coplanar clusters presented in Figure \ref{alg:planarClustering}.

\begin{figure}[htpb]
\centering
\subfloat[Average offsets]{
	\includegraphics[width=\linewidth]{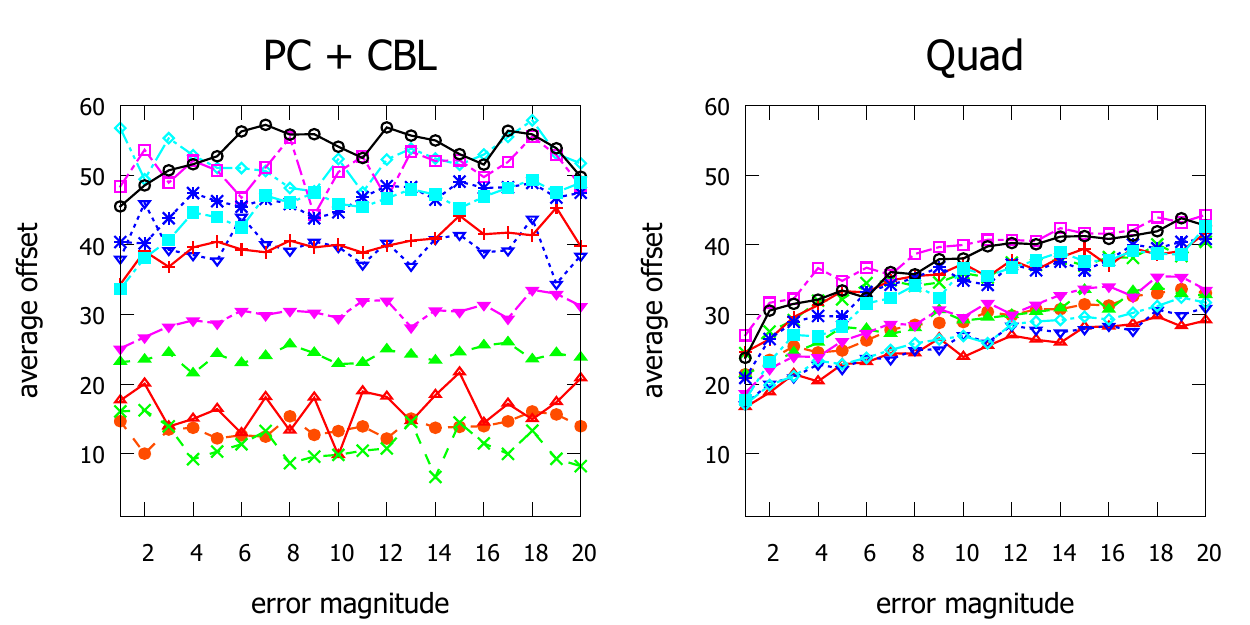}
	\label{fig:pcVSquadOffset}
}

\subfloat[Recall percentages]{
	\includegraphics[width=\linewidth]{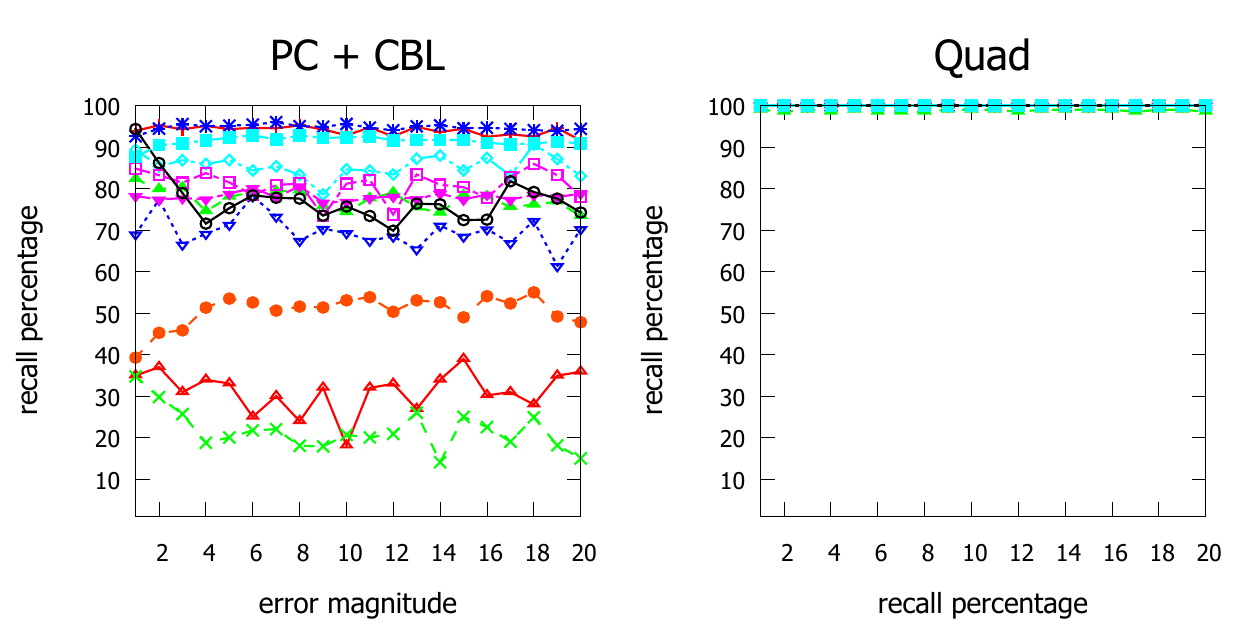}
	\label{fig:pcVSquadRecall}
}

\subfloat{
	\includegraphics[width=0.8\linewidth]{./legend}
}

\caption{Average offsets (a) and recall percentages(b) of CBL and quadrilateration when the information on coplanar clusters is not available}
\label{fig:pcVSquad}
\end{figure}

First, planar clustering algorithm is run to extract coplanar clusters and then CBL is used to localize the network.
We present the results of experiments that we use planar clustering before CBL with title \textbf{PC + CBL}.
The charts where only mere quadrilateration is run are entitled as \textit{Quadrilateration}.
In Figure \ref{fig:pcVSquad}, we see the average offsets and recalls of CBL and quadrilateration when the clustering information is not available.
Figure \ref{fig:pcVSquadOffset} presents the average offsets and Figure \ref{fig:pcVSquadRecall} presents the recall percentages of two algorithms.
The results show us that when CBL is used with the proposed heuristic, we prefer quadrilateration to localize the network.
There is much more room for improvement of this algorithm.
We leave finding the hop distance and volume threshold for an accurate extraction of the coplanar clusters as an open problem.

\chapter{Conclusion}
\label{chap:conclusion}
In this chapter, we give our concluding remarks and future research directions for range-based localization of a 3D WSN that we know to follow a deployment on the planar surfaces.
In Section \ref{sec:relatedwork}, we give the related work. In Section \ref{sec:futurework}, we discuss our work and give the future research topics that we would like to study on.

\section{Related Work}\label{sec:relatedwork}

Range-based localization is basically a graph embedding problem \cite{embed}, which has been shown to be strongly NP-Hard for all dimensions \cite{saxe,pemmaraju}.

In 2004, Aspnes \textit{et al.} \cite{unitdisknphard} showed that the problem is also NP-Hard for unit-disk graphs .
Eren \textit{et. al.} \cite{rigidity} proposed trilateration, which is a polynomial time algorithm that can be used to localize a WSN.
However, trilateration needs exact range measurements to function precisely, which is not the case in real-world due to device errors or environmental noise.
The environmental noise causes the measured distances to be slightly different than the original distances.
Evrendilek and Akcan \cite{nphard} showed that localization through trilateration is NP-Hard when the distance measurements are imprecise.

The imprecise distance measurements cause a node to be localized far from its original position.
This positional offset accumulates at the latter stages of the localization.
In order to reduce this type of ambiguities, Akcan and Evrendilek \cite{reducing} used triangles whose angles are less than a certain value, referred to as robust triangles.

Aspnes \textit{et. al.} \cite{theory} investigated the localization and localizability of a network in 2010.
They defined the term \textit{global rigidity} and showed that it is sufficient and necessary condition for a WSN graph to be localized in 2D.
Even though global rigidity is defined for all dimensions, the sufficient and necessary conditions for a WSN to be localized in 3D have not been found yet.

If the sensor nodes estimate their own positions, \textit{i.e.} self localization is performed, the intractability of the problem signals an excessive amount of computation, potentially depleting the batteries of the self-localizing sensors.
Therefore, not only localizing the sensors, but also completing the process with less number of computations is essential.
Energy efficiency can be achieved either by reducing the energy spent per node \cite{energyefficienthenizelman,efficientlocalizationzhang,energyefficientsrinath}, or by reducing the total number of computations during the localization process \cite{theory,rigidity,reducingstepcountding}.
A review of different approaches of node localization discovery in wireless sensor networks can be found in \cite{survey}.

We study the WSN localization problem using clustering, which is an efficient way to approach the problem \cite{clusteringsurvey}.
In 2000, Amis \textit{et. al} \cite{minmaxdcluster} present a heuristic to
form multi-hop clusters in a dynamic WSN.
In 2002, Chatterjee \textit{et al.} \cite{weightedclustering} proposed a weighted clustering algorithm, that clusters the nodes in a WSN based on the neighbor distances.
In 2004, Demirbas \textit{et. al.}\cite{fastlocalclustering} proposed a clustering method named solid-disc clustering that works in constant time and similar to \cite{weightedclustering}, it is based on the Euclidean distances.
In 2008, Zainalie and Yaghmaee \cite{cfl} designed an algorithm that clusters the sensor nodes with maximum number of nodes in each cluster.
In 2008, Lederer \textit{et. al.} \cite{connectivitybased2008}, developed an algorithm that first partitions the sensor field into Voronoi cells with respect to the apriori given landmark nodes and then extract the combinatorial Delaunay complex as the dual complex of the landmark Voronoi diagram and embed the combinatorial Delaunay complex as a structural skeleton.
In 2009, Yue \textit{et. al.}  \cite{connectivitybased} followed the work done in \cite{connectivitybased2008} and improved the localization quality by a landmark node selection algorithm.

In 2012, Yao \textit{et. al.} \cite{surfacelocalization} studied the localization of a WSN in 3D, in which sensors are deployed on surfaces.
Different than our work, they use a layered approach, that is based upon the nodal height measurements.
Cucuringu \textit{et. al.} \cite{euclideangrouping} developed a non-incremental
non-iterative anchor-free algorithm for localizing sensor networks in 2D, which is experimentally shown to be robust to noisy distance measurements.

We use an extra constraint while clustering, based on the observation that the  sensor nodes usually deploy on surfaces that can be modeled using planes.
In line with the method proposed by Akcan and Evrendilek \cite{dwrl,reducing} in 2013 which uses dual wireless radios, our work localizes the formed structures instead of localizing each node one by one.

\section{Discussion} \label{sec:futurework}
In this section, we conclude this thesis by discussing our contributions to the topic and pointing at the future research directions.

\subsection{Summary of Contributions}
\begin{figure}[htbp]
\centering
\includegraphics[width=0.8\linewidth]{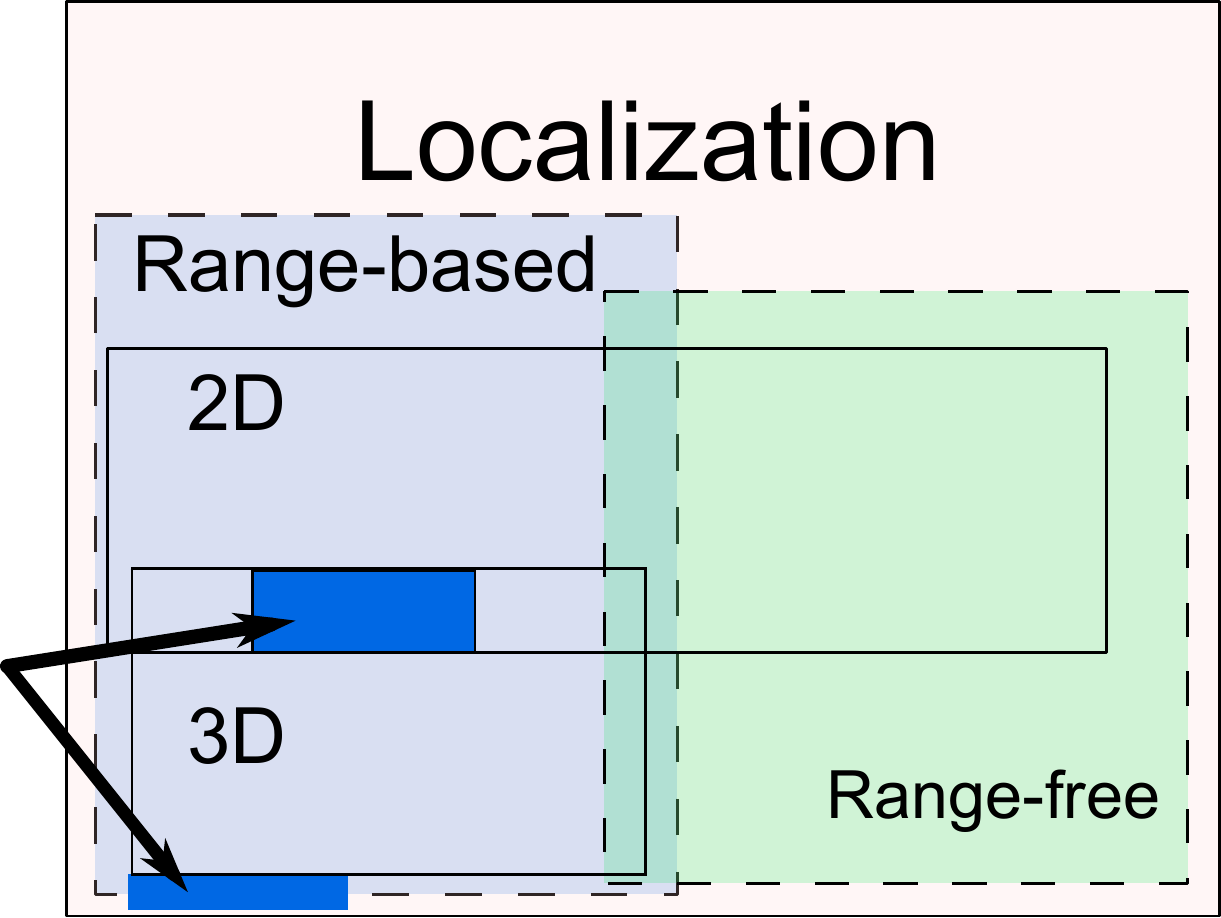}
\caption{This thesis in the state-of-art}
\label{fig:thisthesis}
\end{figure}

This thesis addresses range-based wireless sensor network (WSN)
localization problem in 3D where the sensor nodes sit on planar
surfaces to form rigid 2D structures.
Figure \ref{fig:thisthesis} shows the scope of this thesis in the state-of-art.
We work in the intersection of 2D and 3D localization and make a contribution to 3D localization.

The summary of our contributions can be listed as follows.

\begin{itemize}
\item Our main contribution is the proposed algorithm for range-based WSN localization in Section \ref{sec:localizingclusters}, CBL, that aims to exploit the structural information where the sensor nodes are deployed on planar surfaces in a 3D environment.
CBL can be used as an extension of any localization algorithm.
\item We have shown that if the information on the coplanar clusters is not present, extracting such information is very much needed.
\item Based on the observation above, we have defined the problem of extracting the coplanar clusters and presented a first-attempt heuristic to solve the problem in Section \ref{sec:extractingclusters}. Even though our heuristic fails to extract the information accurately, we have pointed out two parameters, the volume threshold $\kappa$ and the hop distance $\theta$, that need to be set carefully.
\item We have defined a metric called planarity factor to indicate how planar is the deployment of the sensor nodes.
\end{itemize}

\subsection{Future Research Directions}

For our future research, we would like to investigate the following.
\begin{itemize}
\item The most obvious research direction is to discover the relationship among the planarity factor $\mu$, the volume threshold $\kappa$ and the hop distance $\theta$.
\item Another research direction is to investigate the theoretical bounds of the coplanar cluster extraction.
We would like to answer the following questions about the problem:
\begin{itemize}
\item Is the problem NP-Hard?
\item Is the problem approximable?
\item Does the information about the planarity factor change the difficulty of the problem?
\item Given the pairwise distances, can we find values for the volume threshold $\kappa$ and the hop distance $\theta$ based on the average connectivity and the average edge weight of the WSN graph?
\end{itemize}
\item We would like to use CBL with different range-based algorithms than trilateration and quadrilateration to find out if CBL improves the quality of localization with other algorithms.
\end{itemize}

\begin{spacing}{0.9}
\bibliographystyle{ieeetr}
\bibliography{ref}
\end{spacing}
\end{document}